\documentclass[12pt]{amsart}

\usepackage[all]{xy}
\usepackage{amsmath}
\usepackage{amsfonts}
\usepackage{amssymb}
\usepackage{amscd}
\usepackage{amsthm}
\usepackage{latexsym}
\usepackage{amsbsy}
\usepackage{graphicx}
\usepackage{epstopdf}
\usepackage{shuffle}
\usepackage{slashed}
\usepackage{esint}
\usepackage{url}

\input xypic

\usepackage{color}

\numberwithin{equation}{section}

\def\bP{{\mathbb P}}

\def\bS{{\mathbb S}}

\def\N{{\mathbb N}}

\def\Q{{\mathbb Q}}
\def\R{{\mathbb R}}
\def\Z{{\mathbb Z}}

\newcommand{\fh}{{\mathfrak{h}}}

\newcommand{\frf}{{\mathfrak{f}}}

\def\cB{{\mathcal B}}
\def\cC{{\mathcal C}}
\def\cD{{\mathcal D}}

\def\cF{{\mathcal F}}

\def\cI{{\mathcal I}}

\def\cL{{\mathcal L}}
\def\cM{{\mathcal M}}

\def\cP{{\mathcal P}}

\def\cS{{\mathcal S}}

\def\cU{{\mathcal U}}
\def\cV{{\mathcal V}}
\def\cW{{\mathcal W}}

\def\cZ{{\mathcal Z}}

\def\Ker{{\rm Ker}}

\def\Res{{\rm Res}}

\def\Spec{{\rm Spec}}

\def\Tr{{\rm Tr}}

\def\fh{{\mathfrak h}}

\title[Fractality in Cosmic Topology Models]{Fractality in Cosmic Topology Models with Spectral Action Gravity}
\author{Pedro Guicardi and Matilde Marcolli}
\address{Department of Physics and Department of Mathematics, 
Cornell University, Ithaca, NY 14853, USA}
\email{pfg44@cornell.edu}
\address{Department of Mathematics and Department of Computing and Mathematical Sciences, 
California Institute of Technology, Pasadena, CA 91125, USA}
\email{matilde@caltech.edu}

\begin{document}
\maketitle

\begin{abstract}
We consider cosmological models based on the spectral action formulation of (modified) gravity.
We analyze the coupled effects, in this model, of the presence of nontrivial cosmic topology
and of fractality in the large scale structure of spacetime. We show that the topology constrains
the possible fractal structures, and in turn the correction terms to the spectral action due to
fractality distinguish the various cosmic topology candidates, with effects detectable in a slow-roll
inflation scenario, through the power spectra of the scalar and tensor fluctuations. We also
discuss explicit effects of the presence of fractal structures on the gravitational waves equations.
\end{abstract}

\section{Introduction}

The paper is organized in the following way. In this introductory section we review
the (modified) gravity model based on the spectral action functional, and we introduce
the question of cosmic topology and the question of the possible presence of
fractality in the large scale structure of the universe. These two problems have 
so far been investigated as separate questions in theoretical cosmology, while our
goal here is to focus on the {\em combined effects} of the presence of {\em both}
cosmic topology and fractality.

\smallskip

In Section~\ref{FracSpActSec} we present Sierpinski constructions that are
suitable for describing fractality within each of the possible spherical and flat
cosmic topology candidate. These Sierpinski constructions are based on the
polyhedral fundamental domains (in the $3$-sphere or in the $3$-torus) for
these cosmic topology candidates. For each of these cases, we then compute
the spectral action expansion, first in a simplified static model and then in the
actual Robertson--Walker model. We identify the correction terms to the
spectral action that arise due to the presence of fractality and we show the
resulting effect on an associated slow-roll potential arising as a scalar
perturbation of the Dirac operator in the spectral action. We show that the
presence of fractality determines detectable effects that suffice to distinguish
between all the possible cosmic topology candidates. We also discuss other,
Koch-type, forms of fractal growth and we state a general question regarding
the construction of fractal structures on manifolds, to which we hope to return
in future work. 

\smallskip

In Section~\ref{ArithmSec} we discuss the original type of fractal
arrangement of the Packed Swiss Cheese Cosmology model, 
in the case of the sphere topology, which is modelled on Apollonian
sphere packings. We discuss here in particular some arithmetic cases
of packings with integer curvatures, for which the zeta function is
computable explicitly. We show that, in these cases, an interesting
connection emerges between the form of the asymptotic expansion of
the spectral action and series of number theoretic interest, involving
the derivative of the Riemann zeta function at the non-trivial zeros. 
We also discuss how the fractality models based on sphere packings
are not stable under small random perturbations of the geometry,
which would lead to intersecting spheres, while we propose that 
fractal models based on
``cratering" (removal of a random collection of spheres creating
a fractal residual set) can provide a version of these fractal
arrangements of sphere that is stable under the introduction of
some randomness factor. 

\smallskip

In Section~\ref{GrWavesSec} we investigate effects of the presence of
fractality on the gravitational wave equations. In this first section on this
topic, we consider the usual GR equations, without additional
modified gravity terms from the spectral action, and we look at
the kind of model of sphere arrangements that we proposed in
the previous section, where some of the spheres are intersecting.
We show that, in such a model, there is transmission of gravitational
waves between the different spheres across their intersection, 
creating oscillations inside the sphere of intersection. Thus,
multiple gravitational wave sources near the intersection would induce 
a $3$-ball of interfering gravity wave modes on the boundary hypersurface. 

\smallskip

In Section~\ref{GrWavesSec2} we continue investigating effects of
fractality on the gravitational waves. In this section we focus more
specifically on the gravitational waves equation arising from the
spectral action, and we consider the fractal structures investigated in 
Section~\ref{FracSpActSec} for Robertson-Walker spacetimes
associated to the various cosmic topology candidates. We focus on
the effect on the gravitational waves equation of the correction terms
to the spectral action due to fractality, with particular attention to the
contribution of the first term of the series coming from the poles of
the fractal zeta function, the one associated to the pole $s_0$ given by
the self-similarity dimension. We show that the variation of
the metric determines a tensor given by the corresponding variation
of the value at $s_0$ of the zeta function of the Dirac operator on the
Robertson-Walker geometry. This tensor plays the role of an energy-momentum
tensor in the equations, so that the presence of fractality emulates the
presence of a type of matter than only interacts gravitationally but is
otherwise dark.

\subsection{Spectral Action model of gravity}

The spectral action functional as a model of gravity originates in noncommutative geometry \cite{CC},
where it provides a setting for the construction of geometric models of gravity coupled to matter, \cite{CC2}
(see also \cite{WvS} and Chapter~1 of \cite{CoMa} for an overview). 

The main underlying geometric setting is given by {\em spectral triples}, \cite{Co}, which
provide a generalization of Riemannian spin geometry to the noncommutative setting.

A spectral triple $(A,H,D)$ consists of an involutive algebra $A$, with a representation $\pi: A \to B(H)$ as bounded
operators on a complex Hilbert space $H$, and a self-adjoint operator $D$ on a dense domain in $H$, 
with compact resolvent. The compatibility condition between the algebra and the Dirac operator is
expressed as the property that the commutators $[D,\pi(a)]$ are bounded operators for all $a \in A$. 
The {\em spectral action} functional of a spectral triple is defined as
\begin{equation}\label{SpAct}
\cS_\Lambda (D):=\Tr(f(\frac{D}{\Lambda})). 
\end{equation}
Here the Dirac operator is the dynamical variable of the spectral action functional, with
$\Lambda >0$ an energy scale that makes the integral unitless and $f$ is an even smooth
function of rapid decay.

The test function $f$ is a smooth approximation to a cutoff function that regularizes 
the (otherwise divergent) trace of the Dirac operator. 
For example, it is convenient to take the test function to be of the form 
\begin{equation}\label{testf}
 f(x) = \int_0 ^{\infty} e^{-tx^2} \,dm(t) 
\end{equation} 
where $dm(t)$ is some measure on $\R_+$. 

The properties of a spectral triple recalled above include the
requirements that $D$ is self-adjoint and with compact resolvent, which imply the
discreteness of the spectrum $\Spec(D)\subset \R_+$, so that the definition \eqref{SpAct} makes sense.
The main idea here is that gravity is encoded geometrically through the Dirac operator, rather than in the
metric tensor, as in the usual formutation of general relativity. The usual Einstein-Hilbert action
functional of general relativity is recovered from the spectral action through its asymptotic expansion
for large $\Lambda \to \infty$, as we recall in more detail below, along with other higher derivatives terms
that make the spectral action into an interesting {\em modified gravity} model. 

As a prototype (commutative) example of a spectral triple, one should think of $A=\cC^\infty(X)$ as the algebra of smooth 
functions on a compact Riemannian spin manifold $X$, with $H=L^2(X,\bS)$ the Hilbert space of square-integrable spinors, and $D=\slashed{D}_X$ the Dirac operator. This example immediately illustrates two delicate points in the use of the
spectral triples formalism in relation to cosmological model: compactness and Euclidean signature.
Indeed, taken in itself, the spectral action model of gravity that we will summarize in the rest of this
section is a model of Euclidean gravity, which moreover requires a compactness hypothesis. This
seems ill suited for modeling realistic spacetimes. However, as we discuss below, the asymptotic
expansion of the spectral action provides local gravity terms that still make sense when Wick rotated
to Lorentzian signature, and in more general geometries (for example with suitable boundary terms
as discussed in \cite{IoLe}).

Thus, one should view spectral geometry models of gravity in the following way: one uses Wick
rotation to Euclidean signature and compactification 
as a {\em computational device} to achieve the desired 
analytic properties of the Dirac operator that are required for the definition and main features 
of the spectral action functional. The terms then obtained from the expansion of the spectral
action can be interpreted in the original geometry, so that conclusions about cosmological
models can be derived. In this paper we will work with explicit examples constructed using
Riemannian metrics and compactified spacetimes, which will be sufficient to illustrate the
main phenomena we are interested in, through the computation of the spectral action expansion.

The asymptotic expansion of the spectral action is closely related to the associated heat kernel
expansion for the square $D^2$ of the Dirac operator. Assuming that the operator $D^2$ has
a small time heat kernel expansion
\begin{equation}\label{heatker}
\Tr(e^{-tD^2}) \sim_{t\to 0} \sum_{\alpha} a_\alpha (D) t^{\alpha}, 
\end{equation}
using a test function of the form \eqref{testf} gives a corresponding expansion of the
spectral action as
\begin{equation}\label{expandSpAct}
\Tr(f(\frac{D}{\Lambda})) \sim_{\Lambda \to \infty} a_0 f(0) + \sum_\alpha \frf_\alpha c_\alpha \Lambda^{-\alpha},
\end{equation}
with
$$ \frf_\alpha = \left\{ \begin{array}{ll} \int_0^\infty f(v) v^{-\alpha-1} dv & \alpha<0 \\
(-1)^\alpha f^{(\alpha)}(0) & \alpha >0 \, . \end{array} \right. $$
We refer the reader to \cite{CC}, \cite{WvS} for details.

There is a well known Mellin transform relation between the heat kernel and
the zeta function of the Dirac operator,
\begin{equation}\label{Mellin}
|D|^{-s} = \frac{1}{\Gamma(s/2)} \int_0 ^\infty e^{-tD^2} t^{s/2 -1} dt
\end{equation}
so that  
$$\zeta_D (s) = \Tr(|D|^{-s}) \sim \sum_\alpha \frac{a_\alpha}{\Gamma(s/2)} \int_0 ^\infty t^{\alpha + s/2 -1} dt\, . $$
For $\alpha<0$, the values $s=-2\alpha$ correspond to poles of $\zeta_D (s)$ with
$$ \Res_{s=-2\alpha} \zeta_D (s) = \frac{2 a_\alpha}{\Gamma(-\alpha)}. $$
Thus, one can express the coefficients of the leading terms (the case $\alpha<0$) in the spectral action 
expansion in terms of residues of the Dirac zeta function. At each of these residues one has
an associated notion of integration, defined as
 $$ \fint a |D|^{-\beta}:= \Res_{s=\beta} \zeta_{a,D} (s) = \Res_{s=\beta} \Tr(\pi(a) |D|^{-s}), $$
 and the set of points $\beta$ where the zeta functions $\zeta_{a,D} (s)$ have poles is refereed to
 as the {\em dimension spectrum} of the spectral triple. 
In the case of the commutative or almost-commutative geometries used in models of gravity and matter,
these integrals at the points of the dimension spectrum correspond to local terms given by integration
of certain curvature forms on the underlying manifold, which recover the classical form of various
physical action functionals.

In the case of the spectral triple $(\cC^\infty(X), L^2(X,\bS), \slashed{D}_X)$ for a smooth compact Riemannian spin
manifold of dimension $\dim X=4$ the leading terms in the asymptotic expansion of the spectral action are of the forrm
$$Tr(f(\slashed{D}_X/\Lambda)) \sim \frf_2 \Lambda^2 \fint |D|^{-2} + \frf_4 \Lambda^4 \fint |D|^{-4} + f(0)\zeta_D(0)+ o(1),$$
with
$$\frf_2 \Lambda^2 \fint |D|^{-2} = \frac{96 \frf_2 \Lambda^2}{24 \pi^2} \int_X R \,\,  dvol(g), $$
$$\frf_4 \Lambda^4 \fint |D|^{-4} = \frac{48 \frf_4 \Lambda^4}{ \pi^2} \int_X \, dvol(g), $$
$$f(0)\zeta_D(0) = (\frac{\frf_0}{10\pi^2} \int_X \frac{11}{6} R^* R^* - 3C_{abcd}C^{abcd}) \,  dvol(g), $$
where $C_{abcd}$ is the Weyl curvature tensor and $R^* R^*$ is the Gauss-Bonnet term.
Thus, in this case one obtains as leading terms the Einstein-Hilbert action with cosmological term, with
an additional modified gravity term, involving conformal and Gauss--Bonnet gravity. 
The effect of these modified gravity terms on the gravitational waves equations was analyzed in \cite{NOS}.

Cosmological and gravitational models based on the spectral action functional were considered, for instance, in
\cite{BaMa}, \cite{BuFaSa}, \cite{BMK}, \cite{CC4}, \cite{FFM1}, \cite{FFM2}, \cite{FKM}, \cite{FM1}, 
\cite{Ma}, \cite{Ma2}, \cite{Ma3}, \cite{MaPie}, \cite{MPT1}, \cite{MPT2}, \cite{NOS2}, \cite{NeSa}, \cite{Teh}, though
this is certainly not an exhaustive list of references on the subject. The main goal of the present paper is
to continue this investigation by combining the analysis of cosmic topology effects considered in \cite{BMK},
\cite{MPT1}, \cite{MPT2}, with the analysis of the effect of multifractal structures considered in \cite{BaMa} 
and \cite{FKM}.

\subsection{Cosmic Topology}

The problem of cosmic topology in theoretical cosmology aims at identifying possible
signatures of the presence of non-trivial topology on the spatial hypersurfaces of the
spacetime manifold modeling our universe. The main focus of investigation is possible
signatures that may be detectable in the cosmic microwave background (CMB). 
Observational constraints favor a spatial geometry that is either flat or slightly positively 
curved, while the negatively curved case of hyperbolic $3$-manifolds is observationally
disfavoured. Thus, under the assumption of (large scale) homogeneity, the possible
candidate cosmic topologies are either spherical space forms, namely quotients
$Y=S^3/\Gamma$ of the round constant positive curvature $3$-sphere by a finite
group of isometries, or (under the compactness assumption we work with in our
spectral gravity model) the Bieberbach manifolds that are quotients $Y=T^3/\Gamma$ of
the flat $3$-torus by a group of isometries. 

\smallskip

For a general overview of the problem of cosmic topology, we refer the reader to \cite{LaLu},
and also \cite{AuLu}, \cite{Levin}, \cite{LuRou}, \cite{Riaz}. The cosmic topology question
was studied within the framework of spectral action models of gravity in 
\cite{BMK}, \cite{Ma}, \cite{Ma2}, \cite{MPT1}, \cite{MPT2}, \cite{Teh}. 

\smallskip

The main way in which the presence of nontrivial cosmic topology may be detectable,
in the case of a model of gravity based on the spectral action functional, is through a
slow-roll potential for cosmic inflation that arises in the model as a scalar perturbation
of the Dirac operator. 

\smallskip

In a slow-roll inflation model, the power spectra $\cP_s$ and $\cP_t$ of the
scalar and tensor perturbations of the metric have a dependence on the
shape of the slow-roll potential $V$. More precisely, the fluctuations, as a function
of the Fourier modes, obey a power law 
$$ \cP_s(k)\sim \cP_s(k_0) (k/k_0)^{1-n_s+\frac{\alpha_s}{2} \log(k/k_0)} \ \ \
\text{ and } \ \ \  \cP_t(k)\sim \cP_t(k_0) (k/k_0)^{1-n_t+\frac{\alpha_t}{2} \log(k/k_0)}\, , $$
where the exponents $n_s,n_t,\alpha_s,\alpha_t$ are linear functions of the slow-roll 
parameters $\epsilon,\eta,\xi$, whose dependence on the potential $V$ is as follows, up to a
numerical factor that depends on a power of the Planck mass,
$$ \epsilon \sim \left(\frac{V^\prime}{V}\right)^2, \ \ \ 
\eta\sim \frac{V^{\prime\prime}}{V}, \ \ \  \xi\sim \frac{V^\prime V^{\prime\prime\prime}}{V^2} \, . $$
On the other hand, the amplitudes also depend on the shape of the potential as
$$ \cP_s \sim \frac{V^3}{(V^\prime)^2} \ \ \ \text{ and } \ \ \  \cP_t \sim V\, , $$
again up to a constant factor given by a power of the Planck mass, see \cite{SKC} and
also the discussion in Section~1 of \cite{MPT2}.

\smallskip

It is shown in \cite{MPT2} that a
scalar perturbation $\cD^2 \mapsto \cD^2 +\phi^2$ of the Dirac operator in the spectral action
determines a slow-roll potential $V(\phi)$, which exhibits the typical flat plateau of slow-roll
inflation models. The potential is computed explicitly for the candidate cosmic topologies
in both the spherical and the flat case. The result is that the resulting slow-roll 
parameters $\epsilon,\eta,\xi$ are the same within each curvature class, namely
the same for all the spherical space form and the same for all the Bieberbach manifolds,
but different for the two classes, hence they distinguish whether the geometry is flat or
positively curved. Within each class, the amplitudes differ by a factor, which is the
order of the group in the spherical case and a numerical factor depending on the group
and the shape of the fundamental domain in the Bieberbach case. These further amplitude
values do not distinguish all cases. For example, in the spherical case some lens spaces 
can have, by concidence, the same order of the group as one of the other non-isomorphism
spherical space forms.

\smallskip

In this paper, we refine this model by assuming that the universe can exhibit, at the same
time, non-trivial cosmic topology and a fractal structure, where the self-similarity of the
fractal structure now necessarily depends on the topology, through the shape of the
fundamental domain of the candidate topology. We discuss fractality in the following 
subsection. We show in this paper that this combination of cosmic topology and
fractality resolves the ambiguities and the slow-roll potential 
generated by the spectral action now completely distinguished all the possible
cosmic topologies.

\subsection{Multifractal cosmological models}\label{FracIntroSec}

The standard cosmological paradigm, or Cosmological Principle, postulates that (at sufficiently large scales)
the universe should be homogeneous and isotropic. A homogeneous spacetime $(X,g)$ is one that can be foliated by a one-parameter family of spatial hypersurfaces, with the property that any two points in each hypersurface can be transported to the other along an isometry of the metric. In other words, this is the assumption that there is no ``special place" in the universe. The other assumption, isotropy, requires that at any point $p\in X$, given two vectors $v_1,v_2$ orthogonal to a timelike curve passing through $p$, we can find an isometry rotating $v_1$ into $v_2$. 
This corresponds to the idea that there is no ``special direction" in the universe.

A first possible clue that the homogeneity hypothesis may fail even at very large scales came from the distribution 
of galaxy clusters. Statistical analysis in \cite{LMP1} showed fractal correlations up to the observational 
limits. Similarly, in \cite{LMP2} it is shown that all clustering observed have scaling and fractal
properties, which cannot be properly described by small-amplitude fluctuations theory. However, as 
emphasized in \cite{LMP1}, \cite{LMP2}, the presence of fractality in large scale spacetime is still an undetermined 
question in cosmology. A new generation of cosmology probes may be able to give us an answer in larger scales.

A theoretical model of spacetime exhibiting multifractal structures is provided by the Packed Swiss Cheese Cosmology models, see \cite{MuDy}. This model originates from an idea of Rees and Sciama, \cite{ReeSci}, that generated
inhomogeneities by altering a standard Robertson--Walker cosmology through the removal of a configuration of balls. In
the Packed Swiss Cheese Cosmology model this is done so that the resulting spacetime is described by
the geometry of an Apollonian packing of spheres. 

In \cite{BaMa} and \cite{FKM} the Packed Swiss Cheese Cosmology models are analyzed from the point of
view of spectral action gravity and it is shown that the presence of fractality is detectable in the form of
additional terms in the spectral action, coming from poles of the zeta function off the real line and from
the non-integer Hausdorff dimension, as well as in a modification to the effective gravitational and 
cosmological constants of the terms of the action coming from the underlying Robertson--Walker cosmology. 
This analysis is done in \cite{BaMa} in the case of a static model, and in \cite{FKM} in the more general
case of a Robertson--Walker cosmology with a nontrivial expansion/contraction factor. 

One of the mathematical difficulties in describing a model of gravity with fractality lies in the fact that 
one cannot directly describe such spaces in terms of ordinary differential geometry, because of their
fractal nature. One needs a generalization of Riemannian geometry that applies to non-manifold structures 
like fractals. Even though fractals are ordinary commutative spaces, the fact that they are not smooth manifolds
makes them amenable to the tools of noncommutative geometry: although these methods were originally 
designed to apply to noncommutative spaces, they also apply to commutative but non-smooth spaces like fractals.
In particular, fractals such as Apollonian packings of spheres, various Sierpinski-type constructions, Koch curves
and their higher-dimensional analogs, and other such geometries, are well described by the formalism of 
spectral triples, hence they have a well defined gravity action functional given by the spectral action. This shows that
the spectral action model of gravity is the most directly suitable for the description of gravity in the presence of
fractality and of cosmology in fractal spacetimes. 

The main question we focus on in this paper is how the effects of fractality (multifractal cosmologies) 
and of nontrivial topology (cosmic topology) combine in the spectral action model of gravity. We will
consider various relevant examples of fractal geometries, including the Apollonian packings of spheres
and fractal arrangements obtained from various kinds of solids representing fundamental domains of
cosmic topology models, and we will show how the effects of fractality vary in different topological
background and, conversely, how the effect of nontrivial topology of the spacelike hypersurfaces are
affected by the presence of fractality. 

\medskip
\section{Fractality and Topology in static models and in Robertson-Walker models}\label{FracSpActSec}

In this section we consider different candidate cosmic topology models, first in the
spherical and then in the flat case, with two models of spacetime: a simplified static model 
based on a product $Y\times S^1$ with circle compactification, and then a more realistic
Robertson--Walker model on a cylinder $Y\times \R$ with an expansion/contraction
factor $a(t)$. We compute the asymptotic expansion of the spectral action in all of
these cases, along the lines of \cite{BaMa} for the static model and of \cite{FKM} for
the Robertson--Walker case. For every cosmic topology candidate we identify a
corresponding possible fractal structure, based on a Sierpinski construction associated to
the fundamental domain in the $3$-sphere (or the $3$-torus, respectively). We compute how
the presence of fractality affects the spectral action expansion in each of these cases. 
We show that these correction terms coming from fractality
suffice for the spectral action (and the associated slow-roll potential) to distinguish all the
possible cosmic topology candidates. The main mathematical tools in these calculations are
the Poisson summation formula and the Feynman--Kac formula.

\smallskip

At the end of this section we also analyze other models of fractal growth based on
polyhedral domains, modelled on the case of the Koch snowflake. We discuss
Koch-type constructions based on tetrahedra and octahedra and compute the
associated zeta functions. We discuss the related problem of constructing fractal
structures on more general manifolds that are not quotients with polyhedral
fundamental domains. 

\smallskip
\subsection{Static model} The first case we analyze is the static model considered in \cite{BaMa}, where
the spacetime (Euclidean and compactified) is taken to be of the form $S^1_\beta \times \cP$, where
$\beta>0$ is the radius of the circle-compactification and the spacelike hypersurfaces are given by
a fractal arrangement $\cP$. The main case discussed in \cite{BaMa} was with $\cP$ an Apollonian
packing of $3$-spheres, but Section~5 of \cite{BaMa} also discusses an exactly-self-similar model
where $\cP$ is a fractal arrangement of dodecahedra, giving rise to a dodecahedral-space cosmic topology
model with fractality. Our goal in this subsection is to extend the analysis of Section~5 of \cite{BaMa} by
comparing fractal arrangements based on different candidate cosmic topology models and describe
how the different topology affects the contribution of fractality to the spectral action functional. 
From the purely mathematical perspective, the results of this subsection rely essentially on a combination
of earlier results proved in \cite{BaMa}, \cite{BMK}, \cite{MPT1}, \cite{MPT2}. We also correct here 
an inaccuracy in Proposition~5.2 of \cite{BaMa} and we provide a more detailed discussion of the
(weak) convergence of the series of the log-periodic terms.

\medskip
\subsubsection{The fractal arrangements}
The general setting here is the following. Consider a spherical $3$-manifold of the form $Y = S^3 / \Gamma$, where 
$\Gamma$ is a finite group of isometries of $S^3$, which can be identified with the symmetry group of a platonic solid 
$\bP_\Gamma$. 
These manifolds cover the main significant candidate for positively curved cosmic topologies (with the exclusion of
lens spaces, which we will discuss separately).  For each of these candidate models, we 
then construct a Sierpinski fractal, $\mathcal{P}_\Gamma$, using the appropriate construction rules associated with the 
underlying platonic solid $\bP_\Gamma$ with symmetry group $\Gamma$. Each Sierpinski construction yields a {\em scaling factor} and a {\em replication factor}, which we will denote by $\frac{1}{f_\sigma}$ and 
$f_\rho$, respectively. 
We can identify $\bP_\Gamma$ with a fundamental domain of the action of $\Gamma$ on $S^3$ and, with the closed 
$3$-manifold $Y$ obtained by gluing together the faces of this solid according to the action of $\Gamma$. In this way
the fractal arrangement $\cP_\Gamma$ of polyhedra $\bP_\Gamma$ gives rise to a fractal arrangement $\cP_Y$ 
of copies of $Y$.

 \medskip 
 \subsubsection{The Sierpinski construction}
 
\begin{figure}
 \begin{center}
  \includegraphics[scale=0.25]{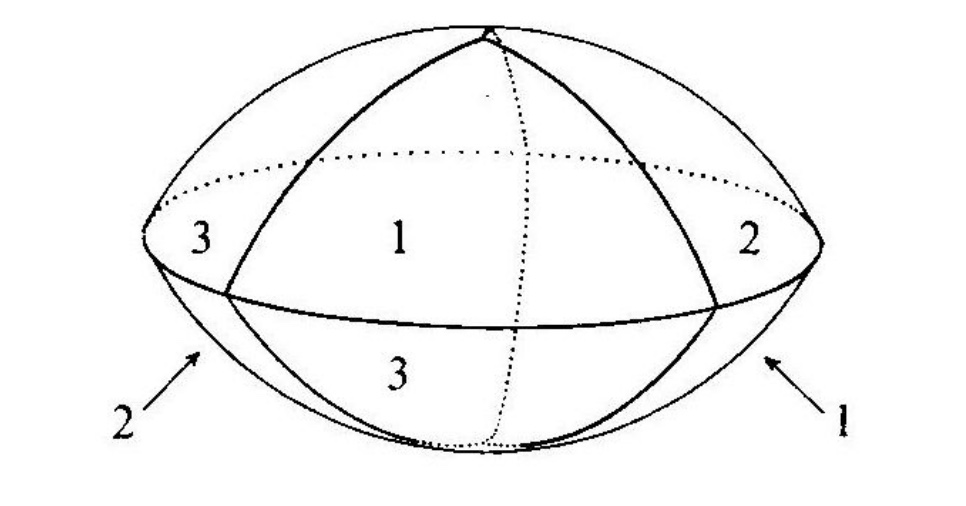}
  \includegraphics[scale=0.25]{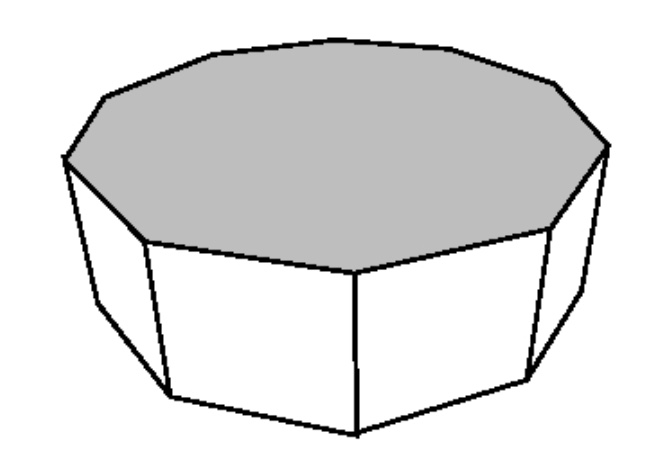} \\
   \includegraphics[scale=0.25]{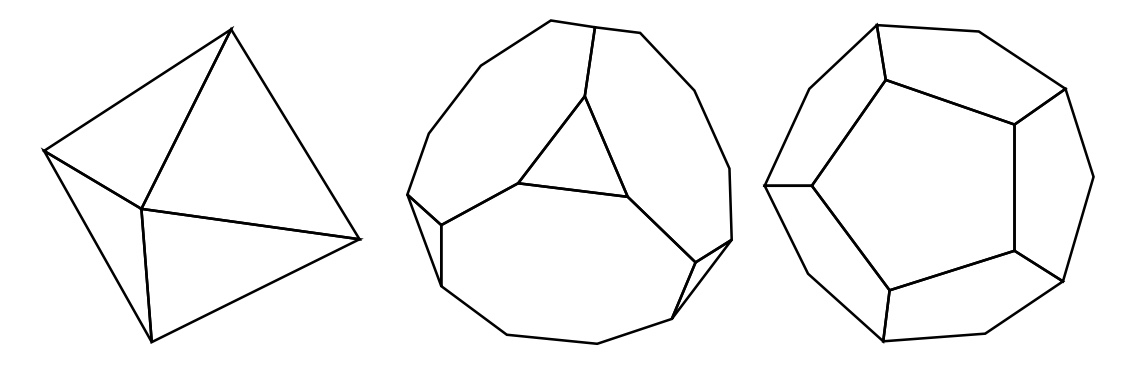}
 \end{center}
 \caption{Fundamental domain for the lens space $L(3,1)$ with face identifications; fundamental domain of 
 the prism manifold of the binary dihedral group $D_5^*$ of order $20$; fundamental
 domains for the binary tetrahedral, binary octahedral, and binary icosahedral group. \label{FigPrism}}
 \end{figure}

 We recall here the main examples of Sierpinski constructions, \cite{JoCa}. Our goal is to compare the
 effects on the spectral action functional and the associated slow-roll potentials of the
 different fractal arrangements associated to different candidate cosmic topologies. 
 We compare the candidate topologies given by spherical forms $Y=S^3/\Gamma$.
 These are quotients by a finite subgroup $\Gamma \subset SU(2)$. According to the
 classification of such subgroups, \cite{Wolf}, there are two infinite families, cyclic groups $Z_n$
 and binary dihedral groups $D^*_n$, and three additional cases, the binary tetrahedral group $T$, 
 binary octahedral group $O$, and binary icosahedral group $\cI$. In the case of cyclic groups $Z_n$ one obtains a lens space $Y=L(n,1)$, with fundamental domain a lens-shaped spherical solid 
 (see Figure~\ref{FigPrism}). The sphere $S^3$ is tiled by $n$ copies of this fundamental domain.
 In the case of the binary dihedral group $D_n^*$, the quotient $Y$ is a prism manifold, where the
 fundamental domain is a prism with $2n$ sides, $4n$ of which tile the sphere (see Figure~\ref{FigPrism}). 
 In the three last cases, the fundamental domain is given, respectively, by a (spherical) octahedron, 
 a truncated cube (cuboctahedron), and a spherical dodecahedron. 
 The sphere $S^3$ is tiled by $24$ copies of the octahedron, by $48$ copies of the truncated cube, 
 and by $120$ copies of the dodecahedron. 
 We describe fractal Sierpinski constructions for various such spaces. We first discuss the
 case of the two regular solids, octahedron and dodecahedron, then the Archimedean case
 (the cuboctahedron), and then the case of lenses and prisms. General Sierpinski
 constructions for polyhedra, and in particular for regular solids, are described in  
 \cite{JoCa}, \cite{Katu}, \cite{Katu2}, \cite{KatuKu}, \cite{Kunn}.
 
 \smallskip
 
  \begin{figure}
 \begin{center}
 \includegraphics[scale=0.25]{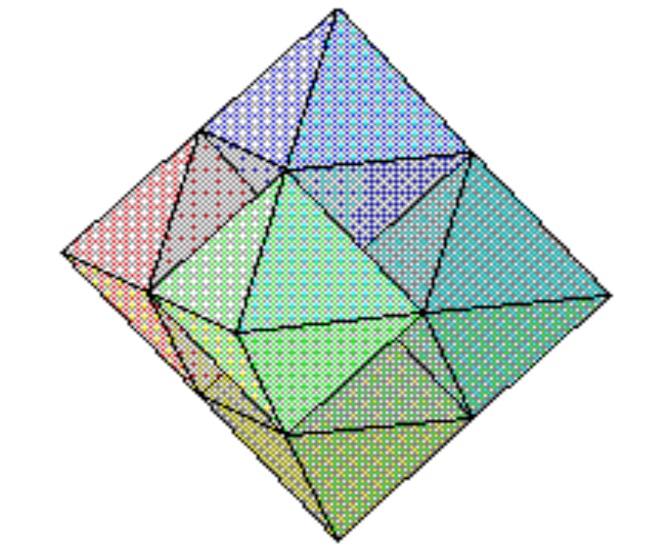}
 \includegraphics[scale=0.25]{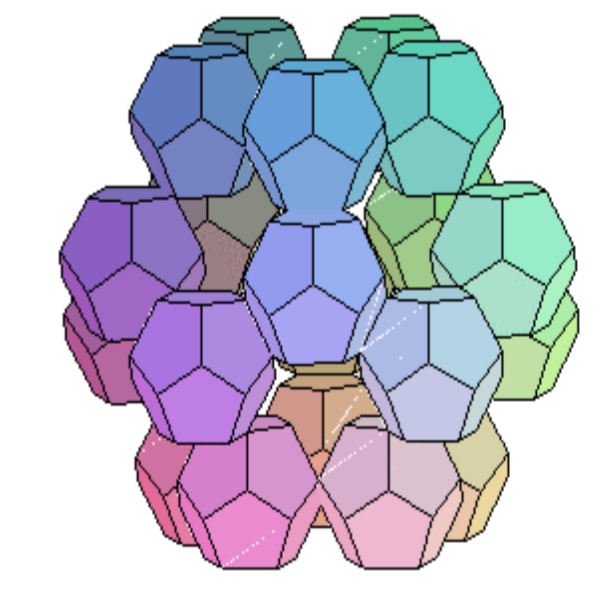}
  \includegraphics[scale=0.28]{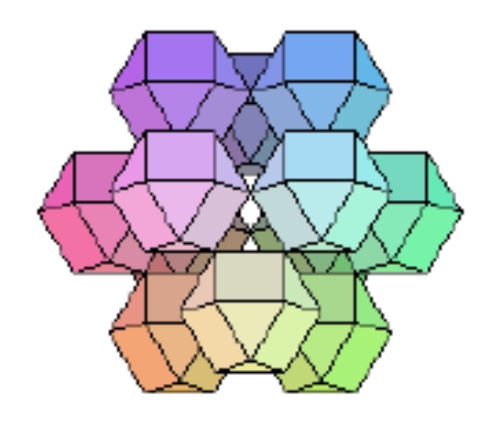}
 \end{center}
 \caption{The first step of the Sierpinski construction for octahedron and dodecahedron, 
 and for the cuboctahedron. \label{FigOcDod}}
 \end{figure}
 
The binary tetrahedral group $T\subset SU(2)$, with $| T |=24$, can be identified with the group of
units in the ring of Hurwitz integers
$$
T= \{\pm 1,\pm i,\pm j,\pm k,{\tfrac {1}{2}}(\pm 1\pm i\pm j\pm k)\}.
$$
Its action on $S^3$ has fundamental domain given by a spherical octahedron.
A fractal arrangement for this case can therefore be obtained by considering 
a Sierpinski construction of octahedra, as illustrated in Figure~\ref{FigOcDod}. 
Each step of this Sierpinski construction replaces each octahedron of the previous
step with $6$ identical copies scaled by a scaling factor of $1/2$. Thus, in this case
we have $f_\rho=6$ and $f_\sigma=2$ and the self-similarity dimension, which for
these regular constructions equals the Hausdorff dimension, is
$$ \dim_H \bP_T = \frac{\log f_\rho}{\log f_\sigma}=\frac{\log 6}{\log 2}  \, . $$
The resulting fractal arrangement $\cP_T$ is the fractal obtained as fixed point of the iteration of this
construction, and the arrangement $\cP_Y$ is then obtained by closing up all the octahedra in $\cP_T$
according to the action of $T$ on its fundamental domain, to give copies of $Y$. 
 
  \smallskip
  
 The other regular solid is the dodecahedron, which is the fundamental domain for
 the binary icosahedral group $\cI$. The first step of a Sierpinski construction of
 dodecahedra is illustrated in Figure~\ref{FigOcDod}. Here at each step the dodecahedra
 of the previous step are replaced with $20$ identical copies scaled by a factor of $(2+\phi)^{-1}$
 where $\phi=\frac{1+\sqrt{5}}{2}$ is the golden ratio. This is the same Sierpinski construction considered in
 \S 5 of \cite{BaMa}. One has $f_\rho=20$ and $f_\sigma=2+\phi$ and
 $$ \dim_H \bP_\cI = \frac{\log f_\rho}{\log f_\sigma}=\frac{\log 20}{\log (2+\phi)} \, . $$
 
 \smallskip
 
 A similar Sierpinski construction can be done for Archimedean solids, \cite{Katu}. In the
 case of the cuboctahedron, which is the fundamental domain of the 
 the binary octahedral group $O$, with $|O|=48$,  the first step of the Sierpinski
 construction is also illustrated in Figure~\ref{FigOcDod}. In this case one replaces
 a cuboctahedron with $12$ identical copies with a scaling factor of $1/3$, so that
 $f_\rho=12$ and $f_\sigma=3$ and
 $$ \dim_H \bP_O = \frac{\log f_\rho}{\log f_\sigma}=\frac{\log 12}{\log 3} \, . $$
 
 \smallskip
 
 \begin{figure}
 \begin{center}
 \includegraphics[scale=0.15]{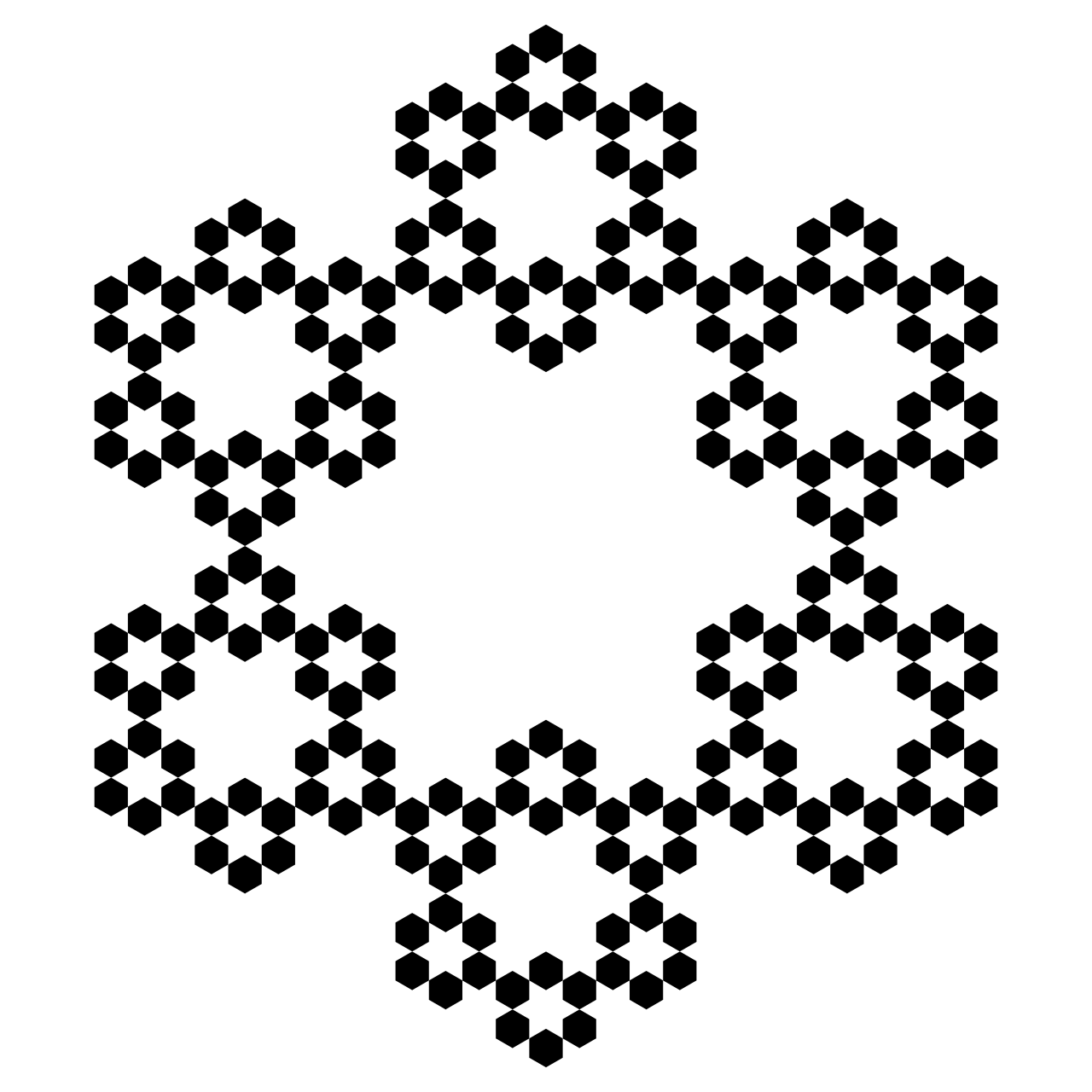}
 \end{center}
 \caption{Second step of the Sierpinski construction of an hexaflake. \label{FigHex}}
 \end{figure}
 
 A Sierpinski construction suitable for handling the case of cyclic groups (lens spaces)
 and of binary dihedral groups (prism manifolds) can be obtained by considering 
 planar Sierpinski constructions for the underlying polygons and obtain from those
 an associated Sierpinski construction for the respective lenses/prisms. 
 In a Sierpinski $n$-polygon (also known as an $n$-polyflake), see \cite{DeSch}, 
 a regular $n$-polygon is replaced by $f_\rho(n)=n$ copies scaled by a factor $1/f_\sigma$ with
 $$ f_\sigma(n) =2 \left( 1+ \sum_{k=1}^{\lfloor n/4 \rfloor} \cos(\frac{2\pi k}{n}) \right). $$
 For example, the second step of the Sierpinski construction of the hexaflake is illustrated
 in Figure~\ref{FigHex}. For the hexaflake $f_\rho(6)=6$ and $f_\sigma(6)=3$. 
 
 Given a Sierpinski $n$-polygon (respectively, $2n$-polygon), 
 we can obtain a fractal arrangement $\bP_{Z_n}$ (respectively, $\bP_{D_n^*}$)
 of lenses/prisms, by building either a lens (double cone) or a prism (product with an interval) over 
 each polygon in the construction and a corresponding fractal arrangement of lens spaces or
 prism manifolds $\cP_Y$, by closing up each lens/prims in the construction according to the face 
 identifications specified by the group action. So we have
 $$ \dim_H \cP_{Z_n}=\frac{\log n}{\log (f_\sigma(n))} \ \ \ \text{ and } \ \ \ 
  \dim_H \cP_{D_n^*} =\frac{\log (2n)}{\log (f_\sigma(2n))} . $$

\medskip
\subsubsection{The spectral triple}
We then construct spectral triples associated to these arrangements $\cP_Y$, as done in \cite{BaMa}, modelled
on the spectral triples of Sierpinski gaskets constructed in \cite{CIL}, \cite{CIS}. 
Namely, to a fractal arrangement $\cP_Y$ where the building blocks are manifolds $Y_{a_n}$ scaled by a
sequence of scaling factors (radii) $a_n$, we assign a spectral triple of the form 
$$(\mathcal{A}_{\mathcal{P}_Y},\mathcal{H}_{\mathcal{P}_Y}, \mathcal{D}_{\mathcal{P}_Y} ) = (\mathcal{A}_{\mathcal{P}_Y},\oplus_n \mathcal{H}_{Y_{a_n}}, \oplus_n \mathcal{D}_{Y_{a_n}}) $$
Where $Y_{a_n} = S^3 _{a_n} / \Gamma$, with $S^3 _{a_n}$ a sphere with radius $a_n$, with the sequence $\{ a_n \}$
of radii determined by the Sierpinski construction.
The algebra $\mathcal{A}_{\mathcal{P}_Y} \subset \oplus_n C(Y_{a_n})$ is given by collections of functions in 
$C(Y_{a_n})$ that agree on all the points of contact between different $Y_{a_n}$ in the packing $\cP_Y$, and each
component Hilbert space is $\mathcal{H}_{Y_{a_n}} = L^2 (Y_{a_n}, \mathbb{S})$. 

\medskip
\subsubsection{The zeta function}
As discussed in \cite{BaMa}, the zeta function of the Dirac operator $\cD_{\cP_Y}=\oplus_n \mathcal{D}_{Y_{a_n}}$
factorizes as a product of the zeta function $\zeta_{\cD_Y}(s)$ of the quotient $Y=S^3/\Gamma$ of the sphere of
radius one, and the zeta function of the fractal string $\cL_\Gamma=\{ a_n \}$ given by the sequence of the 
radii of the packing $\cP_\Gamma$,
\begin{equation}\label{zetaDPY}
\zeta_{\cD_{\cP_Y}}(s) = \zeta_{\cL_\Gamma}(s) \cdot \zeta_{\cD_Y}(s),
\end{equation}
where
\begin{equation}\label{stringzeta}
\zeta_{\cL_\Gamma}(s) := \sum_n a_n^s .
\end{equation}
Indeed, we have
$$ \zeta_{\cD_{\cP_Y}}(s)  = \Tr(|\mathcal{D}_{\mathcal{P}_Y}|^{-s}) 
= \sum_n \Tr( |\cD_{Y_{a_n}} |^{-s})
 = \sum_n a_n^s \, \, \zeta_{\mathcal{D}_{Y}} (s) , $$
 where the last identity follows from the fact that, if the length scales by a factor of $a>0$, the
 Dirac eigenvalues scale by a factor of $a^{-1}$. 
 
 \subsubsection{The fractal string zeta function} 
Thus, in the specific case of the Sierpinski constructions $\cP_\Gamma$ associated to the polyhedra $\bP_Y$,
the sequence of radii of the packing can be described as a set 
$$ \cL_\Gamma =\{  a_{n,k} \,|\, n\in \N, \, \, k\in \{ 1,2,..., f_\rho ^n \} \}, $$
with $f_\rho$ the replication factor of the Sierpinski construction. Moreover, because these
are exact self-similar constructions, where the scaling is uniform across any $n$, we have, at level $n$, 
$$ a_{n,k}= a\cdot f_\sigma^{-n} $$
for some $a>0$ and for all $k=1, \ldots, f_\rho^n$, so that we obtain the very simple expression
\begin{equation}\label{zetaLSierp}
 \zeta_{\cL_\Gamma}(s) = \sum_{n,k} a_{n,k}^s = a^s \sum_{n=1}^\infty f_\rho^n f_\sigma^{-sn} =\frac{a^s}{1- f_\rho f_\sigma^{-s}}  .
\end{equation}
 
 \medskip
 \subsubsection{The manifold zeta function and the spectral action}
 First consider the zeta function $\zeta_{\mathcal{D}_{Y}} (s)$ for the Dirac operator on
 the $3$-manifold $Y=S^3/\Gamma$. Note that, while on $S^3$ there is a unique (trivial)
 spin structure, the manifold $Y$ has different choices of spin structures, each with its
 own Dirac operator, and the Dirac spectrum is dependent on the choice of the spin
 structure. However, as shown in \cite{Teh}, there are cancellations that happen when
 summing over the spectrum that make the resulting spectral action independent of
 the spin structure, so that the spectral action simply satisfies, as $\Lambda\to \infty$
 \begin{equation}\label{SpActQuot}
  \cS_\Lambda(\cD_Y) \sim \frac{1}{|\Gamma|} \, \cS_\Lambda (\cD_{S^3}). 
 \end{equation}
 
 In fact, in \cite{BMK} it is shown that this identity for the spectral action follows from
 the relation between the heat kernel of $\cD_Y^2$ and the heat kernel for $\cD_{S^3}^2$,
 which is of the form (Lemma~3.9 of \cite{BMK})
 \begin{equation}\label{heatkerGamma}
  \Tr(e^{-t \cD_Y^2}) =\frac{1}{|\Gamma|} \Tr(e^{-t \\cD_{S^3}^2})+
 \frac{1}{|\Gamma|} \sum_{\gamma\in \Gamma\smallsetminus\{ e \}} \int_{S^3} {\rm tr}(\rho(\gamma)( x \gamma^{-1}) K_{S^3}(t,x\gamma^{-1},x) dvol_{S^3}(x). 
\end{equation} 
 Note that the Mellin transform relation between heat kernel and zeta function implies that
 the zeta functions $ \zeta_{\mathcal{D}_{Y}} (s)$ and $\frac{1}{|\Gamma|} \,  \zeta_{\mathcal{D}_{S^3}} (s)$
 similarly differ by a term determined by the Mellin transform of the last term in the heat kernel relation.
 As shown in \cite{BMK} (Proof of Theorem~3.6), these terms decay faster than any power of $\Lambda$
 when integrated against a rapidly decaying test function $\phi\in \cS(0,\infty)$
 $$ \int_0^\infty \left( \int_{S^3} {\rm tr}(\rho(\gamma)( x \gamma^{-1}) K_{S^3}(\frac{s}{\Lambda^2},x\gamma^{-1},x) dvol_{S^3}(x)\right) \phi(s) ds = O(\Lambda^{-\infty}), $$
 which implies the identity \eqref{SpActQuot} of the spectral actions for a test function $f(x)=\cL(\phi)(x^2)$,
 written as Laplace transform of a rapidly decaying $\phi$,
 $$ \Tr(f(\frac{D}{\Lambda}))=\int_0^\infty \Tr(e^{-s D^2/\Lambda^2}) \phi(s) ds. $$
 
 We then consider the relation between the leading terms in the expansion of the spectral
 actions  $\cS_\Lambda(\cD_Y)$ and $\cS_\Lambda (\cD_{S^3})$ and the
 poles and residues of the respective zeta functions $\zeta_{\mathcal{D}_{Y}} (s)$
 and $\zeta_{\mathcal{D}_{S^3}} (s)$, namely
 \begin{equation}\label{SpActExpYS3}
 \begin{array}{ll} \displaystyle{
  \cS_\Lambda(\cD_Y) \sim } &  \displaystyle{ \sum \frac{\frf_\alpha \Lambda^\alpha}{2} \Res_{s=\alpha}\zeta_{\mathcal{D}_{Y}}
 + f(0) \zeta_{\mathcal{D}_{Y}}(0) } \\[3mm]
   \displaystyle{ \cS_\Lambda(\cD_{S^3}) \sim } &  \displaystyle{ \sum \frac{\frf_\alpha \Lambda^\alpha}{2} \Res_{s=\alpha}\zeta_{\mathcal{D}_{S^3}} + f(0) \zeta_{\mathcal{D}_{S^3}}(0) \, , } \end{array} 
 \end{equation}  
where the sums are over poles with $\alpha >0$ of the respective zeta functions.
Direct comparison between \eqref{SpActExpYS3} and \eqref{SpActQuot} gives  
$$ \cS_\Lambda(\cD_Y) \sim \sum \frac{\frf_\alpha \Lambda^\alpha}{2 |\Gamma|} \Res_{s=\alpha} \zeta_{\mathcal{D}_{S^3}} + \frac{f(0)}{|\Gamma|} \zeta_{\mathcal{D}_{S^3}}(0)   = \Lambda^3 \frac{\frf_3}{|\Gamma|} - \Lambda \frac{\frf_1}{4|\Gamma|}, $$
with the two terms corresponding to the poles $\alpha=3$ and $\alpha=1$ of
\begin{equation}\label{zetaS3}
  \zeta_{\mathcal{D}_{S^3}}(s)=2\zeta(s-2,\frac{3}{2}) -\frac{1}{2} \zeta(s,\frac{3}{2}), 
\end{equation}  
where $\zeta(s,q)$ is the Hurwitz zeta function. The remaining term is $\zeta_{\mathcal{D}_{S^3}}(0)=0$, since it is equal to $- \frac{2}{3} B_3(3/2)-\frac{1}{2} \zeta(0,\frac{3}{2})=0$, with $B_3(x)$ the third Bernoulli polynomial.

Similarly, we obtain for the spectral action expansion for the Dirac operator of the fractal packing
$\cD_{\cP_Y}$
\begin{equation}\label{SpActPY}
\cS_\Lambda(\cD_{\cP_Y}) \sim
\sum_\alpha \frac{\frf_\alpha \Lambda^\alpha\, \zeta_{\cL_\Gamma}(\alpha)}{2 |\Gamma|} \Res_{s=\alpha} \zeta_{\mathcal{D}_{S^3}} + 
\sum_\beta \frac{\frf_\beta \Lambda^\beta\, \zeta_{\mathcal{D}_{S^3}}(\beta)}{2 |\Gamma|} 
\Res_{s=\beta} \zeta_{\cL_\Gamma} 
\end{equation}
 where we used the identity \eqref{zetaDPY} of the zeta functions. Here the first sum ranges over the
 poles $\alpha >0$ of the zeta function $\zeta_{\mathcal{D}_{S^3}}$ and the second sum ranges over
 the poles with $\Re(\beta)>0$ of the zeta function $\zeta_{\cL_\Gamma}$, where again we do not have the 
 zero-order term as $\zeta_{\mathcal{D}_{S^3}}(0)=0$. This gives
 \begin{equation}\label{SpActPY2}
\cS_\Lambda(\cD_{\cP_Y}) \sim
\Lambda^3 \frac{\frf_3\, \zeta_{\cL_\Gamma}(3)}{|\Gamma|} - \Lambda \frac{\frf_1\, \zeta_{\cL_\Gamma}(1)}{4|\Gamma|} + 
\sum_\beta \frac{\frf_\beta \Lambda^\beta\, \zeta_{\mathcal{D}_{S^3}}(\beta)}{2 |\Gamma|} 
\Res_{s=\beta} \zeta_{\cL_\Gamma} \, ,
\end{equation}
where the terms in the last sum depend on the Sierpinski construction. 

Using the explicit form of the fractal string zeta function from \eqref{zetaLSierp} we
see that the zeta function $\zeta_{\cL_\Gamma}$ as poles where
$$  f_\rho f_\sigma ^{-s} = 1 \Leftrightarrow e^{-s \log f_\sigma + \log f_\rho}=e^{2\pi i n}, \ \ \  n\in \Z, $$
which gives the set of poles
\begin{equation}\label{polesZL}
 \Sigma_{\cL_\Gamma}=\left\{ s_n := \frac{\log f_\rho}{\log f_\sigma} +  \frac{2\pi i n}{\log f_\sigma}\, \bigg| \, n\in \Z \right\}.
\end{equation}
The pole on the real line is also the value of the Hausdorff dimension (self-similarity dimension) of the fractal string
$$ s_0 = \frac{\log f_\rho}{\log f_\sigma} =\dim_H \cL_\Gamma. $$
The residues are given by
\begin{equation}\label{ResZL}
 \Res_{s=s_n} \zeta_{\cL_\Gamma} =\Res_{s=s_n} \frac{a^s}{(s-s_n)\log f_\sigma} = \frac{a^{s_n}}{\log f_\sigma}.
\end{equation} 

Thus, we obtain an overall expression for the spectral action expansion of the form
\begin{equation}\label{SpActPYall}
\cS_\Lambda(\cD_{\cP_Y}) \sim
\Lambda^3 \frac{\frf_3\, \zeta_{\cL_\Gamma}(3)}{|\Gamma|} - \Lambda \frac{\frf_1\, \zeta_{\cL_\Gamma}(1)}{4|\Gamma|} + \sum_{n\in \Z} \Lambda^{s_n} \,  \frac{\frf_{s_n} \, a^{s_n} \,\zeta_{\mathcal{D}_{S^3}}(s_n)}{2 |\Gamma|\, \log f_\sigma} 
\end{equation}
$$ \sim  \frac{ \Lambda^3\, a^3\, \frf_3}{|\Gamma| (1-f_\rho f_\sigma^{-3})} -  \frac{ \Lambda\, a\, \frf_1}{4|\Gamma|(1-f_\rho f_\sigma^{-1})} + \sum_{n\in \Z}  \frac{\Lambda^{s_n} \, \frf_{s_n} \,a^{s_n} \,\zeta_{\mathcal{D}_{S^3}}(s_n)}{2 |\Gamma|\, \log f_\sigma} \, . $$

\smallskip
\subsubsection{Weak convergence}\label{WeakConvSec}

Note that in general asymptotic series are {\em not} convergent series and one relies on
truncations to get approximations (see Section~4 of \cite{BaMa} for a discussion of truncations). 
Nonetheless, we can analyze
more closely the behavior of the series on $n\in \Z$ in \eqref{SpActPYall}.
We can equivalently write the series in the form
$$ \frac{(\Lambda a)^{s_0} }{2 |\Gamma|\, \log f_\sigma} \, \, \sum_{n\in \Z} \frf_{s_n} \, \zeta_{\mathcal{D}_{S^3}}(s_n)\, 
\exp \left( 2\pi i n \frac{ \log (\Lambda a) }{ \log f_\sigma }  \right)  \, . $$
We can assume the momenta $\frf_{s_n}$ of the test function are uniformly bounded. Thus,
by \eqref{zetaS3}, the behavior of this series depends on the behavior of the Hurwitz zeta function
$\zeta(s,3/2)$ along the vertical lines $$ L=\{ s=s_0+it\,|\, t\in \R \} \ \ \text{ and } \ \  L'=\{ s=s_0-2+it \,|\, t\in \R \}. $$
The asymptotic behavior of the Hurwitz zeta function along vertical lines satisfies (see Lemma~2 of \cite{Kat})
\begin{equation}\label{HurAs}
\mu_{s_0,q} = \limsup_{t\to \pm \infty} \frac{|\zeta(s_0+it,q)|}{\log |t|} \leq \left\{ \begin{array}{ll}
\frac{1}{2}-s_0 & s_0<0 \\[2mm] \frac{1}{2}(1-s_0) & 0\leq s_0 \leq 1 \\[2mm] 0 & s_0\geq 1 . 
\end{array}\right. 
\end{equation}
In particular, the sequences $\zeta(s_n,3/2)$ and $\zeta(s_n-2,3/2)$ are bounded 
when the self-similarity dimension is in the range $s_0\geq 3$. In all cases, because of
the estimate \eqref{HurAs} the coefficients of the series satisfy, for sufficiently large $|n|$,
$$ | \zeta_{\mathcal{D}_{S^3}}(s_n) |\leq 2 \bigg| \zeta(s_0-2 + i \frac{2\pi}{\log f_\sigma} n , \frac{3}{2}) \bigg|+\frac{1}{2} \bigg|
 \zeta(s_0 + i \frac{2\pi}{\log f_\sigma} n , \frac{3}{2}) \bigg|    $$
$$  \leq \frac{5}{2} \left( \frac{2\pi}{\log f_\sigma} |n| \right)^\mu \leq C |n|^N $$
with $\mu=\max\{ \mu_{s_0,3/2}, \mu_{s_0-2,3/2} \}$ and $N\geq \lceil \mu \rceil$. 

We can interpret the series 
$$ \sum_{n\in \Z} \zeta_{\mathcal{D}_{S^3}}(s_n)\, 
\exp \left( 2\pi i n \frac{ \log (\Lambda a) }{ \log f_\sigma } \right) $$
as a Fourier series $\sum_{n\in \Z} c_n e^{i n x}$ with $x=\frac{2\pi \log (\Lambda a) }{ \log f_\sigma }$. 
The condition
$$ |c_n| \leq C\, |n|^N $$
for some $C>0$ and some $N\in \N$ ensures the {\em weak convergence} of the Fourier series
to a periodic distribution (see Theorem~9.6 of \cite{Foll}). 
For example, the periodic delta distribution is the weak limit of the Fourier series 
$$ \delta_{per}(x):=\sum_{n\in \Z} \delta(x-2\pi n) \stackrel{\text{weak lim}}{=}
 \frac{1}{2\pi} \sum_{n\in \Z} e^{in x}. $$
 
 Thus, we can write 
\begin{equation}\label{SpActPYall2}
\cS_\Lambda(\cD_{\cP_Y}) \sim
\Lambda^3 \frac{\frf_3\, \zeta_{\cL_\Gamma}(3)}{|\Gamma|} - \Lambda \frac{\frf_1\, \zeta_{\cL_\Gamma}(1)}{4|\Gamma|} +  \frac{(\Lambda a)^{s_0} }{2 |\Gamma|\, \log f_\sigma} \, \, \Theta_{\cP_\Gamma}( \frac{\log (\Lambda a) }{ \log f_\sigma }) \, , 
\end{equation}
where we write
\begin{equation}\label{ThetaP}
\Theta_{\cP_\Gamma}( \frac{\log (\Lambda a) }{ \log f_\sigma })  \stackrel{\text{weak lim}}{=} \sum_{n\in \Z} \zeta_{\mathcal{D}_{S^3}}(s_n)\, 
\exp \left( 2\pi i n \frac{ \log (\Lambda a) }{ \log f_\sigma } \right) 
\end{equation}
for the log-periodic distribution defined by the weak limit of the Fourier series. 
 
 \medskip
 \subsubsection{The static spacetime action} 
 
 We consider a product geometry of the form $S^1_\beta \times \cP_Y$, with $Y=S^3/\Gamma$ and 
 $\cP_Y$ the fractal arrangement resulting from the Sierpinski construction for the polyhedra $\bP_Y$,
 as discussed above, and with $S^1_\beta$ a circle of some (large) compactification radius $\beta>0$. 
 The Dirac operator on this product geometry is of the form
\begin{equation}\label{DiracS1prod}
\mathcal{D}_{S^1_\beta \times \cP_Y} = \begin{pmatrix} 0&\mathcal{D}_ {\cP_Y}\otimes1 + i \otimes  \mathcal{D}_ {S^1_\beta} \\ \mathcal{D}_ {\cP_Y}\otimes1 - i\otimes \mathcal{D}_ {S^1_\beta}&0 \end{pmatrix} \, .
\end{equation}

The spectral action for this product geometry can be computed with the same method used in \cite{CC3}, 
given the previous computation of the spectral action for $\cP_Y$.  The heat kernel of the operator
$\mathcal{D}^2_{S^1_\beta \times \cP_Y}$ of \eqref{DiracS1prod} satisfies
\begin{equation}\label{kerprod}
 \Tr(e^{-s \mathcal{D}^2_{S^1_\beta \times \cP_Y}/\Lambda^2} )= 2 \, \Tr(e^{-s \mathcal{D}^2_{S^1_\beta}/\Lambda^2}) \,\,
\Tr(e^{-s \mathcal{D}^2_{\cP_Y}/\Lambda^2} ) \, .
\end{equation}
 The spectrum of $\mathcal{D}_ {S^1_\beta}$ is $\beta^{-1}(\Z+1/2)$, and the leading part of the heat kernel expansion is given by, for all $k>0$ 
$$  \Tr(e^{-s \mathcal{D}^2_{S^1_\beta}/\Lambda^2}) \sim  \sqrt{ \frac{\pi}{s} } \beta \Lambda 
+ O(\Lambda^{-k})  \, . $$
Thus, we have, for all $k>0$
$$ \Tr(e^{-s \mathcal{D}^2_{S^1_\beta \times \cP_Y}/\Lambda^2} )= 2 \beta \Lambda \, \Tr(\sqrt{\pi/s}\,\, e^{-s \mathcal{D}^2_{\cP_Y}/\Lambda^2} ) + O(\Lambda^{-k+3}) \, .$$
Thus, we obtain 
\begin{equation}\label{SA4d3d}
\Tr(h(\frac{\mathcal{D}_{S^1_\beta \times \cP_Y}^2}{\Lambda^2})) \sim 2\beta \Lambda \Tr(\kappa (\frac{\mathcal{D}_{\cP_Y}^2}{\Lambda^2}))\, , 
\end{equation}
for test functions $h(x)=e^{-sx}$ and $k(x)=\sqrt{\pi/s}\,\, e^{-s x}$ and more generally for a test
function $h(x)$ and $k(x)=\int_0^\infty v^{-1/2} h(x+v) dv$, see Lemma~2 of \cite{CC3}.

Thus, we obtain, as in Proposition~3.6 of \cite{BaMa}, an expansion of the spectral
action of $\mathcal{D}_{S^1_\beta \times \cP_Y}$ of the form
\begin{equation}\label{SpActS1P}
\cS_\Lambda(\mathcal{D}_{S^1_\beta \times \cP_Y}) \sim \frac{\beta}{|\Gamma|} 
\left(\Lambda^4 \,2\, \zeta_{\cL_\Gamma}(3)\, \fh_3 -  \frac{\Lambda^2}{2} \zeta_{\cL_\Gamma}(1)\, \fh_1 + 
\sum_{n\in \Z}  \frac{\Lambda^{s_n +1}\,a^{s_n}\, \zeta_{\cD_{S^3}}(s_n)}{\log f_\sigma}\, \fh_{s_n} \right) 
\end{equation}
$$ \sim  \frac{\beta}{|\Gamma|} 
\left( \,\frac{2 \Lambda^4\, a^3}{(1-f_\rho f_\sigma^{-3})}\, \, \fh_3 -   \frac{\Lambda^2\, a}{2 (1-f_\rho f_\sigma^{-1})} \, \fh_1 + 
\sum_{n\in \Z}  \frac{\Lambda^{s_n +1}\,a^{s_n}\, \zeta_{\cD_{S^3}}(s_n)}{\log f_\sigma}\, \fh_{s_n} \right)\, , 
$$ 
where
$$ \fh_3:=\pi \int_0^\infty h(\rho^2) \rho^3 \,d\rho, \ \ \ \ \fh_1:=2\pi  \int_0^\infty h(\rho^2) \rho \,d\rho, \ \ \ \
\fh_{s_n}= 2 \int_0^\infty h(\rho^2) \rho^{s_n} d\rho. $$

\medskip
\subsubsection{Slow roll potentials}

Correspondingly, if we consider a scalar perturbation of the Dirac operator, of the form
$$ \cD^2 \mapsto \cD^2 + \phi^2, $$
the spectral action computed above and the one for this modified Dirac operator differ
by a Lagrangian density for the scalar field $\phi$, which in the case of these static
models only contains a potential term of the form 
\begin{equation}\label{sphLagPhi}
 \cL(\phi)=A\,\, \Lambda^4\,\, \cV(\phi^2/\Lambda^2) + B \,\, \Lambda^2\, \cW(\phi^2/\Lambda^2) 
+\sum_{n\in \Z} C_n \,\, \Lambda^{s_n+1}\,\, \cU_n(\phi^2/\Lambda^2) , 
\end{equation}
where 
\begin{equation}\label{ABCn}
 A= \frac{\pi \beta\, a^3}{|\Gamma| (1-f_\rho f_\sigma^{-3})}, \ \ \ \
B =- \frac{\pi \beta\, a}{2|\Gamma| (1-f_\rho f_\sigma^{-1})}, \ \ \ \ 
C_n = \frac{2\beta \, a^{s_n}\, \zeta_{\cD_{S^3}}(s_n)}{|\Gamma| \log f_\sigma} 
\end{equation}
and 
\begin{equation}\label{slowrollS3}
 \cV(x):=\int_0^\infty u(h(x+u)-h(u))du, \ \ \ \ \cW(x):=\int_0^x h(u)du, 
\end{equation}
\begin{equation}\label{slowrollfrac}
 \cU_n(x)=\int_0^\infty u^{(s_n-1)/2} (h(x+u)-h(u)) du, 
\end{equation} 
see \cite{BaMa},  and also \cite{BMK}, \cite{CC3}, \cite{MPT1}, \cite{MPT2}.

\smallskip

One can see the behavior of the potentials $\cV(\phi^2/\Lambda^2)$, 
$\cW(\phi^2/\Lambda^2)$, $\cU_n(\phi^2/\Lambda^2)$, and of the resulting
Lagrangian $\cL(\phi)$, in the regime where $\phi/\Lambda$ is small, in
the following way. 

\smallskip

Consider the potential $\cU_n(x)$ as in \eqref{slowrollfrac}. 
A variable substitution yields 
$$\int_0^\infty (v-x)^{(s_n-1)/2} h(v) dv - \int^x_0 (v-x)^{(s_n-1)/2} h(v) dv - \int_0^\infty u^{(s_n-1)/2} h(u) du \, . $$
If we assume we have a function that is approximately constant on the interval $[0,x]$, such as a smooth 
approximation to a cutoff function in the range where $x$ is sufficiently small, the middle term simplifies to
$$\frac{2 h(0) e^{i\pi (s_n+1)/2}}{s_n +1} x^{(s_n+1)/2}\, . $$

\smallskip

Additionally, using the binomial expansion for complex coefficients, the first and third terms can be written as
$$\sum_{j=1}^\infty {(s_n -1)/2\choose j}(-1)^j \,\cF_j \, x^j \, , $$
$$\cF_j = \int_0^\infty v^{(s_n -1)/2 -j} \, h(v)\,  dv \, . $$
The factor of $\Lambda^{s_n+1}$ implies that, in the limit $\Lambda \rightarrow \infty$, we may write
$$\cU_n (\phi^2 /\Lambda^2) \sim \sum_{j=1}^{\left \lfloor{(s_n+1)/2}\right \rfloor} {(s_n -1)/2\choose j}\cF_j \cdot (-1)^j  (\phi/\Lambda)^{2j} + \frac{2 h(0) e^{i\pi (s_n+1)/2}}{s_n +1} (\phi/\Lambda)^{(s_n+1)}\, . $$
The contribution of the real pole $s_0$ of the fractal string zeta function then becomes
$$\frac{2\beta a^{s_0} \zeta_{\mathcal{D}_{S^3}}(s_0) \Lambda^{s_0 +1}}{|\Gamma| \cdot log f_\sigma} \mathcal{U}_0 (\phi^2 / \Lambda^2) \sim  \sum_{j=1}^{\left \lfloor{(s_0+1)/2}\right \rfloor} \mathfrak{a}_j \Lambda^{s_0+1-2j} \phi^{2j} + \mathfrak{b}_0 \phi^{s_0+1} \, , $$
$$\mathfrak{a}_j = C_0 {(s_0 -1)/2\choose j} (-1)^j \mathcal{F}_j  , \ \ \ \ \ 
\mathfrak{b}_0 = C_0 \frac{2 h(0) e^{i\pi (s_0+1)/2}}{s_0 +1}\, , $$
with $C_0$ as in \eqref{ABCn}. Under the same assumptions, denoting $\int^\infty_0 h(v)dv = |h|$, we also have
$$\cV(\phi^2/\Lambda^2) =-|h| \frac{\phi^2}{\Lambda^2} -\frac{h(0)}{2 \Lambda^4} \phi^4$$
$$\cW(\phi^2/\Lambda^2) = h(0) \frac{\phi^2}{\Lambda^2} \, ,$$
in the range where $\phi/\Lambda$ is sufficiently small. 

\smallskip

Thus, in this range the Lagrangian can be written in the form
$$\cL(\phi) \sim (B-A |h|) \Lambda^2 \phi^2 -\frac{h(0)}{2} \phi^4 + \mathfrak{b}_0 \phi^{s_0+1} +  \sum_{j=1}^{\left \lfloor{(s_0+1)/2}\right \rfloor} \mathfrak{a}_j \Lambda^{s_0+1-2j} \phi^{2j} +\mathcal{O}(\Lambda,\phi)\, , $$
where $\mathcal{O}(\Lambda,\phi)$ is the contribution to the complex poles of fractal zeta function, which as discussed in \S \ref{WeakConvSec} converges in a weak sense, 
$$\mathcal{O}(\Lambda,\phi) = \sum_{n\in\mathbb{Z}-\{0\}} C_n \Lambda^{s_n +1} \mathcal{U}_n(\phi^2/\Lambda^2)\, . $$

\smallskip

The behavior of the Lagrangian in the small $\phi/\Lambda$ allows us to compute the effective mass term of the field $\phi$,
by comparing the quadratic term in our Lagrangian to the mass term $-\frac{1}{2}m\phi^2$ and taking the cut-off function 
to be normalized (namely, $|h|=1$). We obtain an effective mass of the form 
$$m_{{\rm eff}} = \frac{\pi \beta}{|\Gamma|} (\frac{2 a^3}{1-f_\rho f_\sigma^{-3}} + \frac{a}{1-f_\rho f_\sigma^{-1}}) \, . $$

\smallskip

In the range where $\phi^2/\Lambda^2$ is large, the approximations considered above no longer apply, and
the potentials $\cV$, $\cW$, and $\cU_n$ flatten out to an asymptotic plateau (see the corresponding discussion
in \cite{CC3}  and in \S 3.2 of \cite{MPT2}). It is this plateau behavior for large $\phi^2/\Lambda^2$ that makes
the resulting $\cL(\phi)$ suitable for describing a slow-roll inflation model in cosmology. 

\smallskip
\subsubsection{Fractality and cosmic topology in static spherical models}
In \cite{MPT1} it was shown that the different candidates for spherical cosmic
topology are distinguished, in a spectral action model of gravity, only through the
different scaling factors $|\Gamma|$ in the spectral action, which in turn gives
a different scaling factor in the amplitude of the power spectra determined by
the slow-roll potential. However, there are some ambiguities that cannot be
resolved by this simple dependence on the scaling factors $|\Gamma|$. For
example, the lens space $L(24,1)$, the prism manifold of the binary dihedral
group $D^*_6$, and the spherical manifold of the binary tetrahedral group
will all have the same scaling factor $1/|\Gamma|=1/24$. So the action
functional with slow-roll potential given by the spectral action will not
distinguish between these cases, and other similar situations. However, in
the presence of {\em both} cosmic topology {\em and} fractality, the dependence
on the topology of the spectral action model of gravity is more subtle, as the
type of Sierpinski fractal arrangement itself depends on the topology, so that
the contributions of fractality to the spectral action and the slow-roll potential
will themselves depend explicitly on the topology. 

\smallskip

To see the effect explicitly, it suffices to consider the contribution of the
real pole $s_0$ of the zeta function $\zeta_{\cL_\Gamma}$ to the spectral
action and the slow-roll potential. These are, respectively, given by the
expressions
\begin{equation}\label{s0terms}
\frac{\Lambda^{s_0} \zeta_{\cD_{S^3}}(s_0)}{2 |\Gamma| \log f_\sigma} \ \ \ \text{ and } \ \ \ 
\frac{2\beta \Lambda^{s_0+1} \zeta_{\cD_{S^3}}(s_0)}{|\Gamma| \log f_\sigma} \int_0^\infty u^{(s_0-1)/2} (h(\frac{\phi^2}{\Lambda^2}+u)-h(u))du  
\end{equation}

The following table of values shows that in this case, even for cases where $|\Gamma$ has
the same value, other terms are different, so that the terms above distinguish the different
spherical topologies. Notice moreover, that the shape of the potential $\cU_0(\phi^2/\Lambda^2)$ 
given by the integral above, also depends explicitly on the value of $s_0$. This means that
not only the amplitude of the power spectra determined by
the slow-roll potential has a dependence on the topology, but also the slow-roll parameters
will now depend on the topology, due to the presence of fractality.

\bigskip

\noindent
{\small
\begin{tabular}{|c||c|c|c|c|} \hline 
 $\Gamma$ &  $|\Gamma|$ &  $s_0$ &  $\zeta_{\cD_{S^3}}(s_0)$ &  $\log f_\sigma$  \\ 
\hline
$Z_n$ & $n$ & $\frac{\log n}{\log f_\sigma(n)}$ & $2\zeta( \frac{\log n}{\log f_\sigma(n)}-2,\frac{3}{2})-\frac{1}{2} \zeta(\frac{\log n}{\log f_\sigma(n)},\frac{3}{2})$ & $\log (f_\sigma(n))$  \\ 
\hline
$D_n^*$ & $4n$ & $\frac{\log n}{\log f_\sigma(2n)}$ & $2\zeta( \frac{\log n}{\log f_\sigma(2n)}-2,\frac{3}{2})-\frac{1}{2} \zeta(\frac{\log n}{\log f_\sigma(2n)},\frac{3}{2})$ & $\log (f_\sigma(2n))$  \\ 
\hline
$T$ & $24$ & $\frac{\log 6}{\log 2}\sim 2.5849$ & $\zeta_{\cD_{S^3}}(\frac{\log 6}{\log 2})  \sim-5.1396$ & $\log(2)\sim 0.6931$ \\
\hline
$O$ &$48$ & $\frac{\log (12)}{\log 3}\sim  2.2618$ & $\zeta_{\cD_{S^3}}(\frac{\log (12)}{\log 3})\sim-3.0905$ & $\log(3)\sim  1.0986$ \\
\hline
$\cI$ & $120$ & $\frac{\log (20)}{\log (2+\phi)}\sim 2.3296$  & $\zeta_{\cD_{S^3}}(\frac{\log (20)}{\log (2+\phi)})\sim -3.3486$ & $\log (2+\phi)\sim 1.2859$  \\
\hline
\end{tabular} }

\medskip

\subsection{Robertson-Walker model} 

We consider here a more realistic Robertson--Walker model of multifractal cosmology,
where the individual manifolds in the fractal configuration are copies of $4$-dimensional
Robertson--Walker spacetimes. This is the same type of model considered, in the case
of Apollonian packings of spheres, in \cite{FKM}. 

\smallskip

A (Euclidean) Robertson--Walker metric on a spacetime $X=\R\times S^3$ is of the form
$$ ds^2_{RW} =dt^2 + a(t)^2 d\sigma^2 $$
with $d\sigma^2$ is the metric on the unit sphere $S^3$. The scale factor $a(t)$
describes the expansion/contraction of the spatial sections of the $4$-dimensional spacetime.
In \cite{FKM} the full asymptotic expansion of the spectral action functional was computed
for a Robertson--Walker spacetime $(X=\R\times S^3, ds^2_{RW})$ and for a multifractal
cosmology given by a product $\R\times \cP$, with $\cP$ an Apollonian packing of $3$-spheres,
with a metric that is of the form
\begin{equation}\label{RWspheres}
 ds^2_{n,k} = a^2_{n,k} (dt^2 + a(t)^2 d\sigma^2 ) 
\end{equation} 
on each sub-cylinder $\R \times S^3_{n,k}$ of the packing $\R\times \cP$.
The case of a metric of the form $dt^2 + a(t)^2 a^2_{n,k} d\sigma^2$ is also discussed in
\cite{FKM}, but for simplicity we will focus here on the analog of \eqref{RWspheres} for other
fractal geometries. 

\smallskip

The explicit computation of the spectral action expansion for the Robertson--Walker metric
is obtained in \cite{FKM} using a technique originally introduced in \cite{CC4}, based on
representing the heat kernel in terms of the Feynman--Kac formula 
and Brownian bridge integrals (see \cite{Simon}). 

\smallskip

We summarize here briefly the main steps of the calculation of  \cite{FKM} that we
will extend to our fractal geometries. The Dirac operator $\cD_{S^3}$
has eigenvalues $\Spec(\cD_{S^3})=\{ (\ell+\frac{3}{2}) \}_{\ell\in \Z}$ with
multiplicities $\mu(\ell+\frac{3}{2}) = 4(\ell+1)(\ell+2)$. On a basis of eigenspinors $\{ \psi_{\ell,j} \,|\, 
\ell \in \Z, \, j=1,\ldots, 4(\ell+1)(\ell+2) \}$, with $\cV_\ell\subset L^2(S^3,\bS)$ the
eigenspaces $\cV_\ell ={\rm span}\{ \psi_{\ell,j} \,|\, j=1,\ldots, 4(\ell+1)(\ell+2) \}$, 
we decompose the operator as
$$\cD^2_{S^3} |_{\cV_\ell} = (\ell+\frac{3}{2})^2 \, . $$
Similarly, as shown in \cite{CC4}, we can decompose the operator 
$$ \cD^2_{\R\times S^3} = \oplus_\ell H_\ell $$
$$ H_\ell = -\frac{d^2}{dt^2} + V_\ell (t) $$
$$ V_\ell (t)=  \frac{(\ell+\frac{3}{2}) }{a(t)^2} \left( ((\ell+\frac{3}{2}) - a^\prime(t) \right)\, , $$
so that the heat kernel can be written in the form
$$ \sum_\ell \mu(\ell+\frac{3}{2})\, \Tr(e^{-s H_\ell})\, . $$ 
This can be evaluated using the Feynman--Kac formula, which gives 
$$ e^{-s H_\ell }(t,t) =\frac{1}{2\sqrt{\pi s}}\int e^{-s\int_0^1 V_\ell (t+\sqrt{2s}\,\alpha(u))du}\, D[\alpha] $$
where $D[\alpha]$ denotes the Brownian bridge integral. In \cite{FKM}, after the change of variables 
\begin{equation}\label{UVvars}
U = s\int_0^1 B(t+\sqrt{2s}\alpha(u))du, \ \ \ \ \ V = s\int_0^1 A'(t+\sqrt{2s}\alpha(u))du \, ,  
\end{equation}
with $A(t) = a(t)^{-1}$ and $B(t) = A(t)^2$, and $x = \ell+\frac{3}{2}$, which gives 
\begin{equation}\label{intVellUV}
 -s \int_0^1 V_\ell(t+\sqrt{2s} \alpha) dv = - x^2 U -x V \, , 
\end{equation} 
the Poisson summation formula is applied to the function 
$$ f_s(x)=(x^2-\frac{1}{4}) \, e^{-x^2U-xV}\,  $$
over the lattice $L=\{ \ell +3/2\,|\, \ell\in \Z \}$. 
The summation of the Fourier dual $\hat f_s$ localizes at the origin and gives
$$ \sum_{x\in L} f_s(\ell)=\sum_{y \in \hat L} \hat f_s(y) \sim \int_\R f_s(x)\, dx $$
$$ \int_\R f_s(x)\, dx = \frac{\sqrt{\pi} e^{\frac{V^2}{4U}} (-U^2+2U+V^2)}{4 U^{5/2}}\, $$
so that one obtains the generating function for the spectral action expansion as 
$$ e^{s H_\ell }(t,t) = \frac{1}{2\sqrt{\pi s}} \int (\sum_{x=\ell +3/2,\, \ell\in \mathbb{Z}} \mu(x) e^{-x^2  U - x  V})D[\alpha] = \frac{1}{2\sqrt{\pi s}} \int (\sum_{w\in \mathbb{Z}}\hat{f_s}(w))D[\alpha]  $$
$$ \sim  \frac{1}{\sqrt{s}} \int \frac{ e^{\frac{V^2}{4U}}   (-U^2+2U+V^2)   }{   4 U^{5/2}   }    \, D[\alpha] \, . $$
By further expanding $U$ and $V$ in powers of $\tau=\sqrt{s}$, it is shown in \cite{FKM} that
one has an expansion 
\begin{equation}\label{UVtaylor}
 e^{\frac{V^2}{4U}} U^r V^m = \tau^{2(r+m)} \sum_{M=0}^\infty C^{(r,m)}_M \tau^M 
\end{equation} 
where the coefficients $C^{(r,m)}_M$ have an explicit expression in terms Bell polynomials. 
Thus, one can write the heat kernel expansion as
$$ \Tr(e^{-\tau^2\, \cD_{\R\times S^3}^2}) \sim \sum_{M=0}^\infty \tau^{2M-4} \int \bigg( \int \big (\frac{1}{2}C^{-3/2,0}_{2M} + \frac{1}{4}(C^{-5/2,2}_{2M-2} - C^{-1/2,0}_{2M-2})\big)\, D[\alpha] \bigg)dt \, . $$
For a further discussion of how to evaluate the Brownian bridge integrals, see \S 4, especially
Lemma~4.14 and and Theorem~4.15, of \cite{FKM}. 

\smallskip

We need a simple modification of the computation above that takes into account replacing
the $3$-sphere $S^3$ with a spherical form $Y=S^3/\Gamma$ and forming a fractal
arrangement $\cP_\Gamma$ of scaled copied of $Y$ with scaling factors 
\begin{equation}\label{listLGamma}
 \cL_\Gamma = \{ a_{n,k} = f_\sigma^{-n}\,|\, n\in \N, \, k=1,\ldots, f_\rho^n \} 
\end{equation} 
for the same Sierpinski constructions considered in the previous section. 

\smallskip

The replacement of $S^3$ with $Y=S^3/\Gamma$ in the argument above affects the
multiplicities of the spectrum and the corresponding Poisson summation formula argument.
This is the same argument used in \cite{MPT1} and \cite{Teh} to compute the
spectral action on $Y$. It was shown in \cite{Teh} that, for each $3$-dimensional spherical
space form $Y=S^3/\Gamma$, the spectrum of the Dirac 
operator $\cD_Y$ admits a decomposition into a union of arithmetic progressions 
$$ \Spec(\cD_Y) = \bigcup_i L_i $$
where the multiplicities are polynomial functions 
$$ \mu(\lambda) = P_i(\lambda), \ \ \ \text{ for } \lambda\in L_i \, . $$ 
These polynomials satisfy
\begin{equation}\label{sumPi}
 \sum_i P_i(x) = \frac{1}{R} \left(  x^2 - \frac{1}{4} \right)\, , 
\end{equation} 
for some $R\in \N$, 
while the corresponding arithmetic progressions are of the form
\begin{equation}\label{arprogLi}
 L_i = \{ b_i + N m \,|\, m \in \Z \} \subset \{ \frac{3}{2}+\ell\,|\, \ell\in \Z \} 
\end{equation} 
for some $b_i\in \Q$, where the integers $N$ and $R$ are related by 
\begin{equation}\label{NRGamma}
N\cdot R=|\Gamma| .
\end{equation}
While the specific form of the $L_i$ and $P_i$ and the values of $N,R$ depend
on which $\Gamma$ one is considering, in all cases one obtains the following
identity for the Poisson summation. Let $P_i(x)=\sum_j c_j x^j$, and $g_i(x)=P_i(x) f_s(x)$
with $h_i(x)=g_i(b_i+N x)=P_i(b_i+N x) f_s(b_i+N x)$. We also set $P(x)=(x^2 -1/4)$ and 
$g(x)=P(x) f_s(x)$.
The Poisson summation formula gives 
$$  \sum_i \sum_{\lambda\in L_i} P_i(\lambda) f_s(\lambda) =\sum_i \sum_{m \in \Z} 
P_i(b_i+N m) f_s(b_i + mN) $$ $$ = \sum_i \sum_{k\in \Z} \frac{1}{N} e^{2\pi b_i k/N} \widehat{g_i}(k/N), $$
since 
\begin{equation}\label{Poissonshift}
 \widehat{h_i}(y)=\int h_i(x) e^{-2\pi i x y} dx = \int P_i(b_i+N x) f_s(b_i+N x) e^{-2\pi i x y} dx 
\end{equation} 
$$ = \frac{1}{N} e^{2\pi i b_i y/N} \widehat{g_i}(y/N) .$$
Thus, if we retain only the contribution at the origin of the Fourier dual summation we have
$$  \sum_i \sum_{\lambda\in L_i} P_i(\lambda) f_s(\lambda) \sim  
\sum_i \frac{1}{N} \sum_j c_j \hat f_s^{(j)}(0) \, . $$
Using then \eqref{sumPi} and \eqref{NRGamma} we can write the above equivalently as 
\begin{equation}\label{PoissonY}
  \sum_i \sum_{\lambda\in L_i} P_i(\lambda) f_s(\lambda) \sim \frac{1}{|\Gamma|}  \hat g(0)
\end{equation}
since
$$ \frac{1}{|\Gamma|}  \hat g(0) =\frac{1}{R\cdot N} \widehat{P \cdot f_s} = \frac{1}{N} 
\sum_i \widehat{P_i \cdot f_s} = \sum_i \frac{1}{N} \sum_j c_j \hat f_s^{(j)}(0) . $$

Thus, we consider the decomposition $\cD_{\R \times Y}^2$  along eigenspaces $\cV_\ell$ of $\cD_Y$,
$$ \cD_{\R \times Y}^2 = \oplus_{i,\ell} H_{i,\ell} $$
$$ H_{i,\ell} = -\frac{d^2}{dt^2} + V_{i,\ell} (t) $$
$$ V_{i,\ell} (t)=  \frac{(b_i + N\ell) }{a(t)^2} \left( (b_i + N\ell) - a^\prime(t) \right)\, , $$
with the heat kernel
$$ \Tr(e^{-s \cD_{\R \times Y}^2 })=\sum_{i,\ell} P_i(b_i + N\ell) \Tr(e^{-s H_{i,\ell} })\, . $$
The Poisson summation argument above then provides an overall factor of $1/|\Gamma|$ in the 
expansions
$$ e^{s H_{i,\ell} }(t,t) \sim \frac{1}{|\Gamma|} \frac{1}{\sqrt{s}} \int \frac{ e^{\frac{V^2}{4U}}   (-U^2+2U+V^2)   }{   4 U^{5/2}   }    \, D[\alpha] \, , $$
and consequently in the expansion
\begin{equation}\label{RYheatTr}
 \Tr(e^{-s \cD_{\R \times Y}^2 }) \sim \frac{1}{|\Gamma|} \sum_{M=0}^\infty \tau^{2M-4} \int \bigg( \int \big (\frac{1}{2}C^{-3/2,0}_{2M} + \frac{1}{4}(C^{-5/2,2}_{2M-2} - C^{-1/2,0}_{2M-2})\big)\, D[\alpha] \bigg)dt \, ,
 \end{equation}
where the coefficients $C^{(r,m)}_M$ on the right-hand-side of this expansion are those computed for the case
of the heat kernel of the operator $\cD_{\R\times S^3}^2$. 
 
\smallskip

Note that the presence of this overall factor $1/|\Gamma|$ can also be deduced directly from the
heat kernel argument of \cite{BMK} that we recalled in the previous section,  
applying  \eqref{heatkerGamma} and showing
as in \cite{BMK} that it is only the first term on the right-hand-side of 
\eqref{heatkerGamma}  that contributes to the spectral action expansion, as the other
terms give rise to contributions smaller than any power of $\Lambda$. 

\smallskip

The fractal arrangement of scaled copies of $Y$ into the corresponding Sierpinski construction 
can be accounted for in the calculation of the spectral action expansion by the same method used
in \cite{FKM} for the packing of spheres. As above let $\cL_\Gamma=\{ a_{n,k} \}$ be the
list of radii of the Sierpinski construction for $Y=S^3/\Gamma$, as in \eqref{listLGamma}.
As in \cite{FKM}, we use the following general result of complex analysis.
Let $h(\tau)$ be a function with a small $\tau$ asymptotic expansion
$$ h(\tau)\sim_{\tau\to 0^+} \sum_N c_N\, \tau^N $$
and let $\cM(h)(z)$ be the (non-normalized) Mellin transform,
$$ \cM(h)(z)=  \int_0^\infty h(\tau) \tau^{z-1} d\tau \, .  $$
Let $g_\Gamma(\tau)$ be the series
$$ g_\Gamma(\tau):=\sum_{n,k} h(a_{n,k}\, \tau), \ \ \ \text{ with } \cL_\Gamma=\{ a_{n,k} \}\,  . $$
This has Mellin transform $\cM(g_\Gamma)(z)=\zeta_{\cL_\Gamma}(z)\cdot \cM(h)(z)$.
Then, from the relation small-time between asymptotic expansion of a function and singular 
expansion of its Mellin transform  $g_\Gamma(\tau)$ has a small-time asymptotic expansion
$$ g_\Gamma(\tau) \sim_{\tau\to 0^+}  \sum_N c_N \,  \zeta_{\cL_\Gamma}(-N)\, \tau^N +
\sum_\alpha\,\, {\rm Res}_{z=\alpha}\zeta_{\cL_\Gamma}\,\cdot \, \cM(h)(\alpha) \, \cdot \, \tau^{-\alpha} \, $$
where $\alpha$ ranges over the poles of $\zeta_{\cL_\Gamma}(z)$. 

\smallskip

We apply this directly to $h(\tau)=\Tr(e^{-\tau^2\, \cD_{\R\times Y}^2})$ with the asymptotic expansion
\eqref{RYheatTr}. Under the scaling $ds^2_{n,k} = a_{n,k}^2 (dt^2+a(t)^2 d\sigma^2)$ 
of the Robertson--Walker metric on $\R\times Y$, the spectrum of the Dirac operator 
$\cD_{(\R\times Y)_{n,k}}^2$ scales by $a_{n,k}^{-2}$, so that we can identify 
$\Tr(e^{-\tau \cD_{\R\times \cP_\Gamma}})$ obtained with these scaled metrics with
$$ g_\Gamma(\tau)=\sum_{n,k} \Tr(e^{-\tau^2\, \cD_{(\R\times Y)_{n,k}}^2}) =
\sum_{n,k} f(a_{n,k}\tau)\, , $$
with corresponding asymptotic expansion
$$ \Tr(e^{-\tau \cD_{\R\times \cP_\Gamma}}) \sim 
\frac{1}{|\Gamma|} \sum_{M=0}^\infty \tau^{2M-4}\, \zeta_{\cL_\Gamma}(-2M+4) \, \int \bigg( \int \big (\frac{1}{2}C^{-3/2,0}_{2M} + \frac{1}{4}(C^{-5/2,2}_{2M-2} - C^{-1/2,0}_{2M-2})\big)\, D[\alpha] \bigg)dt $$
$$ + \sum_{n\in \Z} {\rm Res}_{s=s_n} \zeta_{\cL_\Gamma} \,\,  \cM(h)(s_n)\,  \tau^{-s_n} , $$
where $\Sigma_\Gamma=\{ s_n \,|\, n\in \Z \}$ is the set of poles of $\zeta_{\cL_\Gamma}$. We now
use what we explicitly know about $\zeta_{\cL_\Gamma}$ and $\cM(h)$. Using the Mellin
transform relation \eqref{Mellin} and $\cM(h)(s)=\frac{1}{2}\cM(f)(s/2)$ for $h(\tau)=f(\tau^2)$, we obtain
$$ \cM(h)(s)=\frac{1}{2}\, \Gamma(s/2)\, \zeta_{\cD_{\R\times Y}}(s) \, .  $$
Thus, using \eqref{polesZL} and \eqref{ResZL} for the poles and residues of $\zeta_{\cL_\Gamma}$ we
obtain
$$ \Tr(e^{-\tau \cD_{\R\times \cP_\Gamma}}) \sim 
\frac{1}{|\Gamma|} \sum_{M=0}^\infty \tau^{2M-4}\, \zeta_{\cL_\Gamma}(-2M+4) \, \int \bigg( \int \big (\frac{1}{2}C^{-3/2,0}_{2M} + \frac{1}{4}(C^{-5/2,2}_{2M-2} - C^{-1/2,0}_{2M-2})\big)\, D[\alpha] \bigg)dt $$
$$ + \frac{1}{2 |\Gamma| \log f_\sigma}  \sum_{n\in \Z}   \, \Gamma(s_n/2)\, \zeta_{\cD_{\R\times Y}}(s_n) \, \tau^{-s_n} \, . $$
Correspondingly, the spectral action expansion is of the form 
\begin{equation} \label{SPActRWpack}
\begin{array}{l}
\Tr( f(\cD_{\R\times \cP_\Gamma}/\Lambda))   \sim \\[4mm]
    \displaystyle{\frac{1}{|\Gamma|} \sum_{M=0}^\infty}  \displaystyle{ \frac{\Lambda^{4-2M}\, \frf_{4-2M}} 
  {(1-f_\rho f_\sigma^{2M-4})} \cdot } 
     \displaystyle{\int \bigg( \int \big (\frac{1}{2}C^{-3/2,0}_{2M} + \frac{1}{4}(C^{-5/2,2}_{2M-2} - C^{-1/2,0}_{2M-2})\big)\, D[\alpha] \bigg)dt} \\[5mm]
    \displaystyle{+ \frac{1}{2 |\Gamma| \log f_\sigma}  \sum_{n\in \Z} } \frf_{s_n}\,  \Gamma(s_n/2)\, \zeta_{\cD_{\R\times Y}}(s_n) \Lambda^{s_n} \, .
    \end{array}
\end{equation} 

\smallskip

Again we see, as in the discussion of the static model in the previous section, that the different
Sierpinski constructions for the different topologies $Y=S^3/\Gamma$ give rise to different 
correction terms to the spectral action, through the different values of $f_\sigma$, $s_0$, and
the values $\zeta_\cL(4-2M)$, $M\in \N$, and $\zeta_{\cD_{\R\times Y}}(s_n)$. Different candidate
spherical cosmic topologies are fully detectable by the spectral action model of gravity in the
presence of fractality, in the form of the corresponding Sierpinski constructions. 

\medskip
\subsection{Fractality in flat cosmic topologies} 

In this section we have focused primarily on the spherical space forms as candidate cosmic
topologies, and we have shown that the possible presence of fractality and the explicit form
of the contributions of fractality to the spectral action functional depend on the choice of
spherical space form for the underlying cosmic topology. It is natural to ask whether the
same effect persists in the case of the flat cosmic topologies, given by the Bieberbach
manifolds, obtained as quotients $Y=T^3/\Gamma$ of a torus by a finite group of isometries. 
The spectral action for Bieberbach manifolds
was computed in \cite{MPT2}, \cite{OlSit}. 

 \begin{figure}
 \begin{center}
 \includegraphics[scale=0.25]{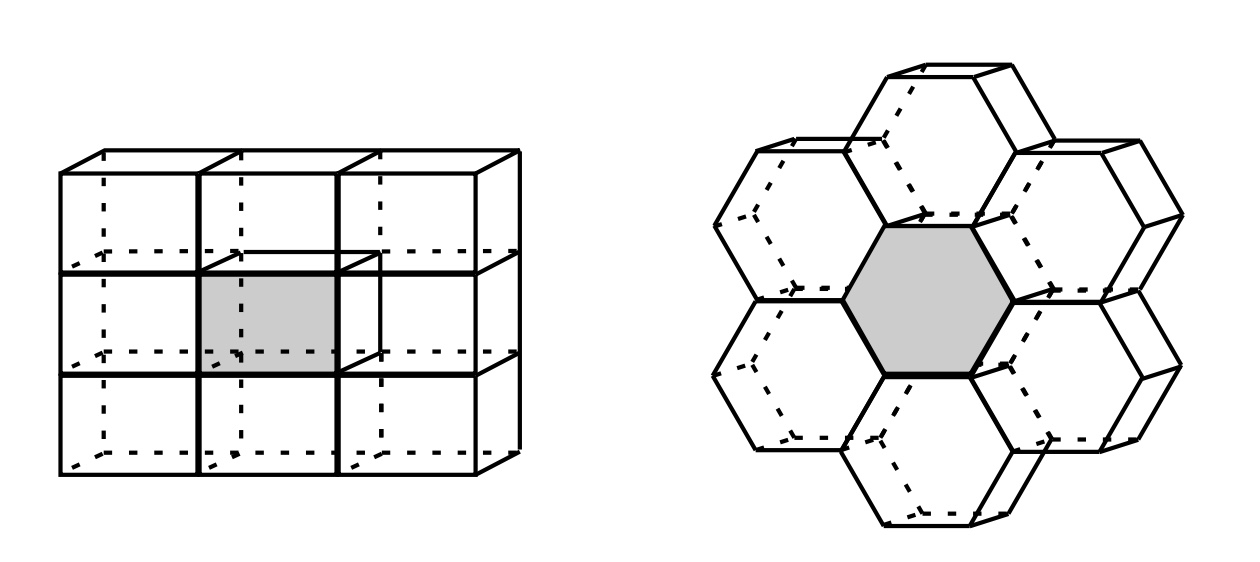}
 \includegraphics[scale=0.15]{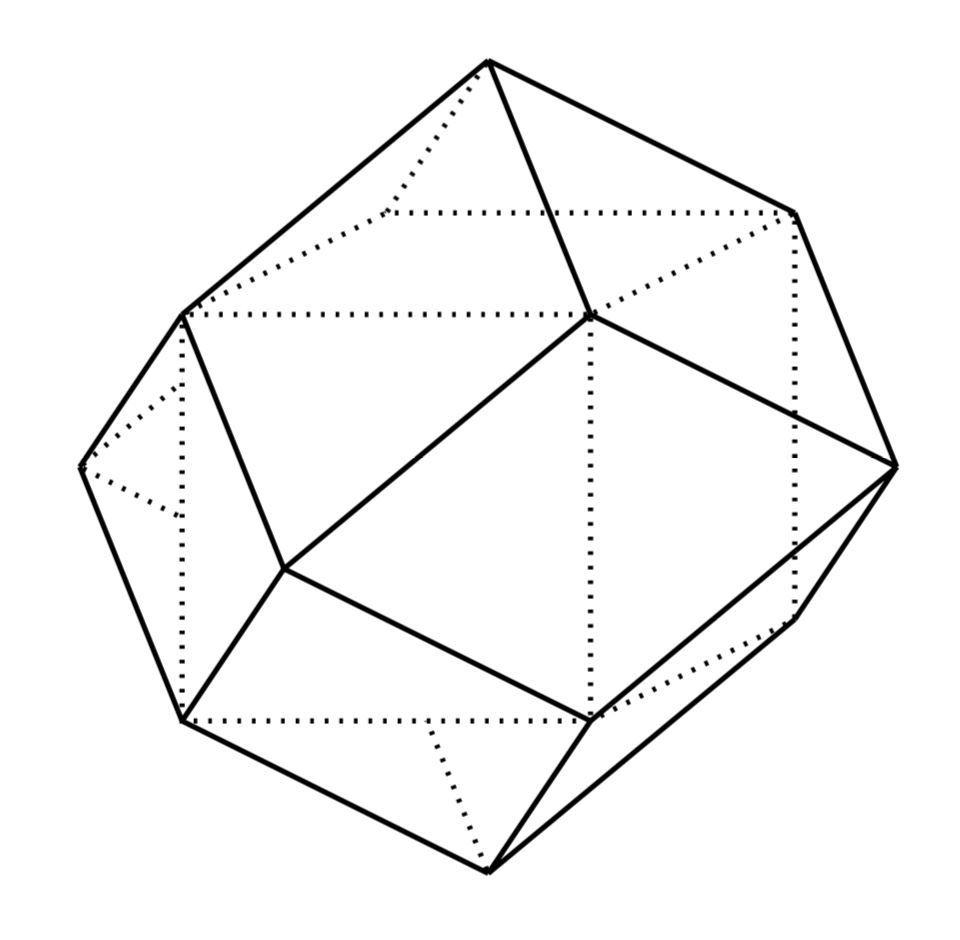}
 \end{center}
 \caption{The fundamental domains of Bieberbach manifolds correspond to
 tilings of $\R^3$ by parallelepipeds ($G_1$, $G_2$, $G_4$) or hexagonal prisms ($G_3$, $G_5$),  
 while the Hantzsche--Wendt manifold $G_6$ has a fundamental domain given by a
 rhombic dodecahedron. \label{FigR3Tiles}}
\end{figure}

\smallskip

There are six affine equivalence classes of (compact and orientable) Bieberbach manifolds, usually labelled
$G_1, \ldots, G_6$, with $G_1=T^3$. The other classes are, respectively known
as {\em half-turn space} $G_2$, {\em third-turn space} $G_3$, {\em quarter-turn
space} $G_4$, {\em sixth-turn space} $G_5$, and {\em Hantzsche--Wendt space} $G_6$, which
corresponds to a half-turn along each coordinate axis.
The corresponding groups, which we denote by $\Gamma_{G_i}$, 
are of the form (with $t_i$ implementing the translation by the vector $a_i$):
\begin{itemize}
\item with $a_1=(0,0,H)$,  $a_2=(L,0,0)$, and $a_3=(T,S,0)$, $H,L,S \in \R^*_+$ and $T\in \R$
$$ \Gamma_{G_2} = \langle \alpha, t_1, t_2, t_3\,|\,  \alpha^2 =t_1, \,  \alpha t_2 \alpha^{-1} =t_2^{-1}, \,  \alpha t_3 \alpha^{-1} = t_3^{-1} \rangle \, , $$
\item with $a_1=(0,0,H)$, $a_2=(L,0,0)$
and $a_3=(-\frac{1}{2}L, \frac{\sqrt{3}}{2}L, 0)$, for $H$ and $L$ in $\R^*_+$
$$ \Gamma_{G_3}= \langle \alpha, t_1, t_2, t_3\,|\, \alpha^3 = t_1, \, \alpha t_2 \alpha^{-1} = t_3,  \,  
\alpha t_3 \alpha^{-1} = t_2^{-1} t_3^{-1} \rangle $$
\item with $a_1=(0,0,H)$, $a_2=(L,0,0)$, and
$a_3=(0,L,0)$, with $H,L>0$
$$ \Gamma_{G_4}= \langle \alpha, t_1, t_2, t_3\,|\,
\alpha^4 = t_1, \,  \alpha t_2 \alpha^{-1} = t_3, \,  \alpha t_3 \alpha^{-1} = t_2^{-1} \rangle $$
\item with $a_1=(0,0,H)$,
$a_2=(L,0,0)$ and $a_3=(\frac{1}{2}L, \frac{\sqrt{3}}{2}L,0)$, $H,L>0$
$$ \Gamma_{G_5}=\langle \alpha, t_1, t_2, t_3\,|\, \alpha^6=t_1, \,  \alpha t_2 \alpha^{-1}=t_3, \,
\alpha t_3 \alpha^{-1}= t_2^{-1} t_3 \rangle $$
\item with $a_1=(0,0,H)$, $a_2=(L,0,0)$, and $a_3=(0,S,0)$, with $H,L,S>0$
$$ \Gamma_{G_6} =\langle \alpha,\beta,\gamma, t_1,t_2,t_3\,|\, $$
$$ \begin{array}{ccc}
\alpha^2 =t_1, & \alpha t_2 \alpha^{-1} = t_2^{-1}, & \alpha t_3 \alpha^{-1} = t_3^{-1}, \\
\beta^2 = t_2, & \beta t_1 \beta^{-1} =t_1^{-1}, & \beta t_3 \beta^{-1} = t_3^{-1}, \\
\gamma^2 = t_3, & \gamma t_1 \gamma^{-1} = t_1^{-1}, & \gamma t_2 \gamma^{-1} = t_2^{-1}, \\
& \gamma\beta\alpha = t_1 t_3 \rangle & 
\end{array}  $$
\end{itemize}.

\smallskip

The fundamental domains for the Bieberbach manifolds are parallelepipeds (for $G_1$, $G_2$, $G_4$) 
or hexagonal prisms (for $G_3$, $G_5$), see Figure~\ref{FigR3Tiles}. The Hantzsche--Wendt 
manifold has a more interesting structure (for its role in Cosmic Topology see \cite{AuLu}) and
a fundamental domain given by a rhombic dodecahedron, see Figure~\ref{FigR3Tiles}. For a general
discussion of these manifolds in the context of the Cosmic Topology problem see 
\cite{Levin}, \cite{LuRou}, \cite{Riaz}). 

\smallskip

 \begin{figure}
 \begin{center}
 \includegraphics[scale=0.45]{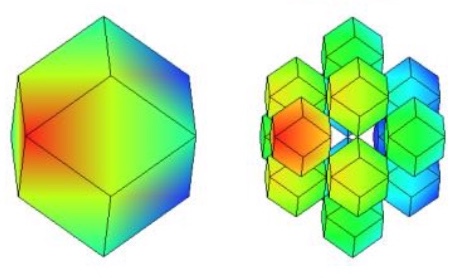}
 \end{center}
 \caption{The first step of the Sierpinski construction for the rhombic dodecahedron, \cite{Katu2}. \label{FigRhDod}}
\end{figure}

Thus, the presence of fractality
would, in this case, require a Sierpinski construction based on these 
parallelepipeds and hexagonal prisms. The case of parallelepipeds, as in the
cube case, leads to Menger-sponge type fractals, while the hexagonal prisms
have a Sierpinski construction based on the hexaflake, see Figure~\ref{FigHex}. 
In the case of the Hantzsche--Wendt manifold, we need to use a Sierpinski
construction for the rhombic dodecahedron. We have already analyzed
the Sierpinski construction for a prism based on the hexaflake fractal in the
previous section, so we focus here on the case of a parallelepiped (for which
it suffices to consider the case of a cube) and of a rhombic dodecahedron. 
For the latter, Sierpinski constructions for Catalan solids have been developed in \cite{Katu2}. 
In the case of the rhombic dodecahedron, the first step of the Sierpinski construction is
shown in Figure~\ref{FigRhDod}. 

\smallskip

 \begin{figure}
 \begin{center}
 \includegraphics[scale=0.15]{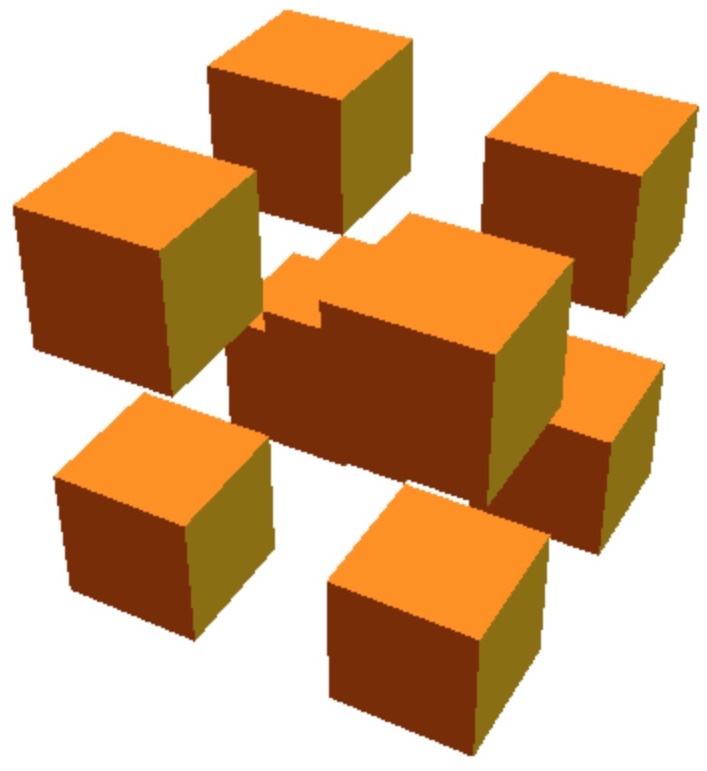}
 \includegraphics[scale=0.15]{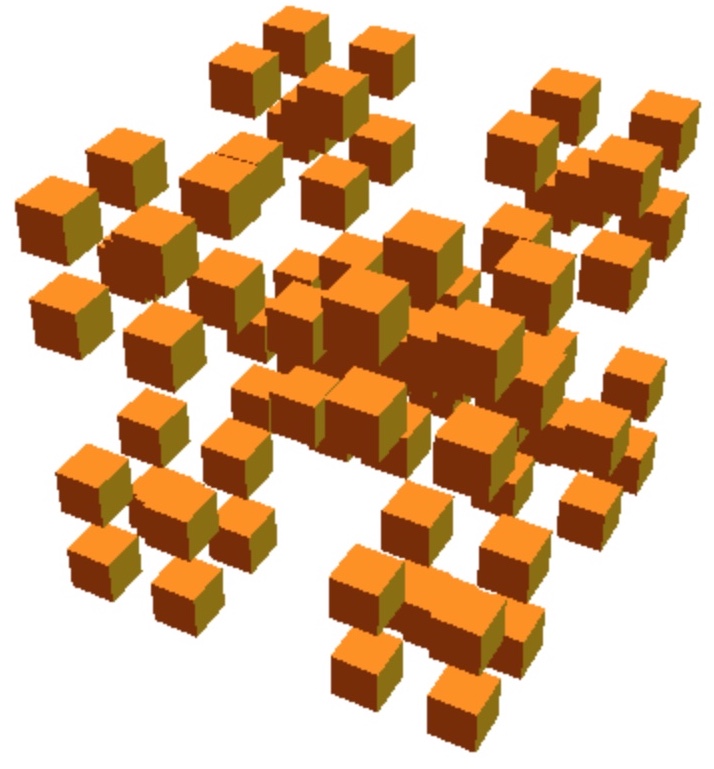}
 \includegraphics[scale=0.15]{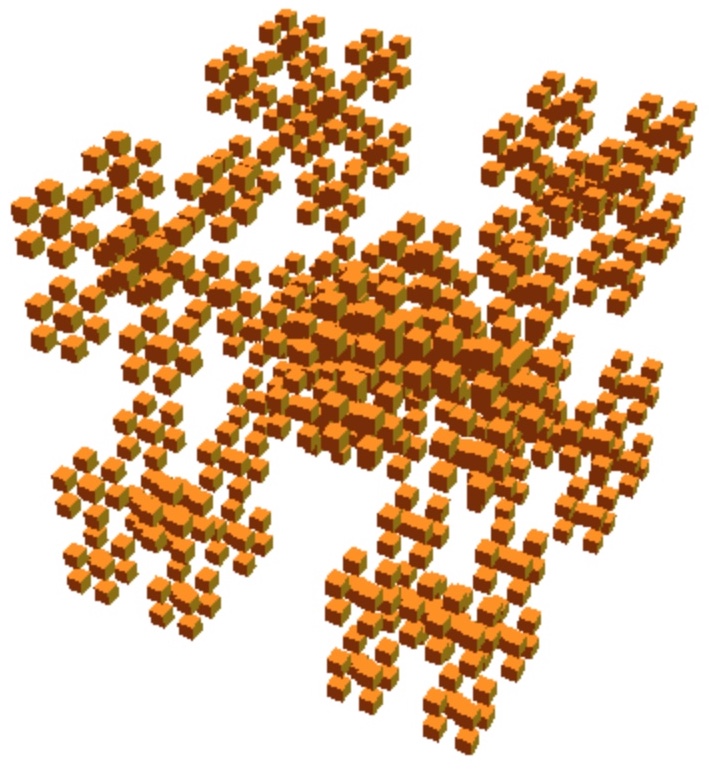}
 \end{center}
 \caption{The first steps of the Sierpinski construction for the Sierpinski-Menger snowflake. \label{FigMenger}}
\end{figure}

The case of a cube (or parallelepiped) is slightly more subtle than it seems at first.
If one applies the same type of Sierpinski construction that we have used for the
regular, Archimedean, and Catalan solids, \cite{Katu}, \cite{Katu2}, \cite{KatuKu}, 
then one does not obtain a fractal, as eight copies of a cube scaled by $1/2$
just fill an identical copy of the same cube, so this Sierpinski construction just
replicated the cube itself without introducing fractality. One can instead construct
Menger-sponge type fractals based on cubes, where a cube is subdivided into
scaled copies of itself and only some of these copies are retained. However, not
all such Menger constructions are suitable for our purpose, since we need each
cube in the construction to be a fundamental domain of a Bieberbach manifold,
which means that the faces of the cube should be free to be glued to other faces
according to the group action, and not attached to faces of other cubes in the
Menger solid. This means that cubes need to be removed so that the remaining
ones are only adjacent along edges or vertices, but not along faces. All the
Sierpinski constructions we considered so far have this property. 
A Menger construction that satisfies this property is given by the
Sierpinski-Menger snowflake (Figure~\ref{FigMenger}).
 
\smallskip

In the Sierpinski-Menger snowflake construction of Figure~\ref{FigMenger}, 
at every step each of the previous cubes
is replaced by eight corner cubes and one central cube of $1/3$ the size. Thus, for this
Sierpinski construction the replication factor is $f_\rho=9$ and the scaling factor is $f_\sigma =3$,
with
$$ \dim_H \bP_{G_a} = \frac{\log f_\rho}{\log f_\sigma} = \frac{\log 9}{\log 3}=2 \, , 
 \ \ \ \text{ for } \ a\in \{ 1, 2, 4 \} \, . $$
 It is worth pointing out that, among the Sierpinski constructions associated to the
 candidate cosmic topologies in both positive and flat curvature this is the only
 case where the self-similarity dimension agrees with the value observed from
 the large scale distribution of galaxy clusters, \cite{LMP2}, \cite{LMP1}.
 
\smallskip
 
 In the case of the hexaflex prism, we have, as discussed in the previous section,
 $$ \dim_H \bP_{G_a} = \frac{\log 6}{\log 3} \sim 1.63093 \, , 
 \ \ \ \text{ for } \ a\in \{ 3,5 \} \, , $$
 while in the case of the rhombic dodecahedron, at each step of the Sierpinski construction each
rhombic dodecahedron is replaced by $14$ identical copies scaled by a factor of $1/3$ so that
$f_\rho=14$ and $f_\sigma=3$ and
$$ \dim_H \bP_{G_6} = \frac{\log (14)}{\log 3} \sim 2.40217 \, . $$

\smallskip
\subsection{Static model in the flat case}

The computation of the spectral action, for both the static and the Robertson--Walker model,
follow the same pattern that we discussed in the previous section, but with $S^3$ replaced
with the flat torus $T^3$. As shown in \cite{MPT1}, \cite{MPT2}, the spectral actions of the torus and the 
Bieberbach manifolds are related by a scalar factor
$$ \Tr(f(\cD_{G_a}/\Lambda))\sim \lambda_{G_a} \cdot \Tr(f(\cD_{T^3}/\Lambda)) =\frac{\lambda_{G_a}  \,\Lambda^3}{4\pi^3} \int_{\R^3} f(u^2+v^2+w^2)\, du\,dv\,dw  $$
up to $O(\Lambda^{-\infty})$ terms. In this case, the factor $\lambda_{G_a}$ also depends
on the continuous parameters $H,L,S,T$ in the generators of the groups $\Gamma_{G_a}$
(see the explicit presentations recalled above), so they are not just a rational number 
$1/|\Gamma|$ as in the spherical case. Nonetheless, the argument is otherwise very
similar. On the product $S^1_\beta\times T^3$ the spectral action is of the form (see \cite{MPT1})
$$ \Tr(h(\cD_{S^1_\beta \times T^3}/\Lambda)) = \frac{\Lambda^4 \beta}{4\pi} \int_0^\infty u h(u) du + O(\Lambda^{-\infty})\, .  $$
Also, as shown in \cite{MPT1}, in the flat torus case, the corresponding slow-roll potential is 
only given by the term
$$ \cV(\phi^2/\Lambda^2)=\int_0^\infty u(h(u+\phi^2/\Lambda^2)-h(u))\, du $$
without the part $\cW(\phi^2/\Lambda^2)$ of the spherical case \eqref{slowrollS3}. 
For the fractal arrangement $\cP_{G_a}$ we will again obtain
$$
\cS_\Lambda(\cD_{\cP_{G_a}}) \sim
\sum_\alpha \frac{\frf_\alpha \Lambda^\alpha\,  \lambda_{G_a}\, \zeta_{\cL_\Gamma}(\alpha)}{2} \Res_{s=\alpha} \zeta_{\mathcal{D}_{T^3}} + 
\sum_\beta \frac{\frf_\beta \Lambda^\beta\, \lambda_{G_a}\, \zeta_{\mathcal{D}_{T^3}}(\beta)}{2} 
\Res_{s=\beta} \zeta_{\cL_\Gamma}
$$
and the same argument based on \eqref{kerprod}, \eqref{SA4d3d} gives then
\begin{equation}\label{SpActS1Ga}
\cS_\Lambda(\mathcal{D}_{S^1_\beta \times \cP_{G_a}}) \sim \beta\, \lambda_{G_a}\, 
\left(\Lambda^4 \,2\, \zeta_{\cL_\Gamma}(3)\, \fh_3  + 
\sum_{n\in \Z}  \frac{\Lambda^{s_n +1}\,a^{s_n}\, \zeta_{\cD_{T^3}}(s_n)}{\log f_\sigma}\, \fh_{s_n} \right) 
\end{equation}
$$ \sim \beta\, \lambda_{G_a}\, 
\left(\frac{\Lambda^4 \,2}{(1-f_\rho f_\sigma^{-3})}\, \, \fh_3  + 
 \sum_{n\in \Z}  \frac{\Lambda^{s_n +1}\,a^{s_n}\, \zeta_{\cD_{T^3}}(s_n)}{\log f_\sigma}\, \fh_{s_n} \right)\, ,
$$
where
$$ \fh_3:=\pi \int_0^\infty h(\rho^2) \rho^3 \,d\rho \ \ \text{ and } \ \ 
\fh_{s_n}= 2 \int_0^\infty h(\rho^2) \rho^{s_n} d\rho. $$
The slow-roll potential is of the form
\begin{equation}\label{flatLagPhi}
 \cL(\phi)=A\,\, \Lambda^4\,\, \cV(\phi^2/\Lambda^2)  
+\sum_{n\in \Z} C_n \,\, \Lambda^{s_n+1}\,\, \cU_n(\phi^2/\Lambda^2) , 
\end{equation}
$$ \text{ with } \ \ \ A= \frac{\pi \beta\,  \lambda_{G_a}}{(1-f_\rho f_\sigma^{-3})}\,  , \ \ \ \text{ and } \ \ \
C_n = \frac{2\beta \, \lambda_{G_a}\,  a^{s_n}\, \zeta_{\cD_{T^3}}(s_n)}{\log f_\sigma}\, , $$
and with $\cV(\phi^2/\Lambda^2)$ as above and $\cU_n(\phi^2/\Lambda^2)$ as in
\eqref{slowrollfrac}.

\smallskip
\subsection{Robertson--Walker model in the flat case}

We sketch here the argument for the Robertson--Walker model $(\R\times T^3, dt^2+a(t)^2 ds_{T^3})$
and the corresponding fractal arrangements $(\R\times \cP_{G_a}, a_{n,k}^2(dt^2+a(t)^2 ds_{T^3}))$. 
We will not provide a full computation in this paper. 

\smallskip

We start by considering the Dirac operator $\cD_{T^3}$ on the flat $3$-torus $T^3=\R^3/\Z^3$. 
We take the standard torus for simplicity, but one can also consider more generally tori $\R^3/L$, 
for other lattices $L\subset \R^3$. The spectrum of $\cD_{T^3}^2$ consists of 
$$ \lambda^2_{(n,m,k)}= 4\pi \| (n,m,k)+(n_0,m_0,k_0) \|^2,  \ \ \ \text{ for } \ (n,m,k) \in \Z^3\, , $$
where each $(n,m,k) \in \Z^3$ contributes with an additional multiplicity $2$, and 
where the vector $(n_0,m_0,k_0)$ depends on the choice of one of the $8$ possible spin
structures. Since the spectral action itself is independent of the choice of spin structure (see 
\cite{MPT1}), we can fix any of the eight choices of $v_0=(n_0,m_0,k_0)$. 
The eigenspinor spaces decompose as
$$ \cV_\lambda =\bigoplus_{v\in \Z^3\,:\, \pm 2\sqrt{\pi} \| v+v_0 \|=\lambda} \cV_{v,\pm}\, . $$ 
We then decompose the operator $\cD^2_{\R\times T^3}$ with the Robertson--Walker metric
$dt^2+a(t)^2 ds_{T^3}$ on $\R\times T^3$ as
$$ \cD^2_{\R\times T^3} = \oplus_{v\in \Z^3} \, H_{v,\pm} $$
$$ H_{v,\pm} = -\frac{d^2}{dt^2} + V_v(t) $$
$$ V_v(t) = \frac{\lambda_v}{a(t)^2} \left( \lambda_v -  a^\prime(t)\right) \, , \ \ \ \text{ with } \ \
\lambda_v= \pm 2\sqrt{\pi} \| v+v_0 \| \, , $$
so that the trace of the heat kernel is written as
$$  \sum_{v\in \Z^3} \Tr(e^{-s H_{v,\pm}}) \, . $$
We again then use the Feynman--Kac formula to express the heat kernel as 
$$ e^{-s H_{v,\pm}} (t,t) =\frac{1}{2 \sqrt{\pi s}} \int e^{-s \int_0^1 
V_{v,\pm}(t + \sqrt{2s}\, \alpha(u))\, du} \, D[\alpha]\, . $$
We use the same variables $U$ and $V$ of \eqref{UVvars} as in the sphere case, so that we have the
same expression \eqref{intVellUV} with $x$ replaced by 
$\lambda_v= \pm 2\sqrt{\pi}\, \| v+v_0 \|$. Thus, we now
apply the Poisson summation to the functions
$$ f_{s,\pm}(v)=  e^{-\lambda_v^2 \, U \pm |\lambda_v| \, V}\, , $$
and again we select the term at $w=(0,0,0)$, which is the
main term contributing to the spectral action expansion. We obtain
$$ \sum_{v\in \Z^3} (f_{s,+} + f_{s,-})(v) = \sum_{w\in \Z^3} (\hat f_{s,+}+ \hat f_{s,-})(w) 
\sim \int_{\R^3} (f_{s,+}+f_{s,-})(x,y,z) dx\, dy\, dz \, $$ 
$$ =  {\rm Vol}(S^2) \,\left( \int_0^\infty  e^{-4\pi r^2 \, U -2\sqrt{\pi}\, r\, V} \,  r^2\, dr +
\int_0^\infty  e^{-4\pi r^2 \, U + 2\sqrt{\pi}\, r\, V} \,  r^2\, dr \right) $$
$$ = \frac{1}{\sqrt{\pi}} \int_\R e^{- x^2 \, U - x\, V} \, x^2\,  dx = \frac{ e^{\frac{V^2}{4U}} (2U +V)}{4 U^{5/2}} \, .$$
Note that this expression differs from the case of the sphere only in the absence of the term $-U^2$
in the numerator, and in the normalization factor of $\sqrt{\pi}$. 

\smallskip

Thus, we then have a very similar argument to \cite{FKM} for the full expansion of the spectral action for
the Robertson--Walker metric on $\R\times T^3$. Using the same expansion \eqref{UVtaylor} 
in powers of $\tau=s^2$ we obtain
$$ \frac{1}{\sqrt{\pi} \tau} \frac{e^{\frac{V^2}{4U}} (2U+V^2)}{4 U^{5/2}}=\frac{1}{4} \sum_{M=0}^\infty C^{(-5/2,2)}_M \tau^{M-2}+ \frac{1}{2} \sum_{M=0}^\infty C^{(-3/2,0)}_M \tau^{M-4}, $$
which differs from the case of $S^3$ of \cite{FKM} in the absence of the $C^{(-1/2,0)}_M$ terms.
Thus, we obtain the heat kernel expansion
$$ \Tr(e^{-\tau^2\, \cD_{\R\times T^3}^2}) \sim \sum_{M=0}^\infty \tau^{2M-4} \int \bigg( \int \big (\frac{1}{2}C^{-3/2,0}_{2M} + \frac{1}{4} C^{-5/2,2}_{2M-2} \big)\, D[\alpha] \bigg)dt \, . $$
The evaluation of the Brownian bridge integrals can then be done as in the sphere case, and we refer the
reader to \cite{FKM} for a detailed discussion. 

\smallskip

Passing from the expansion for the heat kernel of $\cD_{\R\times T^3}^2$ to the expansion for
$\cD_{\R\times G_a}^2$ for the Bieberbach manifolds $G_a$ can be handled in the same way
as we did in the previous section for the spherical space forms, using a decomposition of the
spectrum of $\cD_{G_a}$ into subregions of the lattice $\Z^3$, with appropriate multiplicity functions, 
as shown in \cite{MPT2}, and applying this decomposition to the Poisson summation formula in 
the argument above. The spin structure $v_0$ on $T^3$ needs to be chosen in a subset of those
that descend to the quotient $G_a$. The resulting spectral action is still be independent of this
choice.

\smallskip

We summarize the results of \cite{MPT2} and \cite{OlSit} on the Bieberbach manifolds.
According to the Dirac spectrum computation of \cite{Pfa}, 
the spectrum of $\cD_{G_a}$ decomposes into a symmetric and an asymmetric
component. The asymmetric component is only present for certain spin structures
and it's not present at all in the case of $G_6$. The symmetric part of the spectrum
is parameterized by a finite collection of regions $\cI_i\subset \Z^3$ in the lattice,
with the corresponding eigenvalues given by a function $\lambda_i(v)$, $v\in \cI_i$ with
multiplicity a constant multiple of the number of $v\in \cI_i$ realizing the same
value $\lambda=\lambda_i(v)$. The asymmetric component is an arithmetic progression
of the form $\cB=\{ 2\pi H^{-1} (k\ell +c)\,|\, \ell \in \Z \}$ where $c$ depends on the
spin structure, and $k$ is $2$ for $G_2$, $3$ for $G_3$, $4$ for $G_4$, and $6$ for $G_5$.
Arguing as in \cite{MPT2} and \cite{OlSit}, the Poisson summation formula for the $T^3$ case,
$$ \sum_{v\in \Z^3} (f_{s,+} + f_{s,-})(v) = \frac{1}{\sqrt{\pi}} \int_\R e^{- x^2 \, U - x\, V} \, x^2\,  dx $$
is replaced by the summation
$$ \sum_i \sum_{v\in \cI_i} f_{s,i}(v) + \sum_{\ell\in \Z} f_s(\frac{2\pi c}{H} + \frac{2\pi k}{H}\ell)\, , $$
where $f_s(x)=e^{-x^2 U -x V}$ and $f_{s,i}(v)=e^{-\lambda_i(v)^2 U -\lambda_i(v) V}$. 
As in \eqref{Poissonshift}, the second summation gives 
$$ \sum_{\ell \in \Z^3} \frac{H}{2\pi k} e^{2\pi i c \ell/k}    \hat f_s(\frac{H \ell}{2\pi k}) \sim \frac{H}{2\pi k} \int_\R f_s(x)\, dx \, , $$
while the first term in the Poisson summation, coming from the symmetric component of
the spectrum can be dealt with as in the individual cases discussed in \cite{MPT2} and \cite{OlSit},
which we do not recall explicitly here.
In all cases the Fourier transformed sides of the Poisson summation add up so that the
zero-term of the combined summation is simply a multiple of the same term in the torus case, 
$$ \sum_i \sum_{v\in \cI_i} f_{s,i}(v) + \sum_{\ell\in \Z} f_s(\frac{2\pi c}{H} + \frac{2\pi k}{H}\ell)
\sim \lambda_{G_a} \,   \int_{\R^3} (f_{s,+}+f_{s,-})(x,y,z) dx\, dy\, dz \,  $$ 
$$ = \lambda_{G_a} \,\, \frac{ e^{\frac{V^2}{4U}} (2U +V)}{4 U^{5/2}} \, ,$$
where, as mentioned above, the factor $ \lambda_{G_a}$ depends on the 
continuous parameters $H,L,S,T$ in the groups $\Gamma_{G_a}$ but not on
the spin structure. 

\smallskip

Thus, we obtain the same expansion for the expression above
and a corresponding heat kernel expansion of the form
\begin{equation}\label{RWheatkerGa}
 \Tr(e^{-\tau^2\, \cD_{\R\times G_a}^2}) \sim \lambda_{G_a} \,\, \sum_{M=0}^\infty \tau^{2M-4} \int \bigg( \int \big (\frac{1}{2}C^{-3/2,0}_{2M} + \frac{1}{4} C^{-5/2,2}_{2M-2} \big)\, D[\alpha] \bigg)dt \, . 
\end{equation} 
We can then apply the same method as in the spherical cases to pass to the
heat kernel expansion for the fractal arrangement, and we obtain 
$$ \Tr(e^{-\tau \cD_{\R\times \cP_{G_a}}}) \sim 
\lambda_{G_a} \,\, \sum_{M=0}^\infty \tau^{2M-4} \, \zeta_{\cL_\Gamma}(-2M+4) \,
\int \bigg( \int \big (\frac{1}{2}C^{-3/2,0}_{2M} + \frac{1}{4} C^{-5/2,2}_{2M-2} \big)\, D[\alpha] \bigg)dt $$
$$ + \frac{\lambda_{G_a}}{2 \log f_\sigma}  \sum_{n\in \Z}   \, \Gamma(s_n/2)\, 
\zeta_{\cD_{\R\times G_a}}(s_n) \, \tau^{-s_n} \, . $$
The spectral action is correspondingly of the form

\begin{equation} \label{SPActRWpackGa}
\begin{array}{rl}
\Tr( f(\cD_{\R\times \cP_{G_a}}/\Lambda))  & \sim \\
  &  \displaystyle{\lambda_{G_a}\, \sum_{M=0}^\infty}  \displaystyle{ \Lambda^{4-2M}\, \frf_{4-2M} \,\zeta_{\cL_\Gamma}(4-2M) \cdot  } \\
    & \displaystyle{\int \bigg( \int \big (\frac{1}{2}C^{-3/2,0}_{2M} + \frac{1}{4} C^{-5/2,2}_{2M-2} \big)\, D[\alpha] \bigg)dt} \\
 &   \displaystyle{+ \frac{\lambda_{G_a}}{2 \log f_\sigma}  \sum_{n\in \Z} } \, \frf_{s_n} \,  \Gamma(s_n/2)\, \zeta_{\cD_{\R\times G_a}}(s_n) \Lambda^{s_n} 
    \end{array}
\end{equation} 
$$  \sim  \lambda_{G_a}\, \sum_{M=0}^\infty   \frac{\Lambda^{4-2M}\, \frf_{4-2M}}
 {(1-f_\rho f_\sigma^{2M-4})} \int \bigg( \int \big (\frac{1}{2}C^{-3/2,0}_{2M} + \frac{1}{4} C^{-5/2,2}_{2M-2} \big)\, D[\alpha] \bigg)dt $$
$$  + \frac{\lambda_{G_a}}{2 \log f_\sigma}  \sum_{n\in \Z} \, \frf_{s_n}\,  \Gamma(s_n/2)\, \zeta_{\cD_{\R\times G_a}}(s_n) \Lambda^{s_n}  \, . $$

\smallskip

\subsection{Fractal structures on manifolds}

In our analysis in the previous sections of the possible fractal structures associated 
to the different cosmic topologies, we have used the fact that these are all homogeneous
manifolds obtained as quotients by group actions, which admit a nice polyhedral fundamental
domain (either spherical or flat Euclidean). This allowed us to use Sierpinski constructions
based on the model polyhedron, and apply them to build a fractal arrangement of copies
of the same manifold.

\smallskip

More generally, one can ask a broad question about fractal constructions based on
manifolds with decompositions into polyhedra. These more general manifolds will not
provide standard candidate cosmic topologies as they fail to have the desired 
homogeneity property. However, this appears to be a question of independent 
interest, both for the spectral action model of gravity we are considering and for
more general physical models. 

\smallskip

We will return to this question in a separate paper, where we develop the
necessary mathematical background to address it in the required generality.
However, in this section we outline briefly some aspects of this problem
and some relevant examples. 

\smallskip
\subsubsection{Fractality from triangulations} 

The easiest form of decomposition of a manifold into polyhedral structures is a
triangulation, namely a decomposition into simplices (tetrahedra in a $3$-manifold case,
triangles in a surface case), glued together along lower dimensional faces. 
The algebraic topology of {\em simplicial sets} describes the general topological objects
(not always smooth manifolds) obtained by such arrangements of simplices. Not all
topological manifolds admit a triangulation, but all smooth manifolds do. 

\smallskip

Thus, a first possible idea of how to obtain fractal arrangements of manifolds would
be to consider fractal constructions associated to tetrahedra (and higher dimensional
simplices) and apply them to every tetrahedron (simplex) in the triangulation, by
compatibly performing all the gluing that describe the triangulation at each level of
the fractal construction. A similar idea was used, for instance, in \cite{CaLaBo}.

\begin{figure}
\begin{center}
\includegraphics[scale=0.25]{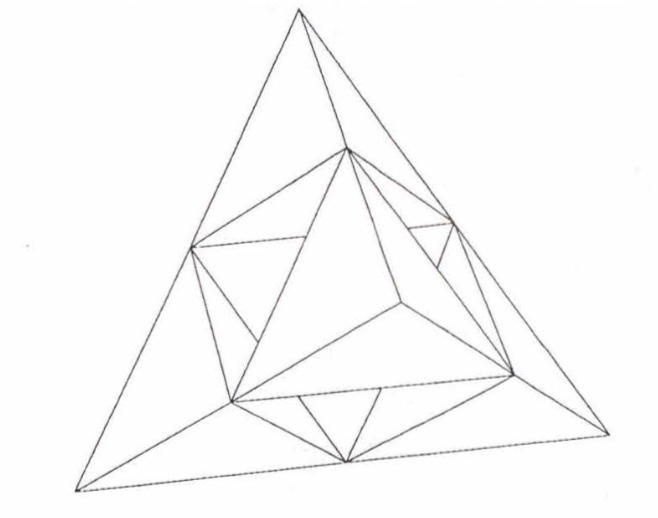}
\includegraphics[scale=0.25]{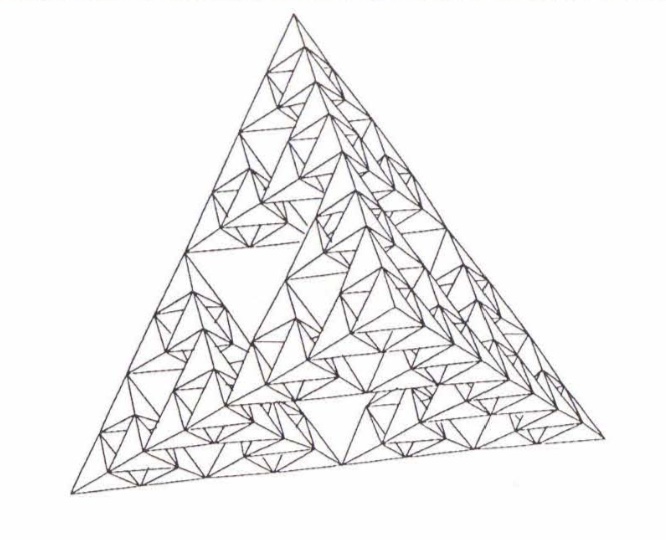}
\end{center}
\caption{The first two stages of the tetrahedral Sierpinski construction, \cite{JoCa}. \label{FigTetra}}
\end{figure}

\smallskip

In the case of a $3$-dimensional manifold $M$ with a triangulation $T$, this 
can be done, for instance, by considering the Sierpinski 
construction for the tetrahedron, as illustrated in Figure~\ref{FigTetra}.
For this tetrahedral Sierpinski construction 
we have $f_\rho=4$ and $f_\sigma =2$, so that the self-similarity dimension is 
$$ \dim_H \bP_T = \frac{\log f_\rho}{\log f_\sigma}=2, $$
where $\bP_T$ is the resulting fractal arrangement of tetrahedra arising from a
given triangulation $T$.  The corresponding fractal string zeta function is of the form
$$ \zeta_{\cL}(s) = \frac{1}{1- 2^{2-s}}\, ,  $$
with poles $s_n = 2 + \frac{2 \pi i n}{\log 2}$, for $n\in \Z$. Note that
in all these cases, the self-similarity dimension is $2$ (as expected for
cosmological reasons, \cite{LMP2}, \cite{LMP1}, though $M$ will in general
not satisfy the requirement of a candidate cosmological model).

\smallskip

In order to be able to perform the gluing of the tetrahedra,
as prescribed by the triangulation, at each level of the fractal construction, we again
need to use the fact that the tetrahedra in each subdivision of a previous level
are only adjacent along vertices and not along faces. 
The resulting fractal arrangement of copies of $M$ will be denoted by $\cP_M$. 
For a higher dimensional simplex $\Delta_n$ generalizations of the Sierpinski
construction have been considered, for instance, in \cite{YFY}.

\smallskip

If the spectral action $\cS_\Lambda(\cD_M)$ for the $3$-manifold $M$ is given by
$$ \cS_\Lambda(\cD_M) =\sim  \sum \frac{\frf_\alpha \Lambda^\alpha}{2} \Res_{s=\alpha}\zeta_{\mathcal{D}_{M}}
 + f(0) \zeta_{\mathcal{D}_{M}}(0) \, , $$
then the spectral action of the fractal arrangement $\cP_M$ is
$$ \begin{array}{rl} \cS_\Lambda(\cD_{\cP_M}) \sim &
\displaystyle{ \sum \frac{\frf_\alpha \Lambda^\alpha\, \zeta_{\cL}(\alpha)}{2} 
\Res_{s=\alpha}\zeta_{\mathcal{D}_{M}}}
 +  f(0) \zeta_{\mathcal{D}_{M}}(0) \zeta_{\cL}(0)  \\[3mm]
 + & \displaystyle{ \sum_{n\in \Z} \frac{\frf_{s_n} \Lambda^{s_n} \zeta_{\cD_M}(s_n) }{2 \log 2} } 
\end{array} 
$$
$$  \begin{array}{rl} 
 \sim & 
  \displaystyle{ \sum \frac{\frf_\alpha \Lambda^\alpha  }{2 (1- 2^{2-\alpha})} 
\Res_{s=\alpha}\zeta_{\mathcal{D}_{M}}}
-  \frac{ f(0) \zeta_{\mathcal{D}_{M}}(0)}{3}   \\[3mm]
+ & \displaystyle{ \sum_{n\in \Z} \frac{\frf_{s_n} \Lambda^{s_n} \zeta_{\cD_M}(s_n) }{2 \log 2} } 
 \, .
\end{array} $$

The asymptotic expansion for the spectral action on a static model $S^1_\beta \times M$
can then be obtained from this along the same lines as in the cases discussed in the
previous section. We don't have in this case a direct analog of the argument for the
Robertson--Walker models, unless something more explicit is known about the
spectrum and eigenspaces of $\cD_M$, which is usually the case only for homogeneous
spaces like those we considered before. 

\smallskip
\subsubsection{Koch-type fractal growth} 

\begin{figure}
\begin{center}
\includegraphics[scale=0.35]{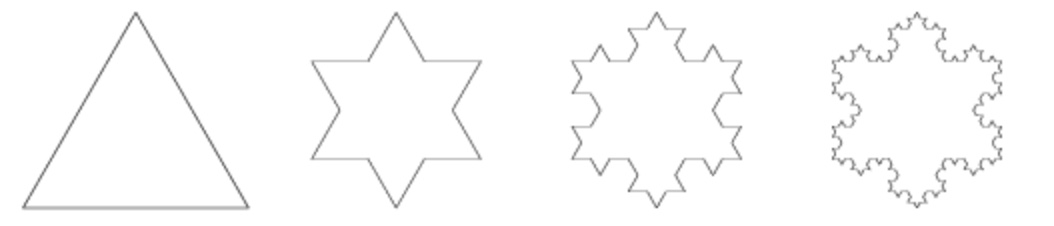}
\includegraphics[scale=0.35]{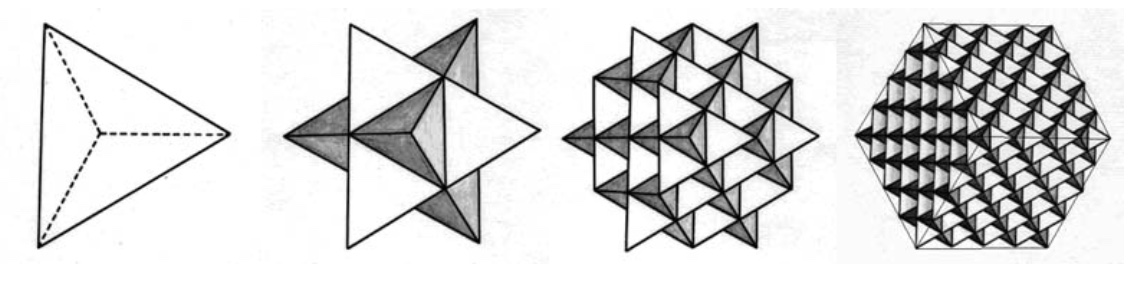}
\includegraphics[scale=0.25]{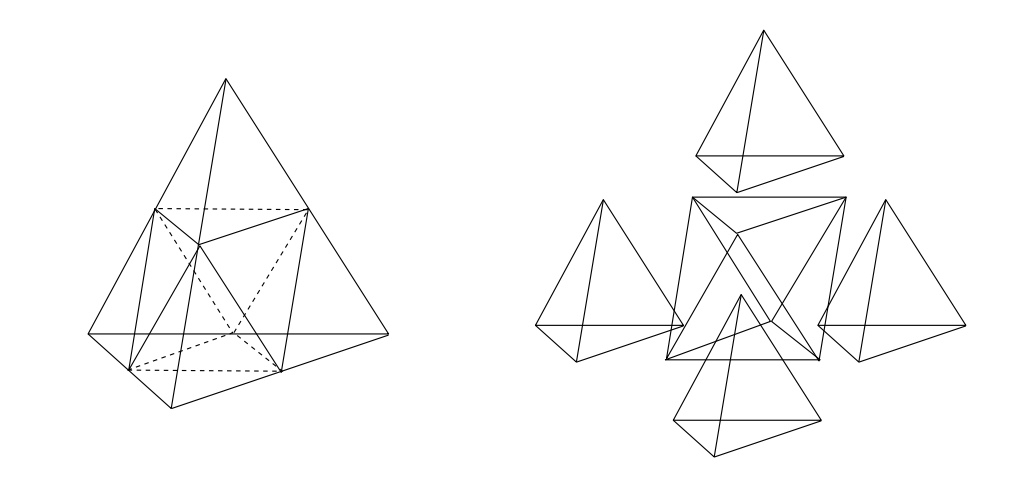}
\end{center}
\caption{The first three steps in the Koch construction for a triangle (Koch snowflake, \cite{Wein}) 
and for a tetrahedron, \cite{tetr}, and the tetrahedron decomposition into tetrahedra 
and an octahedron. \label{FigKoch}}
\end{figure}

There are other natural fractal constructions associated to tetrahedra (and to higher
dimensional simplices) to which we cannot directly apply the argument described
in the previous subsection for the Sierpinski construction. The simplest such example
is the Koch construction. In the case of triangles, the Koch construction produces
the well known Koch snowflake as the boundary curve. At each step of the Koch curve 
construction, one replaces each segment with $4$ segments of length $1/3$ so the 
self-similarity dimension is $\log(4)/\log(3)$. Note that this is a fractal construction
based on the boundary curve. If we consider $2$-dimensional triangles,  then
at the first step of the construction the initial triangle is replaced by 
$12$ triangles with scaling factor $1/3$, as all the triangles filling the triangle
of the previous step are also retained. So at each next step the number $T_n$
of triangles is $T_n=9 \cdot T_{n-1} + S_{n-1}$ where $S_{n-1}$
is the number of boundary sides $S_n = 3\cdot 4^{n-1}$.

\smallskip

In the case of  a tetrahedron, the first steps of
the Koch construction are illustrated in Figure~\ref{FigKoch}. 
At each step of the Koch construction one attaches a new tetrahedron, scaled by $1/2$, 
in the center of each of the triangular faces of the previous step. As in the case of
the Koch snowflake, we can view this as a fractal construction for the boundary
triangles, or for the interior solids. The boundary has exact self-similarity in the
sense that $6$ new  triangles of size $1/2$ replace each triangular face with 
$\dim_H =\log(6)/\log(2)$ and the zeta function is then just given by
$\zeta_{\cL}(s)=(1-6\cdot 2^{-s})^{-1}$. However, the interior solid does not
have exact self-similarity, since the interior region of the first
tetrahedron is replaced in the first step of the Koch construction by a
union of an octahedron and $8$ tetrahedra, but the octahedron does not
further decompose into other regular tetrahedra.

\begin{figure}
\begin{center}
\includegraphics[scale=0.45]{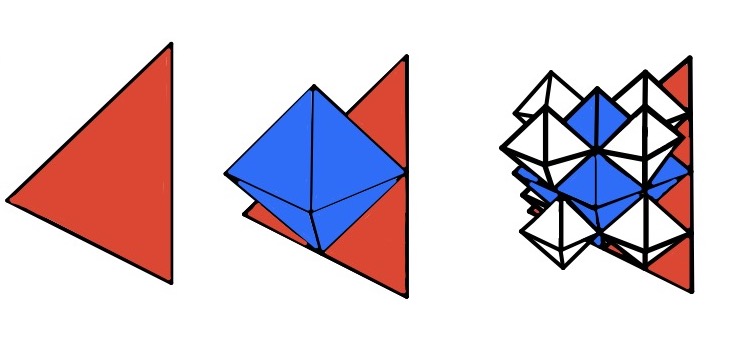}
\includegraphics[scale=0.15]{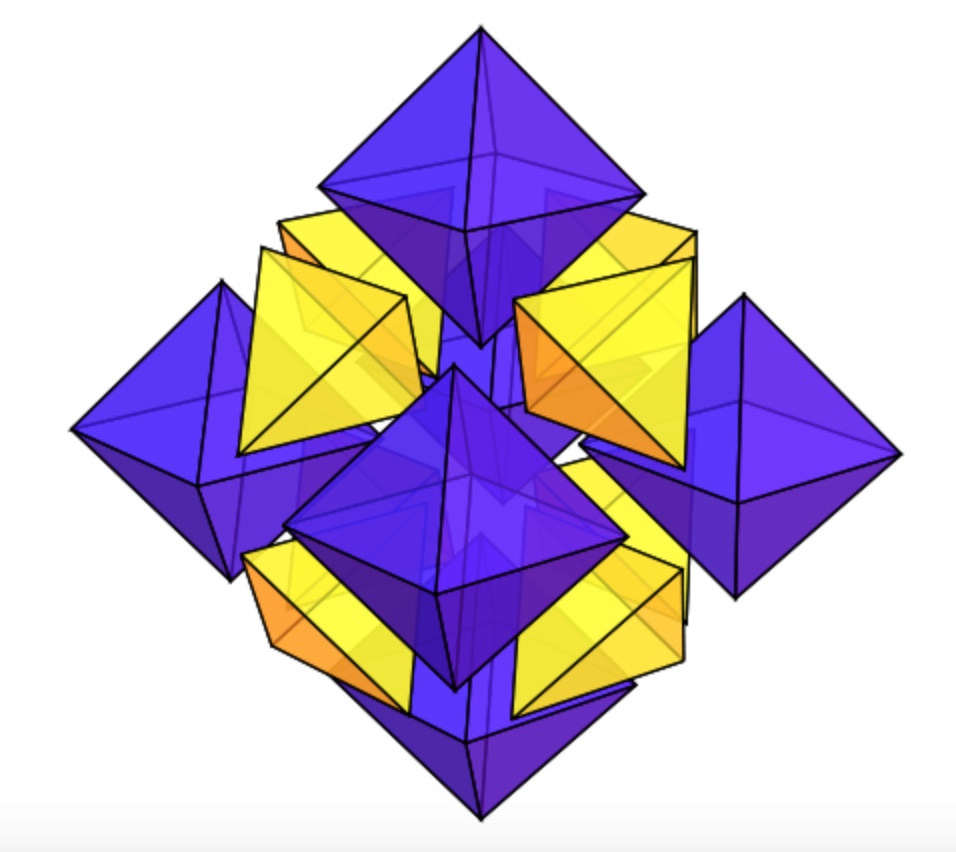}
\end{center}
\caption{The first steps of a Koch octahedron construction, on one of the faces 
of the original solid, and the decomposition of an octahedron into six
octahedra with $1/2$ scale factor and four tetrahedra. \label{OctaKochFig}}
\end{figure}

\smallskip

We can consider a similar case by constructing a Koch-like fractal from an octahedron. 
Starting with an octahedron, in each of its faces we place another octahedron scaled by $\frac{1}{2}$,
see Figure~\ref{OctaKochFig}. Again the situation is the same as in the tetrahedron case.
The fractal construction seen on the $2$-dimensional surface has exact self-similarity with
each triangle being first subdivided into four subtriangles of size $1/2$, of which three are
retained and $7$ new ones are added in place of the fourth one.  
Due to the geometry of the octahedron, in the second
step of the construction, three of the 
scaled copies placed on each face are tangent to faces both in the original face (red) and in 
the octahedron from the previous iteration (blue). Thus, if $T_0$ is the number of
triangles in the original copy, namely $T_0=8$, at the first step we have
$T_1=10\, T_0=80$ (three subtriangles in each original one plus $7$ new ones in place of the fourth one), while
at the second step we have $T_2= 10\, T_1 - 3 T_0$, where the $3 T_0$ accounts for those new faces
that end up matching subtriangles of the original faces. We can repeat this process for each 
of the next iterations: at the $n$-th step the number of triangles is $T_{n+1}= 10\, T_n - 3 \, T_{n-1}$. 
The solution of this recursion with $T_{-1}=0$ and $T_0=8$ is
$$ T_n =(\frac{10}{11} \sqrt{22}+4)\, (5+\sqrt{22})^n+(-\frac{10}{11}\sqrt{22}+4)\, (5-\sqrt{22})^n \, , $$
with the first few terms equal to $T_1=80$, $T_2=776$, $T_3=7520$, $T_4=72872$,
$T_5=706160$, etc. This example is slightly different from the cases of self-similarity 
we analyzed so far, because instead of having a single replication factor $f_\rho$ and
a similarity factor $f_\sigma$, in this case we have a weighted combination of two different
replication factors $f_{\rho,\pm}=5\pm\sqrt{22}$. (Note that here $f_{\rho,\pm}$ are not integers
but their weighted combinations $T_n =a_+ f^n_{\rho,+}+a_- f^n_{\rho,-}$, 
with $a_\pm=4 \pm \frac{10}{11} \sqrt{22}$ are integers. 
Thus, in this case we have
$$ \cL = \{ a_{n,k}=2^{-n} \, |\, k=1, \ldots, T_n \} $$
so that the zeta function is
$$ \zeta_{\cL}(s)=\sum_{n\geq 0}\, T_n\,\, 2^{-ns} = $$
$$ = (\frac{10}{11} \sqrt{22}+4) \, \sum_n  (5+\sqrt{22})^n 2^{-ns} +(-\frac{10}{11}\sqrt{22}+4)\, \sum_n (5-\sqrt{22})^n 2^{-ns} $$
$$ =\frac{ (\frac{10}{11} \sqrt{22}+4) }{ 1-  (5+\sqrt{22}) 2^{-s} } + \frac{(-\frac{10}{11}\sqrt{22}+4)}{1-  (5-\sqrt{22}) 2^{-s} } \, . $$ 

\smallskip

The $3$-dimensional solid octahedra in this construction do not have exact self-similarity.
Indeed, as in the case of the tetrahedron, an octahedron does not decompose exactly into
octahedra of $1/2$ the size, but requires additionally four tetrahedra (see Figure~\ref{OctaKochFig}).

\smallskip

Moreover, in the  tetrahedral case we cannot directly proceed as in the Sierpinski case to glue the tetrahedra
of each step of the construction into a manifold triangulation, as some faces of the tetrahedra are 
already glued to parts of faces of tetrahedra of the previous step. Similarly for the octahedral construction.

\smallskip

However, one still expects that
interesting fractal versions of simplicial objects should be obtainable from this type of fractal growth
model as well. To this purpose, one needs to analyze more in depth the usual notion of simplicial
sets, and the analogous notion of cubical sets, in algebraic topology and enrich it with a notion of
fractality implemented by scaling maps. We will discuss the mathematical structure necessary 
for this construction, and some of its applications, in a forthcoming paper.

\bigskip 
\section{Arithmetic structures in fractal packings}\label{ArithmSec}

In this section we consider again the Packed Swiss Cheese Cosmology models
considered in \cite{BaMa}, \cite{FKM}, based on Apollonian packings of spheres,
as in \cite{MuDy}. As observed in other settings, \cite{FFM1}, \cite{FFM2}, \cite{FM1},
cosmological models based on the spectral action often reveal interesting {\em arithmetic}
structures. Thus, we focus here on a class of Apollonian packing that have
nice arithmetic properties and we investigate how the associated spectral action function 
reflects the presence of an arithmetic structure. 

\smallskip
\subsection{Integral Apollonian packings}

We consider the integral Apollonian packings as in \cite{GLMW}, \cite{GLM}, based on $3$-spheres. 
These are packings where the curvatures of all spheres are integers. Note that the term ``curvature" 
in this setting denotes the {\em oriented curvature}, $a_i = \frac{1}{r_i}$.
We recall the general structure of Apollonian packings from \cite{GLM}. Explicit examples of
constructions of sphere packings of $3$-spheres and $4$-spheres are given in \cite{StaHerr}. 

\smallskip

Apollonian packings of $(n-1)$-spheres in $\mathbb{R}^n$ are characterized by 
$n+2$ tangent spheres obeying the Descartes relation for $n$ dimensions
$$\sum_{i=1}^{n+2} a_i ^2 = \frac{1}{n}(\sum_{i=1}^{n+2} a_i )^2\, ,$$
where $a_i = \frac{1}{r_i}$ is the curvature of each sphere. Following \cite{GLM} we
consider
$$ Q_{D,n} := \textbf{I}_{n+2} - \frac{1}{n} 1_{n+2} 1_{n+2}^T\, , $$
where $\textbf{I}_{n+2}$ is the identity and $(1_{n+2})_j =1, \forall j\in \{ 1,2,...,n+2 \}$ with $ \textbf{I}_{n+2} \in M_{n+2}(\mathbb{R})$ and $1_{n+2}\in \mathbb{R}^{n+2}$. The Descartes relation for $(n-1)$-spheres can
then be written equivalently as
$$ a^T Q_{D,n} a = 0\, . $$
Consider a Lorentz quadratic form:
$$a^T Q_{L,n} a = -a_1^2 + a_2^2 + ... + a_{n+2}^2 = 0 \, . $$
Descartes forms and Lorentz forms are real-equivalent in all dimensions, as discussed in \cite{GLM}.
This leads to a bijection $\{a^\mathcal{D}_{n+2}\} \simeq \{a^\mathcal{L}_{n+2}\}$, which means  
Descartes forms can be counted using the Lorentz forms.
We also need to recall the notion of dual Apollonian packing from  \cite{GLM}. 
The packing is described by the dual Apollonian group
$$ \mathcal{A}^\bot_n = \{S_1^\bot,S_2^\bot ,... , S_{n+2}^\bot \}\, ,$$
where $S_i^\bot = \textbf{I}_{n+2} + 2 \cdot 1_{n+2} e_i^T -4\cdot e_i e_i^T $, with $1_{n+2}$ as before 
and $e_i$ the standard Cartesian basis in $\mathbb{R}^{n+2}$.  Let $\textbf{W}_{\mathcal{D}}$  denote
the curvature-center coordinate matrix of the original Descartes form. The dual Apollonian group acts
on it by $S_i^\bot \textbf{W}_{\mathcal{D}}$.

\smallskip

Here, we will work with the dual Apollonian configuration of sphere packings, since it has been shown to 
not intersect, when taking higher dimensional spheres in the packing, whereas the usual Apollonian construction leads 
to intersections under the same circumstances. Additionally, Theorem~4.1 of \cite{GLM} suggests that Apollonian 
packings in higher dimensions do not necessarily remain integral. However, dual Apollonian packings do retain integral 
curvature for $n>1$. Altogether, Apollonian configurations in higher dimensions are not always an appropriate choice for 
sphere packings, but dual Apollonian configurations retain all the usual qualities associated with circle packings for higher dimensions.

\smallskip

In the construction of the dual Apollonian packing, if the initial configuration of spheres has
integer curvatures, then all the curvatures will remain integer in the whole packing. The packing
zeta function can then be described as
$$ \zeta_{\mathcal{L}}(s) = \sum_{n,k} r_{n,k}^s = \sum_{n\in\mathbb{N}} \frac{r^*_5(a_n^2)}{a_n^s}\, , $$
where $r^*_5(k)$ denotes the number of primitive integer representations of $k$ as a sum of $5$ 
square integers. As shown in \cite{GLM2}, this is given by 
$$ \sum_{n\in\mathbb{N}} \frac{r^*_5(a_n^2)}{a_n^s} = \frac{1}{\zeta(s)} \sum_{n\in\mathbb{N}} \frac{r_5(a_n^2)}{a_n^s}\, ,$$
with $r_5(k)$ the number of general integer representations of $k$. 
It is shown in \cite{Sand} that this sum simplifies to
$$\sum_{n\in\mathbb{N}} \frac{r_5(a_n^2)}{a_n^s} =  \frac{10\,\, \zeta(s)\zeta(s-3)}{\zeta(s-1)(1-2^{1-s})}\, , $$
so that we obtain 
\begin{equation}\label{NTzetaL}
\zeta_{\mathcal{L}}(s) = \frac{10\,\, \zeta(s-3)}{\zeta(s-1)(1-2^{1-s})} \, . 
\end{equation}

\smallskip

The set of poles of the zeta function \eqref{NTzetaL} is the union of the following sets
\begin{enumerate}
\item The point $s=4$, which is a pole of $\zeta(s-3)$ with residue 
$$ {\rm Res}_{s=4} \zeta_\cL = \frac{80}{7\, \zeta(3)}\, . $$
\item The points $s_n=1+ \frac{2\pi i n}{\log 2}$, with $n\in \Z$, which are poles of $(1-2^{1-s})^{-1}$, with
residue
$$ {\rm Res}_{s=s_n} \zeta_\cL = \frac{10\,\, 
\zeta(\frac{2\pi i n}{\log 2}-2)}{\zeta(\frac{2\pi i n}{\log 2})\, \log 2}\, . $$
\item The nontrivial zeros of the zeta function $\zeta(s-1)$, given by the set
$\{ s_\rho = 1+\rho\,|\, \rho\in \cZ(\zeta) \}$, with $\cZ(\zeta)$ the set of non-trivial zeros of the Riemann zeta function, with residues
$$ {\rm Res}_{s=s_\rho} \zeta_\cL = \frac{10\,\, \zeta(\rho-2)}{(1-2^{-\rho})\, \zeta^\prime(\rho)} \, . $$
\end{enumerate}
Note that the trivial zeros of $\zeta(s-1)$, at the points $s_k =1-2k$ with $k\in \N$, do not contribute poles because also $\zeta(-2(k+1))=0$ in the numerator of \eqref{NTzetaL}.

\smallskip

We see that, in the case of a Packed Swiss Cheese Cosmology model based on
an integral Apollonian packing of $3$-spheres, the spectral action expansion will
contain a series over the non-trivial zeros of the Riemann zeta function, of the form
$$ \sum_\rho \frac{5 \, \frf_{s_\rho} \, \zeta_{\cD_{S^3}}(1+\rho) \, \zeta(\rho-2)}{(1-2^{-\rho})\, \zeta^\prime(\rho)}\,      \Lambda^{1+\rho}\, . $$
The behavior of this series depends on delicate number theoretic properties of the Riemann zeta
function and on whether the Riemann hypothesis holds. For example, 
estimates on the behavior of the derivative $\zeta^\prime(\rho)$ of the
Riemann zeta function at non-trivial zeros are derived in \cite{HiOd} and \cite{Fuj}. 
In particular, series of the form
$$ \sum_\rho \zeta^\prime(\rho) \Lambda^\rho \ \ \ \text{ and } \ \ \   \sum_{\rho_n} \zeta^\prime(\rho_n) e^{2\pi i n x} $$
have been studied in  \cite{HiOd} and \cite{Fuj}. The series above, which arises naturally 
in our physical model, appears to be of independent interest, though we will not pursue its
analysis further in the present paper. 

\subsection{Perturbation of sphere tangencies}

The Packed Swiss Cheese Cosmology model, as described in \cite{MuDy}, uses Apollonian
packings of $3$-sphere as a convenient mathematical model for achieving a fractal arrangement
that can be seen as an iterative construction based on the original idea of Rees and Sciama
\cite{ReeSci} on how to introduce inhomogeneities in a Robertson--Walker spacetime. There
are however some limitations to the Apollonian packing model. For example, the condition of
tangency is geometrically not an open condition, in the sense that small random perturbations
of the data will break this condition and will yield intersecting spheres. The latter is a generic
condition that is stable under small perturbations. Moreover, in a realistic physical model one 
would not be working with exact self-similar fractals, but with random fractals as the latter
provide more realistic physical models. Thus, one should think of a physical or cosmological
process that generates fractality as a random process where bubbles randomly open up in
the fabric of Robertson--Walker spacetime, in such a way that what remains acquires  
a fractal geometry. However, if the subtraction of bubbles of spacetime is achieved through
a random process, the resulting geometry will not look like a configuration of tangent spheres
like an Apollonian packing, which is a deterministic fractal, but rather like a configuration of
intersecting spheres with a sequence $\cL=\{ a_{n,k} \}$ of radii. Thus, we can imagine
a model where one starts with an Apollonian packing and maintains the same sequence
of radii but introduces a random perturbation of the positions of the centers of the spheres,
which causes some of the spheres to overlap with intersection along a hypersphere rather
than only at a tangent point. 

\begin{figure}
\begin{center}
\includegraphics[scale=0.2]{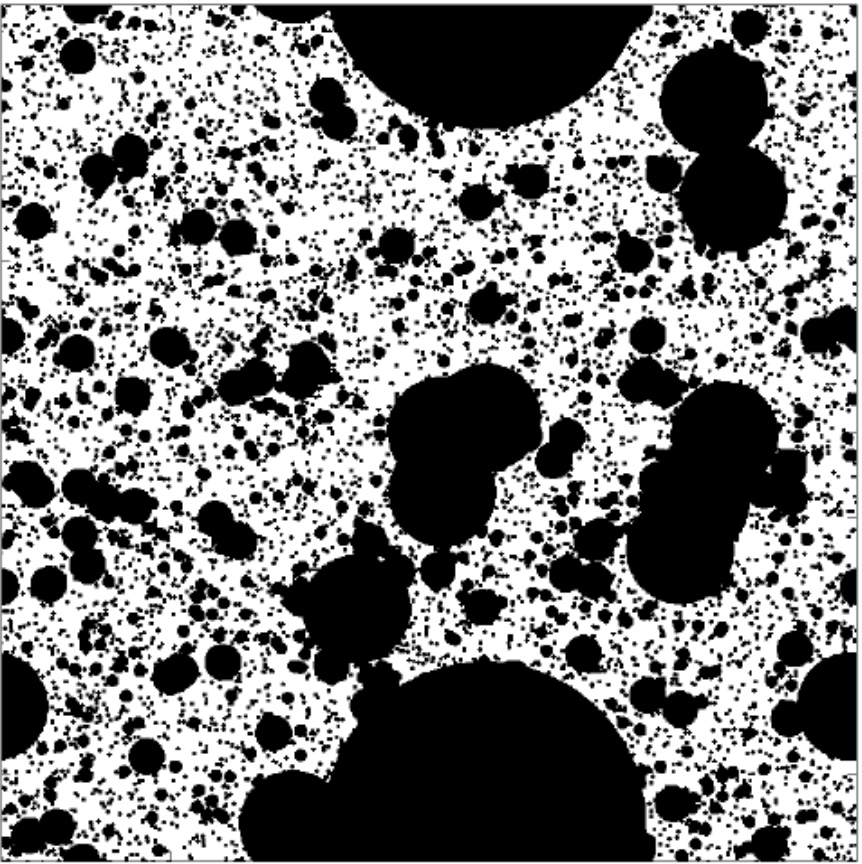}
\end{center}
\caption{Example of a random fractal created by cratering (removal of discs/balls). In the example illustrated here the
sequence of radii is simply taken as $a_{n,k}=3^{-n}$ (from a geophysical model of \cite{WPV}). \label{FigRandom}}
\end{figure}

\smallskip

We can also obtain simple models with intersecting spheres by considering 
Apollonian packings in lower dimensions as the intersection with an equatorial hyperplane
of a configuration of spheres in higher dimension. While for some special choices of
the lower dimensional packing the higher dimensional spheres will also intersect only
at the tangent points, in general the higher dimensional spheres have a larger intersection. 
Such constructions where an Apollonian packing of circles is used to construct a
fractal arrangement of higher dimensional spheres was considered already in \cite{FKM}
in the case of the Ford circles, which also have a nice arithmetic structure. 

\smallskip

We consider here another variant of this model, again by considering configurations
of $3$-spheres obtained starting from an Apollonian packing of circles, where we
consider packings with integer curvatures. We then take each circle and make it the 
equator of a $2$-sphere, and repeat the process one more time by taking each $2$-sphere 
and making it the equator of a $3$-sphere. The resulting configuration of $3$-spheres is
in general no longer an Apollonian packing, as the $3$-spheres now can have intersections
and not only tangency points. The fractal string for this configuration is the same as the 
$S^1$ case for integer curvature, which gives 
$$\zeta_{\mathcal{L}} (s) = \sum_{n,k} a_{n,k}^s = \sum_n \frac{r_3^*(a^2_n)}{a_n^s}\, . $$
The evaluation of this series is similar to the case of integral packings of $3$-spheres
discussed above. 

\smallskip

Let $\eta(s)$ be the Dirichlet $\eta$-function
$$ \eta(s)=\sum_{n\geq 1} \frac{(-1)^{n-1}}{n^s} = (1-2^{1-s}) \zeta(s)\, . $$
Consider then the series 
$$ \tilde\eta(s) = \sum_{n\geq 1} \frac{(-1)^{n-1}}{(2n-1)^s} =
\sum_{n\geq 1}\left(  \frac{(-1)^{n-1}}{n^s} -\frac{(-1)^{n-1}}{(2n)^s} \right) =
(1-2^{-s}) \eta(s)\,. $$
We obtain, as in  \cite{GLM2},  \cite{Sand},
$$\zeta_{\mathcal{L}} (s) = 6 \frac{\zeta(s-1) (1-2^{1-s})}{\tilde\eta(s)} $$
$$ =6 \frac{\zeta(s-1) (1-2^{1-s})}{(1-2^{-s}) \eta(s)} = 6 \frac{\zeta(s-1)}{(1-2^{-s}) \zeta(s) }     \, .$$
Note that \cite{Sand} uses the notation $\eta(s)$ for our $\tilde\eta(s)$. 
In this case again we have poles at $s_n=\frac{2\pi i n}{\log 2}$, at $s=2$, and at the
zeros (in this case both the trivial and the nontrivial zeros) of the Riemann zeta function.
Again this means that the spectral action expansion for this configuration of $3$-spheres
will contain a series with a sum over the nontrivial zeros of the same general form we
discussed above,
$$ \sum_\rho \frac{3 \, \frf_{s_\rho}\, \zeta_{\cD_{S^3}}(\rho) \zeta(\rho-1)}{(1-2^{-\rho}) \zeta^\prime(\rho)} \Lambda^\rho \, . $$

\section{Gravitational waves in fractal models with intersecting spheres}\label{GrWavesSec}

In this section we discuss some aspects of the gravitational waves behavior in fractal
cosmological models. We focus on the case discussed at the end of the previous
section, where instead of packings of spheres that only touch at tangent points, we
allow for perturbations of the geometry that introduce intersections between some of
the $3$-spheres along some $2$-spheres. 

\subsection{Transmission between spheres in perturbed packings}

In this section, we discuss some consequences of being an observer in one of the 
positively curved spaces in a fractal configuration of $3$-spheres obtained as
a perturbation of a sphere packing configuration. Specifically, we consider the intersections 
of the spheres as a transmission site for gravitational waves going from one $3$-sphere to another 
and derive an approximate form of the metric describing the gravitational wave in the receiving sphere. 
As we will see, this calculation implies there exist directly observable phenomena that come 
from having positively curved universes intersect each other.

\begin{figure}
\begin{center}
\includegraphics[scale=0.25]{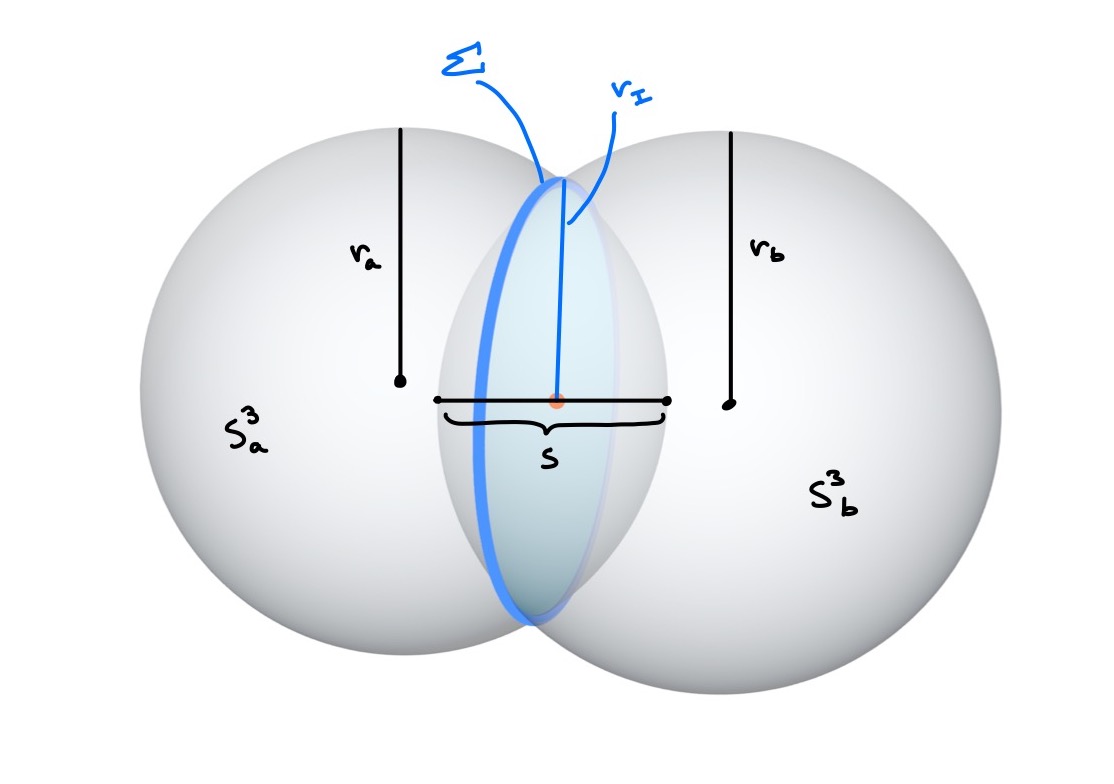}
\end{center}
\caption{Two 3-spheres $S^3_a$ and $S^3_b$ with radii $r_a$ and $r_b$, intersecting at a 2-sphere, $\Sigma$, with separation distance $s$. \label{IntSabFig}}
\end{figure}

\smallskip

We begin by considering the intersection of two 3-spheres, $S^3_a$ and $S^3_b$ along a $2$-sphere
$\Sigma \equiv S^3_a \bigcap S^3_b = S^2_I$.
We write $s$ for the distance between the poles of the two smaller spherical caps subtended
by the $2$-sphere of intersection, see Figure~\ref{IntSabFig}. We refer to $s$ as the ``separation
distance" of $S^3_a$ and $S^3_b$. 
An quick calculation yields the relation between the radii, 
$$r_I ^2 = \frac{(2r_b-s)(2r_a-s)(2r_a+2r_b-s)s}{4(r_a+r_b-s)^2}\, . $$

\smallskip

In the case of a Robertson-Walker cosmology, each $3$-sphere would be undergoing inflation, which 
would affect this calculation. For simplicity, we will consider only the static model, 
where there is no time evolution in the shape of each sphere, so both the radii $r_a,r_b$ and
the relative position of the two $3$-spheres are constant. 
Thus, an observer in each sphere follows the same spatial hypersurface-orthogonal timelike 
vector $\eta^\mu = (1,0,0,0)$. 

\smallskip

Also, for the purpose of the computations presented in this section, we work directly with
Lorentzian signature, unlike for the previous spectral action calculations, where 
we had to Wick rotate to Euclidean signature.

\smallskip

Another assumption for this section is that $r_b>>r_a$. This makes it reasonable to assume that the region surrounding $\Sigma$ in $S^3_b$ can be approximated by Minkowski space $\mathbb{R}^{3,1}$, which we can use to significantly simplify the solution of the metric tensor under linearized General Relativity (GR) conditions. 

\smallskip

We discuss briefly the boundary conditions in GR. Following \cite{Poiss}, if $x^\mu$ and $x'^\mu$ 
are local coordinates on the spheres $S^3_a$ and $S^3_b$, respectively, 
and $y^a$ are the coordinates on $\Sigma$, then the tangent curves and induced 
metric on $\Sigma$ are given by
$$e^\mu_a = \frac{\partial x^\mu}{\partial y^a}\, ,$$
$$h_{ab} = g_{\mu\nu} e^\mu_a e^\nu_b\, .$$
The extrinsic curvature tensor is given by 
$$K_{ab} := \nabla_\nu \eta_\mu e^\mu_a e^\nu_b\, , $$
where $\eta_\mu$ is the orthogonal vector to $\Sigma$ in either $S^3_a$ or $S^3_b$.

\smallskip

As above, let $s$ be the separation distance of the two hypersurfaces
and let $\ell$ be a coordinate in $[-s/2,s/2]$.
Let $\Theta(\ell)$ be the Heaviside function. We consider a 
stress energy-momentum tensor of the form
$$ T_{\mu\nu} = T^a_{\mu\nu}\,\Theta(-\ell) +S_{\mu\nu}\,\delta(\ell)+T^b_{\mu\nu}\,\Theta(\ell)\, , $$
where $S_{\mu\nu}$ is the surface energy-momentum tensor (see \cite{Poiss}) given by
$$ S_{ab} = - \frac{\eta_{\mu} \eta^\mu}{8\pi} ([K_{ab}] - [K]h_{ab})\, .$$
Here the  jump of a quantity across $\Sigma$ is given by
$$[A^{\mu_1 \mu_2 ... \mu_n}_{\nu_1 \nu_2 ... \nu_k}] =A^{\mu_1 \mu_2 ... \mu_n}_{\nu_1 \nu_2 ... \nu_k}|_{\Sigma^+} - A^{\mu_1 \mu_2 ... \mu_n}_{\nu_1 \nu_2 ... \nu_k}|_{\Sigma^-}\, ,$$ 
with $+,-$ for the two hypersurfaces $\Sigma^\pm$ (here given by the spheres $S^3_a$ and $S^3_b$).

\smallskip

The respective unperturbed metrics on $S^3_a$, $S^3_b$ and $\Sigma$ are given by
$$ds_a^2 = -dt^2 + r_a^2 (d\psi_a^2 + \sin^2\psi_a\, (d\theta_a^2 + \sin^2\theta_a \, d\phi_a^2))\, ,$$
$$ds_b^2 = -dt^2 + r_b^2 (d\psi_b^2 + \sin^2\psi_b\, (d\theta_b^2 + \sin^2\theta_b \, d\phi_b^2))\, ,$$
$$ds_\Sigma^2 = -dt^2 + r_I^2 (d\theta_I^2 + \sin^2\theta_I \, d\phi_I^2)\, .$$

\smallskip

We orient each shere so that the $2$-sphere $\Sigma$ is located at a constant value of the angle 
coordinate $\phi$ for each $3$-sphere. The variation of the other two coordinates in each system of
local coordinates on a $3$-sphere maps out a $2$-sphere. With the correct choice of 
$$\phi_a = \phi^a_o \equiv {\rm arcsin}(\frac{r_I}{r_a})\ \ \text{ and } \ \  
\phi_b = \phi^b_o \equiv {\rm arcsin}(\frac{r_I}{r_b}), $$ 
this gives the coordinate description of $\Sigma$ in the respective coordinate systems. 
This implies the vectors orthogonal to $\Sigma$ are $\eta^\mu = (0,0,0,1)$ in both spaces 
and the tangent curves are, respectively, given by
$$e^a_\mu = \begin{pmatrix} 
1 & 0 & 0 \\
0 & \frac{r_I}{r_a} & 0\\
0 & 0 & \frac{r_I \sin\theta_I}{r_a \sin\psi_a} \\
0 & 0 & 0
\end{pmatrix} \, ,$$
$$e^b_\mu = \begin{pmatrix} 
1 & 0 & 0 \\
0 & \frac{r_I}{r_b} & 0\\
0 & 0 & \frac{r_I \sin\theta_I}{r_b \sin\psi_b} \\
0 & 0 & 0
\end{pmatrix}\, .$$

\begin{figure}
\begin{center}
    \includegraphics[scale=0.3]{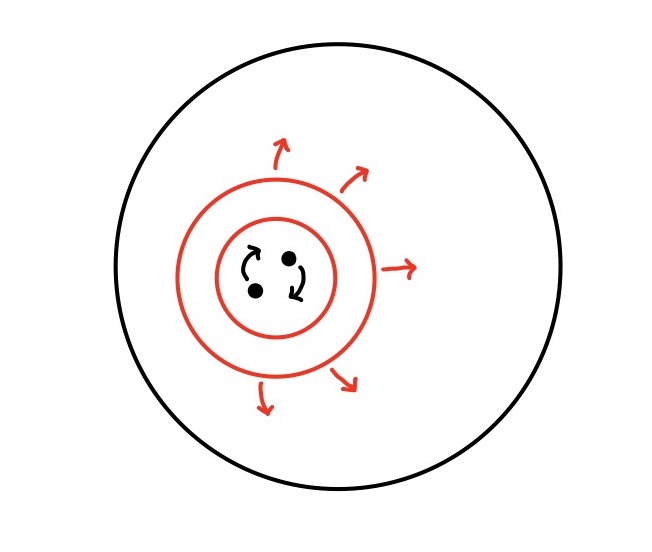}
    \caption{Gravitational waves production on a $2$-sphere. \label{WaveS2Fig}}
\end{center}   
\end{figure}

\smallskip

Note that in a $2$-sphere, plane waves due to the spiralling of two objects in 
spacetime would be produced by distorting the $\theta$-component of the metric while 
moving in the $\phi$-direction, see Figure~\ref{WaveS2Fig}. So, the wave perturbation to the metric looks like 
$$ \Psi_{\mu\nu} = {\rm diag}(0,A \sin(k(\phi - t)),0) \, . $$

\smallskip

Thus, we can consider a perturbation to $S^3_a$ of the form
$$g^a_{\mu\nu} \rightarrow g^a_{\mu\nu} + \epsilon \Psi_{\mu\nu},\ \ \ \text{ with } |\epsilon|<<1 \, ,$$
with $\Psi_{\mu\nu} = {\rm diag}(0,A_\psi \cos(k(\phi_a - t)),0,0)$. 

\smallskip

\begin{figure}
\begin{center}
    \includegraphics[scale=0.2]{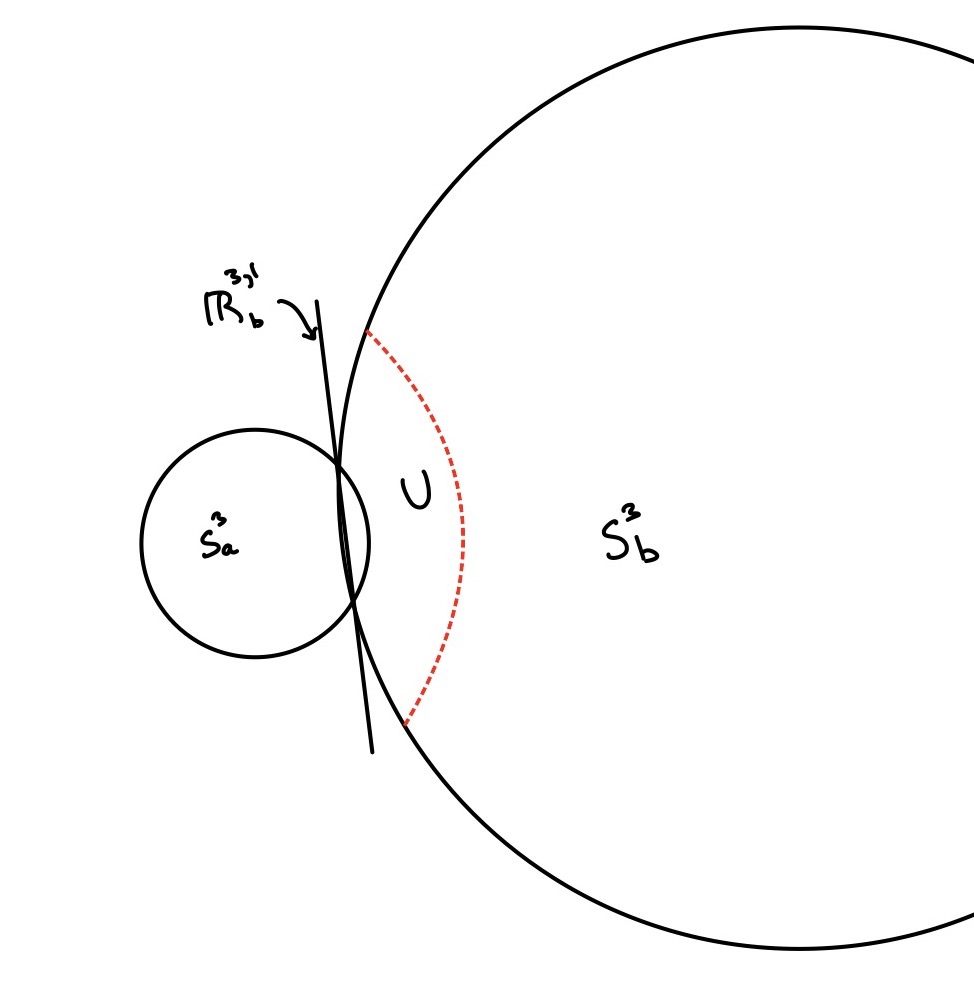}
    \caption{As $r_b \rightarrow \infty$, an open $3$-ball $U$ on $S^3_b$ with $\Sigma \subset U$ can be approximated by some $U'\subset \mathbb{R}^3$. \label{Wave3Fig}}
\end{center}   
\end{figure}

We then obtain the surface energy-momentum tensor induced by this perturbation, of the form
$$ S_{ab} = \begin{pmatrix}
-\frac{A_\psi k \epsilon \sin(k(\phi_o -t))}{16\pi(r_a^2 + A_\psi k \epsilon \cos(k(\phi_o -t)))} & 0 & 0\\
0&0&0\\
0&0&\frac{A_\psi r_I^2 \sin^2\theta_I k \epsilon \sin(k(\phi_o -t))}{16\pi(r_a^2 + A_\psi k \epsilon \cos(k(\phi_o -t)))}
\end{pmatrix} \, . $$
We use the assumption that $r_b>>r_a$ to approximate the space surrounding $\Sigma$ in $S^3_b$ with a flat space. Specifically, we can imagine sending the metric inside some open $3$-ball $U\subset S^3_b$ to 
the limit $$ds^2_b|_U \rightarrow -dt^2 + dr^2 + r^2(d\theta^2 + \sin^2\theta d\phi^2).$$ 
In this limit, the metric component of the azimuth angle $\phi_b$ that we fixed becomes 
associated with the $r$-component for $\mathbb{R}^{3,1}_b$, while the components for the 
other two angles stay the same, since they still map out the same  $2$-sphere in $\mathbb{R}^{3,1}_b$, see
Figure~\ref{Wave3Fig}. 

Then the induced energy-momentum tensor on $U$, described in spherical coordinates, is approximated by 
$$T_{\mu\nu} \approx \begin{pmatrix}
-\frac{A_\psi k \epsilon \sin(k(\phi^a_o -t))}{16\pi(r_a^2 + A_\psi k \epsilon \cos(k(\phi^a_o -t)))} \delta(r-r_I) & 0 & 0&0\\
0&0&0&0\\
0&0&0&0\\
0&0&0&\frac{A_\psi r_I^2 sin^2\theta k \epsilon \sin(k(\phi^a_o -t))}{16\pi(r_a^2 + A_\psi k \epsilon \cos(k(\phi_o^a -t)))}\delta(r-r_I)
\end{pmatrix}\, . $$

\smallskip

Recall that, for a tensor $\gamma_{\mu\nu}$, we have $\gamma=\gamma^\mu_{\,\mu}=\eta_{\mu\nu}\gamma_{\mu\nu}$, with the flat Minkowski metric $\eta_{\mu\nu}$, and the trace-reverse of
$\gamma_{\mu\nu}$ is given by $\bar\gamma_{\mu\nu}=\gamma_{\mu\nu}-\frac{1}{2} \eta_{\mu\nu}\gamma$.
The equation for the trace-reversed pertubation to the metric tensor,
$\bar{\Psi}_{\mu\nu}$, in linearized gravity for the Lorenz gauge condition is then given by
$$ \Box \bar{\Psi}_{\mu\nu} = -16\pi G T_{\mu\nu} \, . $$
Using the known Green function for Minkowski space, 
$$G(x^\mu -x'^\mu) = -\frac{1}{4\pi |\mathbf{x-x'}|} \delta(\mathbf{|x-x'|} - (x^0 - x'^0))\Theta(x^0-x'^0)\, , $$
we can write our solution in the form
$$\bar{\Psi}_{\mu\nu} = 4G \int d^3x' \frac{1}{|\mathbf{x-x'}|} T_{\mu\nu} (t - |\mathbf{x-x'}|,\mathbf{x'})\, . $$

\smallskip

This integral is non-trivial for either $\bar{\Psi}_{\phi\phi}$ or $\bar{\Psi}_{tt}$. 
Since this calculation is meant to give us an overall idea of what happens at this boundary, 
we have opted to approximate the result in two cases:
\begin{enumerate}
\item the solution for the metric inside $\Sigma$ in $U$;
\item the solution for the metric outside and far away from $\Sigma$. 
\end{enumerate}

\smallskip
\subsubsection{First case: internal solution}

We begin with the solution in the first case listed above. 
In this case, to linear order in $\epsilon$, $T_{\phi\phi}$ we have
$$T_{\phi\phi} = \frac{A_\Psi k r_I^2}{16 \pi r_a^2} \sin^2\theta \sin(k(\phi_o^a -t)) \delta(r-r_I) \epsilon + O(\epsilon ^2) \, . $$
Thus, we obtain a solution for $\bar{\Psi}_{\phi\phi}$ given by the integral 
$$\frac{4GA_\Psi k r_I ^2 \epsilon}{16 \pi r_a^2} \int_{\mathbb{S}^2} d\Omega' \, \sin^2\theta'\,\, \frac{\sin(k(|\mathbf{x-x'}|-t))}{|\mathbf{x-x'}|}\, ,  $$
where we have suppressed the term $\phi^a_o$, since it is simply a phase factor. 
The Euclidean distance is given by
$$|\mathbf{x-x'}| = \sqrt{r^2 + r_I^2 - 2r r_I \mathbf{u\cdot u'} }\, , $$
where $\mathbf{u}$ and $\mathbf{u'}$ are the respective unit vectors on a sphere $S^2$. 

\smallskip

This integral can be solved exactly using the Funk-Hecke formula of harmonic analysis 
(see for instance \cite{Seeley})
$$\int f(x\cdot y) Y_k (x) \sigma_{\mathbb{S}^{n-1}}(dx) = \lambda_k Y_k (y)\, , $$
for any bounded measurable function $f: S^{n-1}\to \R$, and any $y\in S^{n-1}$, 
where $Y_k$ are the spherical harmonics on $S^{n-1}$ of degree $k\geq 0$ 
and the constant $\lambda_k$ is given by
$$\lambda_k = \frac{\Gamma(n/2)}{\sqrt{\pi}\, \Gamma((n-1)/2)} \int_{-1}^1 f(t) P_k (t) (1-t^2)^{(\frac{n-3}{2})}\, , $$
with $P_k$ the $k$-th Gegenbauer polynomial. 

\smallskip

Note that polynomials of sinusoids can be written in the basis of spherical harmonics 
on a sphere, so we may write
$$ \sin^2 \theta' = -\frac{1}{3} (4\sqrt{\frac{\pi}{5}} Y_2^0 (\theta',\phi')-2)\, , $$
where
$$Y_2^0 (\theta',\phi') = \frac{1}{4} \sqrt{\frac{5}{\pi}} (3\cos^2\theta -1)\, . $$
We then define
$$ f(x) := \frac{e^{ik\sqrt{\gamma - \beta x}}}{\sqrt{\gamma - \beta x}}\, , $$
where $\gamma = r^2 + r_I^2$ and $\beta = 2rr_I$. The integral can then be solved as
$$\bar{\Psi}_{\phi\phi} = A e^{-ikt} \bigg(- \frac{4}{3} \sqrt{\frac{\pi}{5}} \int d\Omega_{S^2}' Y^0_2 (\theta',\phi') f(\mathbf{u\cdot u'}) + \frac{2}{3} \int d\Omega_{S^2}' f(\mathbf{u\cdot u'}) \bigg)$$
$$= A e^{-ikt} \bigg(\sqrt{\frac{\pi}{5}} \frac{8\Lambda_2(\gamma,\beta)}{\beta^3 k^4} Y_2^0(\theta,\phi) +\frac{2i}{3\beta} (e^{ik\sqrt{\gamma-\beta}}-e^{ik\sqrt{\gamma+\beta}} ) \bigg)\, ,$$
where $A = \frac{4GA_\Psi r_I ^2 \epsilon}{16 \pi r_a^2}$ and $\Lambda_2(\gamma,\beta)$ is given by
$$\Lambda_2 =  e^{ik\sqrt{\gamma+\beta}}\, \mathbb{P}^+ +e^{ik\sqrt{\gamma-\beta}} \, \mathbb{P}^- \, , $$
with the $\mathbb{P}^\pm$ given by 
$$\mathbb{P}^+ = -(k^3\beta - 6k )\sqrt{\gamma+\beta} - i((2\gamma +3\beta)k^2 -\frac{\beta^2k^4}{6}-6)\, , $$
$$\mathbb{P}^- = -(k^3\beta + 6k) \sqrt{\gamma-\beta} + i((2\gamma -3\beta)k^2 -\frac{\beta^2k^4}{6}-6)\, .$$

\smallskip

\begin{figure}
\begin{center}
    \includegraphics[scale=0.137]{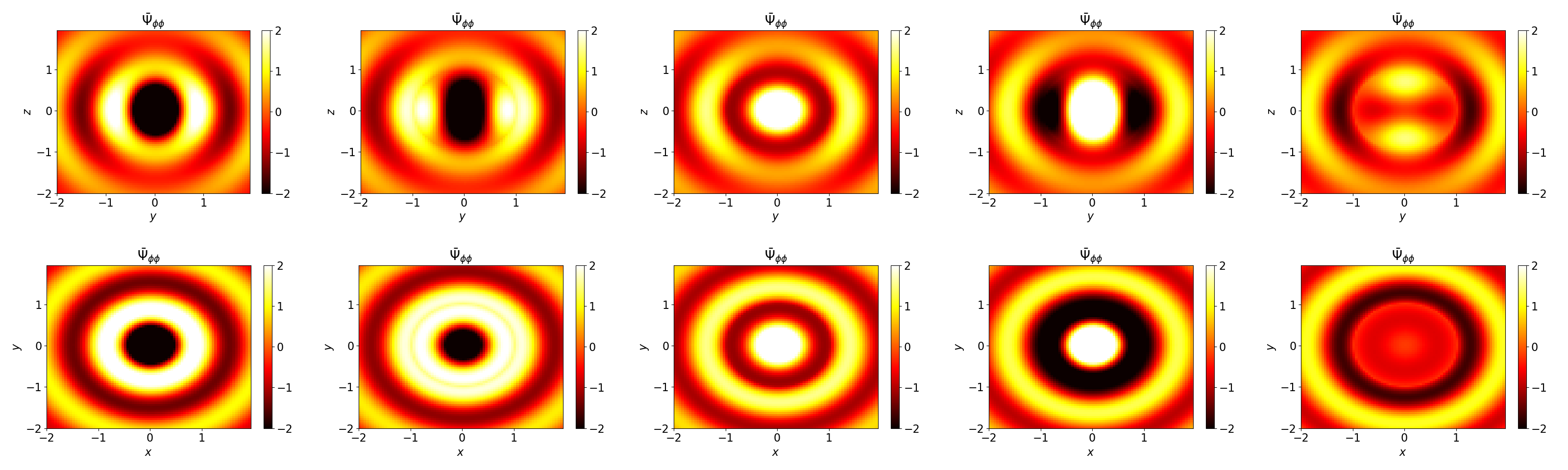}
    \caption{The numerical solution to $\bar{\Psi}_{\phi\phi}$, plotted in the emission plane (2nd row) and the perpendicular plane (1st row) at times $t = (\frac{i\cdot 2\pi}{5k})_{i=0,1,2,3,4}$ for the parameters described. The most obvious consequence is the inward and outward radial emissions in the $\theta = \pi/2$ plane (as $T_{\mu\nu}$ would suggest). The azimuthal symmetry in the solution is also characteristic of plane waves. \label{WavesPlotFig}}
\end{center}   
\end{figure}

Getting the imaginary part of this expression for the metric is then trivial. This solution is quite unwieldy without any approximations and one could easily take different limits to analyze the consequences. Instead, here we opt to numerically graph the solutions. The results are plotted in Figure~\ref{WavesPlotFig} 
for different planes of the sphere intersection.

\smallskip

We choose an arbitrary system for numerical plotting, with 
$$k=5.0, \ \ \ r_I=1, \ \ \ A_\Psi= \bigg(\frac{4G k r_I ^4 \epsilon}{16 \pi r_a^2}\bigg)^{-1}\, ,$$
where $k=5.0$ is chosen to highlight the presence of oscillations in the solution. 
The solution to our theory has the property that the perturbation of the trace reversed metric tensor 
is mainly located in the $z=0$ plane, which can be somewhat expected of a plane wave centered 
at $\theta = \pi/2$, as our energy tensor indicates. It is clear the energy tensor induced by the transmission of gravitational waves between two positively curved spaces produces perturbations inside of the intersection sphere that can directly influence the trajectory of particles passing through it. Specifically, it is clear through the plots that the boundary itself sends waves inwards into the $3$-ball region, as well as outwards.

\smallskip

\smallskip
\subsubsection{Second case: external distant solution}

Now, we discuss the second case listed above, namely the far away case ($r>>r_I$). 
We can use the quadrupole formula
$$ \bar{\Psi}_{i j} \approx \frac{2G}{r} \frac{d^2 I_{ij}}{dt^2} (t_r)\, , $$
and one can easily check that the resulting trace-reversed metric in Cartesian coordinates is then given by
$$\bar{\Psi}_{\mu \nu} \approx \frac{A_\Psi G \epsilon \pi k^3 r_I^2}{8 r_a^2}  \frac{\sin(k(r -t))}{r} \begin{pmatrix}
0&0&0&0\\
0& 1/2 &0&0\\
0&0& 1/2 &0\\
0&0&0&1
\end{pmatrix}\, . $$

\smallskip
\subsubsection{Summary of behavior}

To summarize, we have considered the intersection of two positively curved spaces 
as a possible boundary for the transmission of gravitational waves. We derived an 
expression for the stress-energy tensor on the receiving universe induced by the gravitational 
waves from the emitting universe. We then considered a scenario where radius of the receiving 
space is much larger than the one of the emitting space ($r_b>>r_a$) and we considered what 
the solution to the linearized gravity field equations would look like under such stresses on the $2$-sphere boundary. We found that plane waves can be transmitted at this boundary and create oscillations inside the sphere of intersection as well apparent emission far way from the boundary using the quadruple formula. 

\smallskip

The former result has significant theoretical value, as it predicts the creation of observable universe intersections. 
That is, if one considers the case where we have multiple gravitational wave sources near the intersection on the emitting sphere, our results indicate this configuration would induce a $3$-ball of interfering gravity wave modes on the boundary between the two universes. This $3$-ball will then cause the gravitational lensing of photons if they began their path on a geodesic passing through the intersection and should thus be directly observable. 

\smallskip

Our approximations and idealized circumstances should caution one to the exactness of our results, 
but our result is an indication that this system would follow a behavior similar to that of our solutions, 
and should still create modes inside the intersection as well as likely-not-observable modes 
outside of it, falling at $1/r$ as they travel away from the intersection.

\smallskip
\subsection{Green functions for the non-flat case} 

If we were looking for a solution in the non-flat case, we would have to use the Green 
function for a $3$-sphere crossed with $\mathbb{R}$ (which gives a manifold locally 
homeomorphic to Minkowski space). To do this, we can use a result derived from \cite{BCC}. 
Using $\hat{\phi} (||\mathbf{x}-\mathbf{x'}||) = G(\mathbf{x},\mathbf{x'})$, where $||\mathbf{x}-\mathbf{x'}||$ is the Euclidean distance on $S^n$, we can write the Green function for an $S^n$ sphere as 
$$ \hat{\phi}(x) =\frac{2}{n\, {\rm Vol}(S^n)} \int_{x^2/4}^1 {}_2F_1(1,n,\frac{n}{2}+1,1-s)ds + C\, , $$
where ${}_2F_1$ is the Gauss hypergeometric function. This directly implies that, for a $3$-sphere $S_b^3$
with radius $r_b$ we have 
$$\hat{\phi}(x) =\frac{1}{4\pi^2 r_b^3} \bigg( 1 - \frac{2(x^2-2){\rm arcsin}(\sqrt{1-(x/2)^2})}{\sqrt{x^2(x^2-4)}}\bigg) + C $$
$$=-\frac{1}{4\pi^2 r_b^3} \frac{2(x^2-2){\rm arcsin}(\sqrt{1-(x/2)^2})}{\sqrt{x^2(x^2-4)}} + O(1) \, .$$

\smallskip

\medskip
\section{Gravitational waves and the spectral action with fractality}\label{GrWavesSec2}

It was shown in \cite{NOS} that, when considering the expansion of the 
spectral action functional on a $4$-dimensional spacetime manifold, 
the Euler-Lagrange 
equations take the form (for trivial cosmological constant)
$$ \frac{-1}{2\kappa} (R^{\mu\nu}-\frac{1}{2} g^{\mu\nu} R) +2\alpha (2\nabla_\lambda
\nabla_\kappa C^{\mu\kappa\nu\lambda} + C^{\mu\kappa\nu \lambda} R_{\kappa\lambda}), $$
with $C^{\mu\kappa\nu\lambda}$ the Weyl curvature tensor. Thus, the variational equations
with respect to a perturbation of the metric $g_{\mu\nu}=\eta_{\mu\nu}+\gamma_{\mu\nu}$,
to first order in $\gamma_{\mu\nu}$ 
are of the form
$$ \frac{-1}{2} \partial_\kappa \partial^\kappa \bar\gamma^{\mu\nu} + \frac{1}{2\beta^2} \partial_\kappa \partial^\kappa (\partial_\lambda \partial^\lambda \bar\gamma^{\mu\nu} +\frac{1}{3} (\eta^{\mu\nu} \partial_\lambda \partial^\lambda-\partial^\mu \partial^\nu)\gamma) \, ,$$
where $\bar\gamma_{\mu\nu}$ is the trace-reversed perturbation as
in the previous section and $\beta^2=-1/(32 \pi G \alpha)$. We refer the reader to \cite{NOS} for
a further discussion of the gauge fixing conditions. 

\smallskip

In this derivation of the Euler-Lagrange equations of \cite{NOS}, one considers
only the usual the leading terms of the spectral action expansion on a $4$-manifold,
which correspond to the order $0$, order $2$, and order $4$ terms of the heat kernel
expansion of the squared Dirac operator. As classical action functionals these terme
recover the usual Einstein--Hilbert action (with cosmological term) plus the 
Weyl curvature term and the non-dynamical topological Gauss--Bonnet term, hence
one obtains the variational equations of the form recalled above.

\smallskip

Here we consider the effect on these equations of the presence of fractality. We have
seen that there are two main effects of fractality on the spectral action. One effect corrects
the coefficients of these cosmological, Einstein--Hilbert, and Weyl curvature terms by special 
values of the fractal packing zeta function $\zeta_\cL(s)$ at $s=4$, $s=2$, and $s=0$, respectively, for
the cosmological, the Einstein--Hilbert, and the Weyl term.  This effect can be seen as altering
the effecting gravitational and effective cosmological constants, and the coupling constant 
$\alpha$ of the model. 
In the variational equations above, this change will affect the value of the $\beta^2$
coefficient. The second, more interesting effect, is the
presence of the series of log-periodic terms coming from the poles of the 
fractal packing zeta function $\zeta_\cL(s)$. We focus here on this second effect and we
show that one can interpret these terms as contributing an effecting energy-momentum tensor
to the equations of motion. Thus, through this second effect on the spectral action, 
the presence of fractality is perceived in the gravitational equations as a presence of a type of
matter that only interacts gravitationally and is otherwise dark.

\subsection{Fractality as an effective energy-momentum tensor}

Consider the spectral action expansion for the fractal packings arising from the
Robertson--Walker spacetimes for spherical forms or Bieberbach manifolds, as 
discussed in \S \ref{FracSpActSec}. We focus on the leading terms in the spectral action
expansion including the first term (at $s=s_0$ of the log-periodic series). These terms suffice to
see the effect of fractality on the gravitational waves equation. In the spherical case these terms are of the form
$$ \cS_\Lambda(\cD_{\cP_\Gamma\times \R})\sim \frac{1}{|\Gamma|} \sum_{M=0}^4
\frac{\Lambda^{4-2M}\, \frf_{4-2M}\, A_M}{(1-f_\rho f_\sigma^{2M-4})} +\frac{\Lambda^{s_0}\, \frf_{s_0}\, \Gamma(s_0/2)}{2|\Gamma| \log f_\sigma} \zeta_{\cD_{Y\times \R}}(s_0) $$
where $s_0=\frac{\log f_\rho}{\log f_\sigma}=\dim_H \cL_\Gamma$, and 
$$ A_M := \int \bigg( \int \big (\frac{1}{2}C^{-3/2,0}_{2M} + \frac{1}{4}(C^{-5/2,2}_{2M-2} - C^{-1/2,0}_{2M-2})\big)\, D[\alpha] \bigg)dt \, . $$
In the flat case they are of the form
$$ \cS_\Lambda(\cD_{\cP_{G_a}\times \R})\sim \lambda_{G_a}  \sum_{M=0}^4 
\frac{\Lambda^{4-2M}\, \frf_{4-2M}\, B_M}{(1-f_\rho f_\sigma^{2M-4})} 
+\frac{ \Lambda^{s_0}\, \frf_{s_0}\, \Gamma(s_0/2) \, \lambda_{G_a} }{ 2 \log f_\sigma } \zeta_{ \cD_{Y_a} \times \R } (s_0)\, , $$
with 
$$ B_M:=\int \bigg( \int \big (\frac{1}{2}C^{-3/2,0}_{2M} + \frac{1}{4}(C^{-5/2,2}_{2M-2} )\big)\, D[\alpha] \bigg)dt \, . $$
As shown in \cite{FKM}, the terms $A_M$, $B_M$, 
for $M=0,2,4$ give the cosmological, Einstein--Hilbert, and Weyl term. 
Thus, we see that, when we consider a variation of the metric and the resulting Euler--Lagrange 
equations, the first three terms, for $M=0,2,4$, contribute the same variational equations as in
\cite{NOS} (including the cosmological term), with coefficients scaled by the
$(1-f_\rho f_\sigma^{2M-4})^{-1}$ factors that are an effect of the presence of fractality. On the other hand,
the last term contributes a variation of the form 
$$ K\cdot\,\, \frac{\delta\, \zeta_{\cD_{Y\times \R}}(s_0)}{\delta \gamma_{\mu\nu}} \, , $$
where the factor $K$ is either
$$ K= \frac{\Lambda^{s_0}\, \frf_{s_0}\,\Gamma(s_0/2)}{2|\Gamma| \log f_\sigma} \ \ \ \text{ or } \ \ \
K= \frac{ \Lambda^{s_0}\, \frf_{s_0}\, \Gamma(s_0/2) \, \lambda_{G_a} }{ 2 \log f_\sigma } \, , $$
for the spherical and flat case, respectively. 

\smallskip

To analyze the variation 
\begin{equation}\label{ZetaVar}
T_{\mu\nu}:=  \frac{\delta\, \zeta_{\cD_{Y\times \R}}(s_0)}{\delta \gamma_{\mu\nu}} 
\end{equation} 
we use the following general fact about the zeta function of the Dirac operator.
Let $\cD=\cD_{X,g}$ be the Dirac operator on a $4$-manifold $(X,g)$, which for us will
be a Robertson--Walker spacetime $Y\times \R$ with a candidate cosmic topology $Y$. 
We consider a smooth variation of the metric $g_{\mu\nu}(u)=g_{\mu\nu}+u\gamma_{\mu\nu}$,
in a one-parameter family with parameter $u\geq 0$. We write the
corresponding Dirac operator as $\cD_u=\cD_{Y\times \R, g_{\mu\nu}(u)}$.
We assume that, under this variation of the metric, the eigenvalues $\lambda_u$ of
$\cD_u$ depend smoothly on the parameter $u$. 
We also assume for simplicity that $\Ker(\cD_u)=0$ for all $u\geq 0$.  Consider
the zeta function
$$ \zeta_{\cD_u}(s)=\Tr( |\cD_u|^{-s} )=\sum_{\lambda_u \in {\rm Spec}(\cD_u)} |\lambda_u|^{-s} \, .$$
We assume that this series converges for $\Re(s)>R$ for some sufficiently large constant $R>0$, 
for all $u\geq 0$. With the notation $Q_u=| \cD_u |$  and  $\dot{Q}_u = \frac{d}{du} Q_u$,  
 we then have
$$ \frac{d}{du} \zeta_{\cD_u}(s) =- s \sum_{\lambda_u \in {\rm Spec}(Q_u)} \dot{\lambda}_u \lambda_u^{-(s-1)}  = - s \Tr( \dot{Q}_u  \, Q_u^{-(s-1)} )\, . $$
We can then write the variation \eqref{ZetaVar} as
$$ T_{\mu\nu} = - s_0\, \Tr\left( \frac{\delta Q}{\delta \gamma_{\mu\nu}} \, \, Q^{-(s_0-1)} \right)\, . $$
We can interpret the variations \eqref{ZetaVar} of the zeta value as playing the 
role of an energy-momentum tensor in the variational equations for the spectral action gravity functional.

\bigskip
\bigskip

\subsection*{Acknowledgment} This work was supported by NSF grants DMS-1707882 and DMS-2104330. The
first author acknowledges support from the Caltech WAVE program for undergraduate research.

\end{document}